\documentclass[twocolumn]{aa}
\usepackage{graphicx}
\usepackage{txfonts}
\begin{document}
   \title{Spectroscopic confirmation of high-redshift supernovae with
the ESO VLT \thanks{Based on observations obtained at the European
Southern Observatory using the ESO Very Large Telescope on Cerro
Paranal (ESO programs 265.A-5721(A), 67.A-0361(A), 267.A-5688(A),
169.A-0382(A) and (B)). Based in part on data collected at the Subaru
Telescope, which is operated by the National Astronomical Observatory
of Japan.}\fnmsep\thanks{Figures A.1 to A.39 are only available in
electronic form via http://www.edpsciences.org.}}

   \subtitle{}
\author{C.~Lidman\inst{1}
 \and D.~A.~Howell\inst{2,3}
 \and G.~Folatelli\inst{4}
 \and G.~Garavini\inst{4,5}
 \and S.~Nobili\inst{4,5}
 \and G.~Aldering\inst{2}
 \and R.~Amanullah\inst{4}
 \and P.~Antilogus\inst{5}
 \and P.~Astier\inst{5}
 \and G.~Blanc\inst{2,6}
 \and M.~S.~Burns\inst{7}
 \and A.~Conley\inst{2,8}
 \and S.~E.~Deustua\inst{9}
 \and M.~Doi\inst{10}
 \and R.~Ellis\inst{11}
 \and S.~Fabbro\inst{12}
 \and V.~Fadeyev\inst{2}
 \and R.~Gibbons\inst{2}
 \and G.~Goldhaber\inst{2,8}
 \and A.~Goobar\inst{4}
 \and D.~E.~Groom\inst{2}
 \and I.~Hook\inst{13}
 \and N.~Kashikawa\inst{14}
 \and A.~G.~Kim\inst{2}
 \and R.~A.~Knop\inst{15}
 \and B.~C.~Lee\inst{2}
 \and J.~Mendez\inst{16,17}
 \and T.~Morokuma\inst{10}
 \and K.~Motohara\inst{10}
 \and P.~E.~Nugent\inst{2}
 \and R.~Pain\inst{5}
 \and S.~Perlmutter\inst{2,8}
 \and V.~Prasad\inst{2}
 \and R.~Quimby\inst{2}
 \and J.~Raux\inst{5}
 \and N.~Regnault\inst{2,5}
 \and P.~Ruiz-Lapuente\inst{17}
 \and G.~Sainton\inst{5}
 \and B.~E.~Schaefer\inst{18}
 \and K.~Schahmaneche\inst{5}
 \and E.~Smith\inst{15}
 \and A.~L.~Spadafora\inst{2}
 \and V.~Stanishev\inst{4}
 \and N.~A.~Walton\inst{19}
 \and L.~Wang\inst{2}
 \and W.~M.~Wood-Vasey\inst{2,8}
 \and N.~Yasuda\inst{20}\vspace{2mm}\\(The Supernova Cosmology Project)
}

\offprints{C. Lidman: \email{clidman@eso.org}}

\institute{European Southern Observatory, Alonso de Cordova 3107, Vitacura, Casilla 19001, Santiago 19, Chile
 \and E. O. Lawrence Berkeley National Laboratory, 1 Cyclotron Rd., Berkeley, CA 94720, USA
 \and Department of Astronomy and Astrophysics, University of Toronto, 60 St. George St., Toronto, Ontario M5S 3H8, Canada
 \and Department of Physics, Stockholm University,  Albanova University Center, S-106 91 Stockholm, Sweden
 \and LPNHE, CNRS-IN2P3, University of Paris VI \& VII, Paris, France
 \and Osservatorio Astronomico di Padova, INAF, vicolo dell'Osservatorio 5, 35122 Padova, Italy
 \and Colorado College, 14 East Cache La Poudre St., Colorado Springs, CO 80903
 \and Department of Physics, University of California Berkeley, Berkeley, 94720-7300 CA, USA
 \and American Astronomical Society,  2000 Florida Ave, NW, Suite 400, Washington, DC, 20009 USA.
 \and Institute of Astronomy, School of Science, University of Tokyo, Mitaka, Tokyo, 181-0015, Japan
 \and California Institute of Technology, E. California Blvd, Pasadena,  CA 91125, USA
 \and CENTRA-Centro M. de Astrof\'{\i}sica and Department of Physics, IST, Lisbon, Portugal
 \and Department of Physics, University of Oxford, Nuclear \& Astrophysics Laboratory,  Keble Road, Oxford, OX1 3RH, UK
 \and National Astronomical Observatory, Mitaka, Tokyo 181-0058, Japan
 \and Department of Physics and Astronomy, Vanderbilt University, Nashville, TN 37240, USA
 \and Isaac Newton Group, Apartado de Correos 321, 38780 Santa Cruz de La Palma, Islas Canarias, Spain
 \and Department of Astronomy, University of Barcelona, Barcelona, Spain
 \and Louisiana State University, Department of Physics and Astronomy, Baton Rouge, LA, 70803, USA
 \and Institute of Astronomy, Madingley Road, Cambridge CB3 0HA, UK
 \and Institute for Cosmic Ray Research, University of Tokyo, Kashiwa, 277 8582 Japan
}

\date{Received June 22, 2004; accepted October 4, 2004}

\abstract{We present VLT FORS1 and FORS2 spectra of 39 candidate
high-redshift supernovae that were discovered as part of a
cosmological study using Type Ia supernovae (SNe~Ia) over a wide range
of redshifts. From the spectra alone, 20 candidates are spectrally
classified as SNe~Ia with redshifts ranging from $z=0.212$ to
$z=1.181$. Of the remaining 19 candidates, 1 might be a Type II
supernova and 11 exhibit broad supernova-like spectral features and/or
have supernova-like light curves. The candidates were discovered in 8
separate ground-based searches. In those searches in which SNe~Ia at
$z \sim 0.5$ were targeted, over 80\% of the observed candidates were
spectrally classified as SNe~Ia. In those searches in which SNe~Ia with
$z > 1$ were targeted, 4 candidates with $z > 1$ were spectrally
classified as SNe~Ia and later followed with ground and space based
observatories. We present the spectra of all candidates, including
those that could not be spectrally classified as supernova.

    }

\authorrunning{Lidman, C. et al.}
\titlerunning{Spectroscopic confirmation of high-redshift SNe~Ia}

   \maketitle
%

\section{Introduction}

Over the past decade, observations of SNe~Ia have played a leading
role in measuring the expansion history of the Universe and in
constraining cosmological parameters. It was through these
observations that we discovered that the expansion is currently
accelerating and that the Universe is presently dominated by an
unknown form of dark energy with a negative equation of state
(Perlmutter et al. \cite{Perlmutter98}; Garnavich et
al. \cite{Garnavich98}; Schmidt et al. \cite{Schmidt98}; Riess et
al. \cite{Riess98}; Perlmutter et al. \cite{Perlmutter99}; Tonry et
al. \cite{Tonry03}; Knop et al. \cite{Knop03}, Riess et
al. \cite{Riess04}; Barris et al. \cite{Barris04};  for a review, see
Perlmutter and Schmidt \cite{Perlmutter03}).

When these results are combined with the results that have been
derived from the fluctuations in the cosmic microwave background
(Jaffe et al. \cite{Jaffe01}; Bennett et al. \cite{Bennett03}; Spergel
et al. \cite{Spergel03}), the properties of massive clusters (Allen,
Schmidt \& Fabian \cite{Allen02}; Borgani et al. \cite{Borgani01}) and
the large scale structure of galaxies (Hawkins et
al. \cite{Hawkins02}), a picture of a flat Universe dominated by dark
energy emerges.

Considerable effort has been directed towards extending the redshift
range over which SNe~Ia are observed.  The Hubble diagram of SNe~Ia
with $z \sim 0.5$ is degenerate to a linear combination of
$\Omega_{\rm M}$ and $\Omega_{\Lambda}$. Hence, an independent
determination of these two parameters from SNe~Ia at $z \sim 0.5$ is
not possible.  However, observations of SNe~Ia over a wide range of
redshifts and, in particular, very distant ($z \ga 1$) SNe~Ia can
break this degeneracy (Goobar and Perlmutter \cite{Goobar95}). With
this aim in mind, and following a highly successful pilot search
(Aldering \cite{Aldering}), the Supernova Cosmology Project (SCP)
started a program to discover, spectrally confirm and
photometrically monitor a substantial number of SNe~Ia with redshifts
beyond one.

In this paper, we present VLT FORS1 and FORS2 spectra of 39 candidate
high redshift supernovae. We present all spectra, including those
spectra for which a secure spectroscopic classification could not be
made.  The results of the photometric follow-up, the derived apparent
magnitudes and the implications these measurements have for cosmology
will be reported elsewhere.

\section{Observations}

\subsection{Search and discovery}

The candidates discussed in this paper were discovered during 8
separate, but not fully independent, high-redshift supernovae
searches.  The searches were divided into 4 observing campaigns that
occurred during the Northern Springs of 2000, 2001 and 2002 and the
Northern Fall of 2002. The observing campaigns, the months during
which data were taken and the telescopes used in the searches are
listed in Table \ref{tab:campaigns}.

\begin{table*}
\caption[Campaigns]{The instruments and telescopes used during 4
campaigns. In general, a single campaign consisted of multiple
searches. The prefix is used  in the internal SCP candidate names. The individual searches are numbered for easy reference.}
\label{tab:campaigns}
\begin{tabular}{llllll}
\hline
Campaign    & Months           & Instrument/Telescope   & Search Number & Search type & Prefix \\
\hline
Spring 2000 & April/May        & CFHT12k/CFHT        & 1 & Standard  & C00\\ 
&&&\\
Spring 2001 & March/April      & CFHT12k/CFHT        & 2 & Standard  & S01 \\
            & March/April      & MOSAICII/CTIO 4m Blanco   & 3 & Standard  & S01 \\
&&&\\
Spring 2002 & March to June    & CFHT12k/CFHT        & 4 & Rolling    & C02 \\
            & April/May        & MOSAICII/CTIO 4m  Blanco   & 5 & Standard  & T02 \\
            & March/April      & Suprime-Cam/Subaru   & 6$^1$ & Back-to-back & S02  \\
            & April/May        & Suprime-Cam/Subaru   & 7$^1$ & Back-to-back & S02  \\
&&&\\
Fall 2002   & October/November & Suprime-Cam/Subaru   & 8$^2$ & Standard with additional & SuF02 \\
            &                  &                        &   & wide-field monitoring& \\
\hline
\end{tabular}
\\
$^1$ Searches 6 and 7 were part of the Subaru Deep Field Project (Kodaira et al. \cite{Kodaira03}).\\
$^2$ Search 8 was part of the Subaru XMM/Newton Deep Survey (Sekiguchi et al. in preparation)\\
\end{table*}      

Following the search and discovery techniques described in Perlmutter
et al. (\cite{Perlmutter95}, \cite{Perlmutter97},
\cite{Perlmutter99}), the searches generally consisted of 2 to 3
nights of imaging to take reference images (images in which supernovae
had not yet appeared), followed 3 to 4 weeks later by an additional 2
to 3 nights of imaging to take search images (images with the
supernovae). In this paper, we refer to this type of search as a
``standard'' search, and searches 1,2, 3 and 5 were of this
type. Searches 4, 6, 7 and 8, were a variation on this theme.

The Spring 2002 CFHT search (search 4 in Table \ref{tab:campaigns}),
for example, was a ``rolling'' search, where images were taken once
every few nights during a two week period. This was followed one, two
and three months later by similar observations on the same fields. In
this way, the search images of one month become the reference images
of a later month, and, since images of the search fields are taken
several times in any one month, one automatically gets a photometric
time series without having to schedule follow-up observations
separately, as one must do in a standard search.

The Subaru searches during the Spring and Fall of 2002 (searches 6, 7
and 8 in Table \ref{tab:campaigns}) also differed from the standard
search. Searches 6 and 7 were ``back-to-back'' searches, in which the
search images of the first search  (search 6)became the
reference images for the second search  (search 7). Search
8 was a standard search that was then immediately followed with
additional observations with the same instrument and telescope. This
search offered the advantage of allowing us to follow several
candidates simultaneously, rather than following candidates
individually, as is the case with the standard search.

The data were processed to find objects that had brightened and the
most promising candidates were given an internal SCP name and a
priority.  The priority is based on a number of factors: the
significance of the detection, the percentage increase in the
brightness, the distance from the center of the apparent host, the
brightness of the candidate and the quality of the subtraction. The
candidates were then distributed to teams working at the Gemini, Keck,
Paranal, and Subaru Observatories for spectroscopic
confirmation. The distribution was handled centrally and was done
according to the priority of the candidates, the results from data
that had been taken during previous nights, the capability and
availability of the instruments and the telescopes at each of the
observatories and the weather conditions at the individual
observatories at any one time. Hence, the factors that affect whether
or not a candidate is observed at any one observatory are complex and
such factors would have to be taken into account in any statistical
analysis. Note that these searches for extremely high-redshift
supernovae are in this way more complex than previous searches that
have been reported in our previous papers. 

A preliminary analysis of the spectroscopic data is done within a day
of when the data are taken - a more careful analysis is done
later. Only those candidates that are confirmed as SNe~Ia are
then scheduled for follow-up observations, which consist of
photometric monitoring in at least two broad-band filters during the
first two months immediately following the discovery and final
reference images, which are taken about one year later. These data are
used to measure the peak magnitudes, the light curve widths, which 
are used to correct the peak magnitudes, and the
colours of the SNe~Ia. In some searches, such as searches 4 and
8 in Table \ref{tab:campaigns}, the optical follow-up is integrated
into the search.

The aim of the spectroscopic follow-up is not to confirm as many
SNe~Ia as possible, but to provide a number of spectrally classified
SNe~Ia (typically four SNe~Ia per campaign) to be scheduled for HST
 and ground-based follow-up within one to two days of the end of
the spectroscopic runs. Without exception, we succeeded  in
providing a sufficient number of SNe~Ia for the follow-up.

We are presenting the spectra of all candidates that were observed
with the ESO VLT, so there are a number of candidates that have only
an internal SCP name. The SCP name consists of a prefix, which
indicates at which telescope the candidate was discovered, and a
running number. A list of prefixes is given in Table
\ref{tab:campaigns}.  The spectra of candidates that were not observed
at the ESO VLT will be reported elsewhere.

\subsection{Spectroscopic follow-up}

The long slit spectroscopic modes of FORS1 and FORS2 (Appenzeller et
al. \cite{Appenzeller98}) on the ESO VLT were used to observe high
priority candidates.  For the purpose of long-slit spectroscopy, FORS1
and FORS2 are very similar instruments. The principle difference is
that the detector in FORS1 is a single 2kx2k Tektronix CCD, while the
detector in FORS2 is a mosaic of two 2kx4k red-optimized MIT CCDs. The
FORS2 detector is more sensitive than the FORS1 detector, especially
at red wavelengths. The availability of the red optimized CCDs in
FORS2  after March 2002 made it possible to observe and confirm
candidates at $z\sim1.2$.

The dates during which the VLT spectroscopic observations took place
and the redshift interval over which SNe~Ia were targeted for VLT
follow-up are listed in Table \ref{tab:VLTobs}.

\begin{table*}
\caption[VLT Campaigns]{Instruments and telescopes that were used in the spectroscopic follow-up.} 
\label{tab:VLTobs}
\begin{tabular}{lllll}
\hline
Campaign    & Instrument and Telescope     & Dates                    & Observing Mode & Redshift Interval\\ 
\hline					   
Spring 2000 & FORS1 on Antu (VLT-UT1)      & 12 May 2000              & Service & $z=0.3-0.7$\\
Spring 2001 & FORS1 on Antu (VLT-UT1)      & 21-22 April 2001         & Visitor & $z=0.3-0.7$\\
            &                              & 27 April and 28 May 2001 & Service & $z=0.3-0.7$\\
Spring 2002 & FORS2 on Yepun (VLT-UT4) and & April-August 2002        & Service & $z=0.3-1.2$\\
            & FORS1 on Melipal (VLT-UT3)   & 11-12 May 2002           & Service & $z=0.3-1.0$\\
Fall   2002 & FORS2 on Yepun (VLT-UT4)     & 7-11 November  2002      & Service & $z>1$\\
\hline
\end{tabular}
\end{table*}      

Three grisms (300V, 300I and 600z) and two slit widths (0.7 and 1.0
arc seconds) were used for the observations. In general, the grism was
chosen to match the expected redshift of the candidate and the slit
was matched to the seeing. The 300V grism was used with the GG435
order-sorting filter and the 300I and 600z grisms  were used with the
OG590 order-sorting filter.

Nearly all targets were acquired in the same way. The slit was placed
through the candidate and a relatively bright and nearby pivot star.
There were only three exceptions: SuF02-026 and \object{SN 2002lc}
were observed together and \object{SN 2000fr} was acquired
directly. The observational details are listed in Table
\ref{tab:VLTsummary1}. Generally speaking, three exposures with small
offsets along the slit were taken for each candidate. Exceptions
occurred when observations were aborted because we thought that we had
sufficient data to identify the candidate or when we integrated longer
for the fainter candidates.

\begin{figure}
\centering
\resizebox{\hsize}{!}{\includegraphics{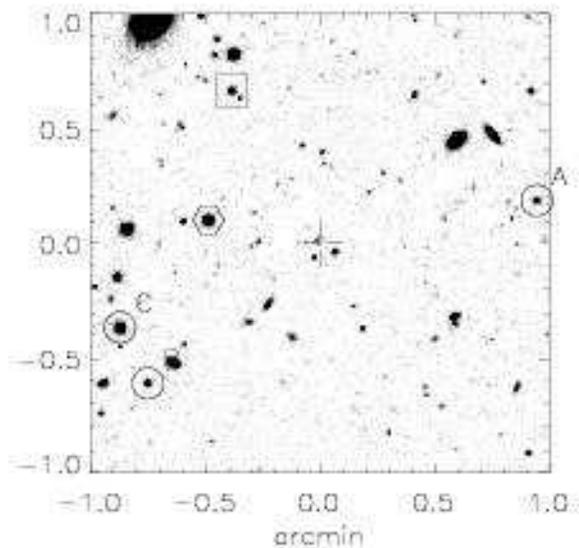}}

\caption{A finding chart centered on \object{SN 2000fr}
(C00-008). North is up and East is to the left. The candidate is
marked with a cross and bright pivot stars are marked with either a
box or a hexagon. Fainter pivot stars are circled and labelled
alphabetically. The pivot star that was used during the acquisition is
recorded in Table \ref{tab:VLTsummary1}. The finding charts of other
candidates are available in the appendix.}

\label{fig:FC_C00-008}
\end{figure}

 Finding charts showing both the candidate and the pivot star are
displayed in Fig. \ref{fig:FC_C00-008} and in Figs. \ref{fig:FC_S01-004} to
\ref{fig:FC_SuF02-083}. Candidates are
marked with a cross and bright pivot stars are marked with either a
box or a hexagon. Fainter pivot stars are circled and labelled
alphabetically. The pivot star that was used during the acquisition is
recorded in Table \ref{tab:VLTsummary1}. In all finding charts, North
is up and East is to the left.

In addition to the 39 candidates that were observed soon after they
were discovered, the spectrum of the probable host galaxy of T02-047,
which was observed several months after it was discovered, is also
reported. The light curve of T02-047 indicates that it is a 
supernova.

\begin{table*}
\caption[VLT SNe]{Summary of the observations. The SCP name is an
internal name used by the SCP and is reported here as not all
candidates have an IAU name.}
\label{tab:VLTsummary1}
\begin{tabular}{lllllrrrrr}
\hline
SCP      &IAU   & Campaign  &Coordinates of           & Pivot & Offset       & PA     & MJD     & Grism &Exp. \\
Name     &Name  &           &the candidate            & Star  &              &        &         &       &(sec)\\
\hline			      			                           
C00-008  &\object{SN 2000fr}& Spring 00 & 13 42 00.14 +04 43 42.4 & -$^1$ & -$^1$        &  40.00 & 51676.2 & 300V  &7200\\
\hline			      			                           			 
S01-004  &\object{SN 2001gl}& Spring 01 & 14 01 16.60 +05 12 48.9 & Hex   & -6.07,  0.16 &  92.41 & 52021.2 & 300V  &3600\\
S01-005  &\object{SN 2001gm}& Spring 01 & 14 01 51.18 +05 05 38.5 & Hex   & 23.92, 24.80 &  43.97 & 52021.3 & 300V  &2400\\
S01-007$^4$&\object{SN 2001go}& Spring 01 & 14 02 00.95 +05 00 59.2 & Hex   & 34.22, -4.46 &  97.42 & 52021.3 & 300V  &2400\\
S01-007$^4$&\object{SN 2001go}& Spring 01 & 14 02 00.95 +05 00 59.2 & Hex   & 34.22, -4.46 &  97.43 & 52027.2 & 300V  &7200\\
S01-007$^4$&\object{SN 2001go}& Spring 01 & 14 02 00.95 +05 00 59.2 & Hex   & 34.22, -4.46 &  97.43 & 52058.2 & 300V  &9000\\
S01-017  &\object{SN 2001gr}& Spring 01 & 10 04 23.27 +07 40 48.3 & Box   & -10.05,-24.64 &  22.19 & 52021.0 & 300V  &3600\\
S01-028  &\object{SN 2001gs}& Spring 01 & 10 00 52.68 +06 07 09.3 & Box   & 11.89,-25.79 & -24.75 & 52022.1 & 300V  &4800\\
S01-031  &\object{SN 2001gu}& Spring 01 & 10 03 28.61 +07 24 38.9 & Hex   & 37.16,  3.32 &  84.89 & 52021.1 & 300V  &4800\\
S01-033  &\object{SN 2001gw}& Spring 01 & 15 43 45.86 +07 57 50.3 & Hex   &-14.09, 32.22 & 156.37 & 52021.4 & 300V  &1200\\
S01-036  &\object{SN 2001gy}& Spring 01 & 13 57 04.54 +04 30 59.8 & Hex   & 21.49,  0.43 &  88.85 & 52021.3 & 300V  &2400\\
S01-037  &-     & Spring 01 & 13 55 51.17 +04 48 06.7 & Hex   &-56.87, 32.41 & 119.68 & 52021.1 & 300V  &3600\\
S01-054  &\object{SN 2001ha}& Spring 01 & 10 06 33.50 +07 38 03.2 & Hex   & 13.51, 22.72 &  30.74 & 52022.0 & 300V  &3600\\
S01-065  &\object{SN 2001hc}& Spring 01 & 09 44 31.52 +08 02 02.8 & Hex   &-14.17, 46.46 & -16.96 & 52022.1 & 300V  &1800\\
\hline			      			                           			 
S02-000  &\object{SN 2002fd}& Spring 02 & 14 03 54.08 +04 59 49.0 & Box   & -6.48,  2.62 & 112.01 & 52376.1 & 300V & 600\\
S02-001  &-     & Spring 02 & 14 03 56.42 +05 23 16.6 & Hex   &-27.85, 39.10 & 144.54 & 52376.3 & 300I &2700\\
S02-002  &\object{SN 2002fe}& Spring 02 & 14 04 18.16 +05 19 25.6 & B     & -8.49,  1.52 & 100.15 & 52376.2 & 300I &2700\\
S02-025  &-     & Spring 02 & 13 57 50.11 +05 17 25.5 & Hex   &  0.09, 14.94 &   0.34 & 52376.2 & 300I &2700\\
S02-075  &\object{SN 2002fg}& Spring 02 & 13 24 25.92 +27 44 33.9 & Hex   & 57.74,-22.44 & -68.76 & 52431.0 & 300V &7200\\
C02-016  &\object{SN 2002fr}& Spring 02 & 14 00 46.40 +04 33 41.4 & Hex   &-12.67  10.57 & 145.65 & 52400.3 & 300V & 900\\
C02-028  &\object{SN 2002fm}& Spring 02 & 14 00 29.75 +04 46 50.1 & B     &-27.76, 21.91 & 128.28 & 52413.0 & 300V &1800\\
C02-030  &\object{SN 2002fp}& Spring 02 & 14 02 18.40 +04 47 05.9 & Hex   &  1.69,-21.86 &  -4.43 & 52407.1 & 300I &3600\\
C02-031  &-     & Spring 02 & 14 01 38.07 +04 38 02.2 & Box   &  0.88, 38.36 & 178.69 & 52406.1 & 300I &3600\\
C02-034  &-     & Spring 02 & 14 00 30.75 +05 13 55.6 & Hex   & 5.82,-22.77 & -14.34 & 52413.0 & 300V &1800\\
T02-015  &\object{SN 2002gi}& Spring 02 & 13 57 12.25 +04 33 16.8 & Hex   &  1.17,-68.78 &  -0.97 & 52407.2 & 300I &7200\\
T02-028  &\object{SN 2002gj}& Spring 02 & 15 36 25.48 +09 28 18.2 & Hex   &-40.55, 62.58 & 147.06 & 52413.2 & 300V &3000\\
T02-029  &\object{SN 2002gk}& Spring 02 & 15 37 07.47 +09 36 18.7 & C     &-20.24, 16.98 & 129.99 & 52413.3 & 300V & 900\\
T02-030  &\object{SN 2002gl}& Spring 02 & 15 43 24.40 +07 53 57.5 & Hex   &  2.32, 43.98 &-176.98 & 52413.1 & 300V &3000\\
T02-047$^3$  &- & Spring 02 & 15 36 29.88 +09 38 42.8 & Hex   & -42.94,-29.25 &  55.74 & 52494.0 & 300V &3000\\
\hline			      			                           			 
SuF02-002&\object{SN 2002kq}& Fall 02   & 02 17 12.24 -04 55 08.7 & Hex   &-21.25, -4.06 &  79.18 & 52586.1 & 300I &3600\\
SuF02-005&-     & Fall 02   & 02 18 35.67 -04 31 11.2 & A     & 18.26,  0.06 & -90.19 & 52586.1 & 300I &3600\\
SuF02-007&-     & Fall 02   & 02 18 52.32 -05 01 14.0 & Hex   &  6.63,-40.66 &  -9.26 & 52588.7 & 300I &13200\\
SuF02-012&\object{SN 2002lc}& Fall 02   & 02 18 51.60 -04 47 25.7 &Hex$^2$&-19.04, 14.75 &   8.81 & 52588.3 & 600z &7200\\
SuF02-017&\object{SN 2002kn}& Fall 02   & 02 16 45.71 -05 09 51.2 & Hex   &-48.24, -0.53 &  89.37 & 52590.2 & 300I &1800\\
SuF02-025&\object{SN 2002km}& Fall 02   & 02 16 23.93 -04 49 29.4 & Box   & -7.30,  5.14 & 125.15 & 52588.1 & 300I &3600\\
SuF02-026&-     & Fall 02   & 02 18 51.90 -04 46 57.4 &Hex$^2$&-19.04, 14.75 &   8.81 & 52588.3 & 600z &7200\\
SuF02-028& \object{SN 2002kz} & Fall 02   & 02 16 56.36 -05 00 58.1 & Hex   & 26.08,-47.36 & -28.84 & 52587.1 & 300I &3600\\
SuF02-051&-     & Fall 02   & 02 17 27.47 -04 40 45.3 & C     &-11.62, -2.10 &  79.76 & 52586.3 & 300I &3600\\
SuF02-060&\object{SN 2002kr}& Fall 02   & 02 17 34.51 -04 53 46.6 & A     & 19.82,-17.49 & -48.75 & 52587.2 & 300I &3600\\
SuF02-065&\object{SN 2002ks}& Fall 02   & 02 17 34.53 -05 00 15.4 & A     &-28.15,-23.05 &  50.69 & 52586.2 & 300I &3600\\
SuF02-081&-     & Fall 02   & 02 20 07.49 -05 08 27.4 & A     & 51.24,-20.89 & -67.82 & 52589.2 & 300I &9600\\
SuF02-083& \object{SN 2002kx}     & Fall 02   & 02 18 06.21 -05 00 38.8 & Box   &-35.39,  1.64 &  92.65 & 52587.1 & 300I &7200\\
 
\hline
\end{tabular}
\\$^1$ Centered on the candidate.\\
$^2$ The slit passed through SuF02-012 and SuF02-026\\
$^3$ T02-047 was observed several months after maximum light\\
$^4$ \object{SN 2001go} was observed at three epochs\\

\end{table*}             

\section{Data reduction and classification}

Standard IRAF\footnote{IRAF is distributed by the National Optical
Astronomy Observatories, which are operated by the Association of
Universities for Research in Astronomy, Inc., under the cooperative
agreement with the National Science Foundation.} procedures were used
to process the data. The bias was estimated by fitting the over-scan
region with a low-order polynomial, flat-fielding was done with lamp
flats that were first cleaned of parasitic light, and wavelength
calibration was performed with arc frames.
 
For observations with the 300V grism, fringing is not a significant
limitation in the data so the two-dimensional spectra were combined
(with suitable clipping to remove cosmic rays) and the sky was removed
by estimating the background flux on either side of the object trace.

For observations with the 300I and 600z grisms, fringing is more
significant. If it is not treated carefully, the systematic error from
fringing residuals can be large. Before combining individual spectra,
a fringe correction was applied to the data. The fringe correction
consists of the following steps:
 
\begin{itemize}
 
\item For a given spectrum in which the object was at a certain slit
position, a group of similar spectra (the same grism, order sorting
filter and slit) with objects at different slit positions was collected.
 
\item Fringe frames are created by clipping object pixels and
averaging the remainder.  Since the intensity of night sky lines can
vary with respect to one another, each column (the spatial direction
of the spectra are along columns) is treated
individually. Instrumental flexure for FORS1 and FORS2 is small, so
some fringe frames were created from data that were taken on different
nights.
 
\item The fringe frames are subtracted from the data after suitable
scaling. Again, each column is treated individually.
 
\item An average sky spectrum (calculated by averaging along columns
of the fringe frame) is added back to the data. This helps with the
clipping of cosmic rays when the two dimensional spectra are combined.
 
\item The data are combined with suitable clipping for cosmic rays and
the sky is removed by estimating the flux on either side of the
object trace.

\end{itemize}
 
The resulting two-dimensional sky-subtracted spectra are free of
fringes at the expense of a slight reduction in the statistical
signal-to-noise ratio.

The spectra of the candidates and, in some cases, the spectra of the
hosts were then extracted and calibrated in wavelength and flux. In all
cases, we also produce error spectra, which  are used to estimate the
significance of spectral features.

The signal-to-noise ratio varies from very low ($\la 1$ per wavelength
element) to moderately good ($ \ga 10$ per wavelength element). The
spectra with the highest signal-to-noise ratios are studied in
more detail in Garavini et al. (in preparation).  The quality of some
of the high-redshift SN~Ia spectra that are presented in this paper,
\object{SN 2002ks} at z=1.181 is one example (see
Fig. \ref{fig:FC_SuF02-065}), matches the quality of spectra that have
been taken with HST (Riess et al. \cite{Riess04}).

\subsection{Classification}

At high redshifts ($z \ga 0.4$), the broad Si~II feature at 6150~\AA,
which is  one of the defining spectral signatures of the SN~Ia
class, is often outside the wavelength range covered by the
spectra.  Therefore, we use other features, such as the Si~II
feature at $\sim 4000~\AA$ and the S~II "W" feature at $\sim
5400~\AA$, which are only seen in SNe~Ia, to spectrally identify SNe~Ia
when the Si~II feature at $\sim 6150~\AA$ is not visible (Hook et
al. in preparation).

We also use a library of galaxy and nearby supernova spectra to fit
the spectra of candidates (Howell and Wang, \cite{Howell02}), and we
use these fits to classify candidates when the Si~II and S~II features
cannot be clearly identified. A representative sample of galaxy
spectral templates ranging from early to late types and more than 250
spectra of nearby supernova of all types currently make up the
library. For a given candidate and a given nearby supernova, the fit
determines the fraction of host galaxy light, the host galaxy spectral
type, the redshift of the supernova and the amount of reddening. The
quality of the fit is quantified with the reduced $\chi^2$, which has
little meaning in an absolute sense. The finite size of the spectral
library and systematic calibration errors in both low and high samples
mean that the reduced $\chi^2$ is always greater than one. However, it
is useful for ordering the fits.

The fit can be constrained by using additional information. For
example, if there is no apparent host - SN 2001ha is one example -
the fraction of light from the host galaxy is set to zero in the
fit. Alternatively, if there is a host and if the redshift of the host
galaxy is known, the redshift of the candidate is fixed to this value
and the redshift is reported to three decimal places. Otherwise, the
redshift is determined from the fit and is reported to two decimal
places. 

The fits are ordered according to the reduced $\chi^2$ and the first
dozen fits are inspected visually. If Silicon or Sulfur are clearly
identified or if the spectrum can be matched with the spectrum of a
nearby SN~Ia, the candidate is assigned the label ``Ia'' and the
classification is considered secure. Less secure candidates are
labelled ``Ia$^*$''. The asterisk indicates some degree of
uncertainty. This usually means that we see spectral features that are
consistent with a SN~Ia classification and can find an acceptable
match with a nearby SN~Ia, but that other types, such as a SN~Ic,
also result in acceptable matches and cannot be excluded. For example,
the spectra of SNe~Ia 10 days after maximum light resemble the spectra
of some SNe~Ic at maximum light, especially around the 4000~\AA\
region. In these cases, the light curve can be used to estimate the
epoch at which the spectra were taken and to distinguish between the
two possibilities.  A good example is the spectrum of \object{SN
2002gj}, which can be matched with either a SN~Ia or a SN~Ic. By using the
light curve to constrain the epoch, \object{SN 2002gj} is clearly a
SN~Ia.

The best matching nearby supernova is chosen by visually examining the
best dozen fits and selecting the best qualitative fit. For candidates that
are classified as ``Ia'' or ``Ia*'', the best matches are listed in
Table \ref{tab:results} and plotted in Fig. \ref{fig:Spec_C00-008} and
in Figs. \ref{fig:FC_S01-004} to \ref{fig:FC_SuF02-083}.

A simple dash indicates that a classification based on the VLT
spectrum alone could not be made. This does not mean that these
candidates are not supernovae.  Some of the unclassified candidates
show broad supernova-like features in their spectra, while others have
well measured light curves.  Candidates that fall in  the former
category include \object{SN 2001gl}, \object{SN 2002lc} and
SuF02-007. Candidates that fall in the latter category include
\object{SN 2002fr}, \object{SN 2002fm},  C02-034, T02-047,
\object{SN 2002kq}, SuF02-007, \object{SN 2002lc},  SuF02-026,
 \object{SN 2002kz}, SuF02-051 and \object{SN 2002kx}.

\section{Results}

The results of the four SCP campaigns are summarized in Table
\ref{tab:results}, where each candidate is identified with the
internal SCP name. The IAU name, the spectral classification, the
redshift and the best template match are also listed if these items
are available.  Beside each classification, we also give the
reason for the classification. If Si~II at 4000~\AA\ or 6150~\AA\ or
S~II at 5400~\AA\ were identified then we attach the label ``Si~II''
to the classification. If the classification was done from the fit, we
attach the label ``SF'', which stands for spectral fit.  The
comments provide additional information. For example, in the cases
where a classification from the VLT spectrum could not be made, we
note down any relevant information from the light curve.

The spectrum of \object{SN 2000fr} is shown in
Fig. \ref{fig:Spec_C00-008} and the spectra of all other candidates
are presented in the appendix\footnote{The appendix is only available in
the electronic version of the journal.}.  In some cases, we have
compensated for telluric absorption by dividing the spectra with a
suitably scaled spectrum of the telluric absorption on Cerro
Paranal. In any case, the location of telluric absorption features
(usually the A and B bands and, for the 300I and 600z grisms, the
telluric feature that starts at 9300~\AA) are marked in all spectra
with the symbol $\oplus$. The location of night sky subtraction
residuals (usually from the bright night sky lines at 5577, 5890, 6300
and 6364~\AA) are marked with  the letters ``NS''. Spectroscopic
features from the host galaxy are marked where appropriate.

In the comparison plots, nearby SNe are shown in blue, while the
observations minus the host galaxy template are shown in black.  
In most cases, the observations have been re-binned to 20 $\rm~\AA$.

The results\footnote{T02-047 is not considered since the spectrum was 
taken many months after it was discovered.} are summarized as follows:
 
\begin{itemize}
 
\item 39 candidates were observed.
 
\item 20 candidates are classified as SNe~Ia.

\item 1 candidate is classified as a possible Type II supernova.

\item Of the remaining 18 unclassified candidates, labelled with a
dash in Table \ref{tab:results},  11 have broad supernova-like
 spectral features and/or have supernova-like light curves. 
One of these 11 candidates - SuF02-026 - has two strong emission lines
that cannot be identified.

\item Of the remaining 7 candidates, 5 have neither clear supernova
features nor sufficient photometric follow-up to measure a light
curve, but posses a galaxy component from which a redshift can be
determined.

\item  The remaining 2 candidates have featureless continua.

\end{itemize}

\begin{figure}
\centering
\includegraphics[width=8cm]{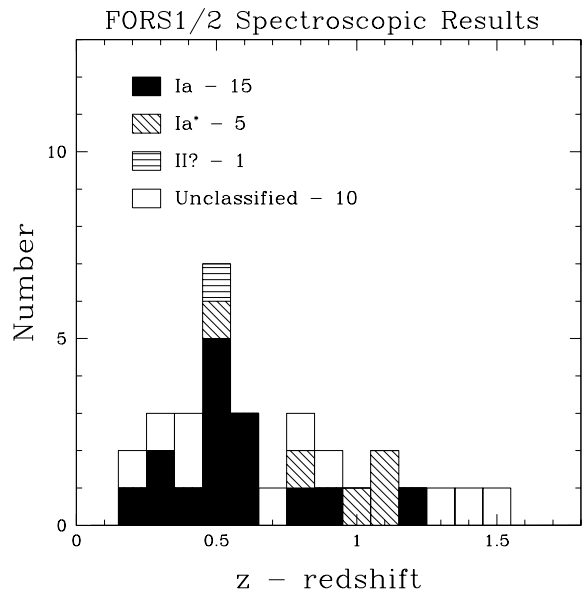}
\caption{A redshift histogram of the candidates.}
\label{fig:hist}
\end{figure}

A redshift histogram is shown in Fig. \ref{fig:hist}. Of the eight
candidates that do not have redshifts, three have broad spectral
features and  two have supernova-like light curves.
 
\begin{table*}
\caption[VLT SNe]{Classifications, redshifts and nearby supernova matches. Redshifts
based on the host are quoted to an accuracy of three decimal
places. Redshifts based on the fit are quoted to two. The comments provide additional information on each candidate.}
\label{tab:results}
\begin{tabular}{lllrllll}
\hline\hline
SCP      &IAU   &\multicolumn{2}{c}{Spectroscopic}  &Redshift & Template    &Comments\\
Name     &Name  &\multicolumn{2}{c}{Classification}  &         & Match       & \\
\hline\hline
C00-008      &\object{SN 2000fr}& Ia & Si~II    & 0.543   & \object{SN 1990N}  -7 days &            \\ 
\hline
S01-004$^2$  &\object{SN 2001gl}& -  & -             & -       & -              & Broad spectral features and no detectable host.$^3$\\
S01-005$^2$  &\object{SN 2001gm}& Ia & Si~II    & 0.478   & \object{SN 1992A}  +5 days & \\
S01-007$^2$  &\object{SN 2001go}& Ia & Si~II    & 0.552   & \object{SN 1992A}  +5 days & \\
S01-017      &\object{SN 2001gr}& Ia & SF       & 0.540   & \object{SN 1996X}  +2 days & \\ 
S01-028      &\object{SN 2001gs}& -  & -             & 0.658   &                & Host dominated. \\ 
S01-031      &\object{SN 2001gu}& Ia & SF       & 0.777   &  \object{SN 1999bp}  +1 day & \\ 
S01-033      &\object{SN 2001gw}& Ia & Si~II    & 0.363   & \object{SN 1989B}  -1 day  & \\
S01-036      &\object{SN 2001gy}& Ia & Si~II      & 0.511   & \object{SN 1990N}  -7 days & \\
S01-037      &-     & -              & -     & -  &                & Featureless blue spectra.\\
S01-054      &\object{SN 2001ha}& Ia & Si~II      & 0.58    & \object{SN 1981B}  Max.    & No detectable host.$^3$\\
S01-065      &\object{SN 2001hc}& Ia & Si~II      & 0.35    & \object{SN 1981B}  Max.    & Faint host.\\
\hline
S02-000      &\object{SN 2002fd}& Ia & Si~II      & 0.279   & \object{SN 1990N} -7days   & \\
S02-001      &-     & -           & -   & 1.424   &                &              \\   
S02-002      &\object{SN 2002fe}& Ia$^*$ & SF   & 1.086   & \object{SN 1999ee} -8 days &              \\   %
S02-025      &-     & -          & -    & -       &                & Featureless             \\   
S02-075      &\object{SN 2002fg}& Ia$^*$ & SF   & 0.78   &  \object{SN 1999bm} +6 days     &              \\   
\hline
C02-016      &\object{SN 2002fr}& - & -             & 0.303?  &                & Blue spectrum. Supernova-like light curve.\\   
C02-028      &\object{SN 2002fm}& - & -             & 0.448   &                & Host dominated. Supernova-like light curve. \\   
             &         &               &   &         &                & Small percentage increase. \\ 
C02-030      &\object{SN 2002fp}& -  & -   & 0.352   &                &              \\   
C02-031      &-        & II?   & SF   & 0.541  &  \object{SN 1999em} Max.& Host dominated. Small percentage increase. \\   
C02-034      &-        & -          & -    & 0.243   &                & Host dominated.  Small percentage increase. \\   
\hline
T02-015      &\object{SN 2002gi}& Ia & Si~II      & 0.912   & \object{SN 1996X}  +2 days &                \\ 
T02-028      &\object{SN 2002gj}& Ia$^*$ & SF  & 0.45    & \object{SN 1992A}  +9 days & Small percentage increase.               \\ 
T02-029      &\object{SN 2002gk}& Ia & Si~II      & 0.212   & \object{SN 1992A}  +6 days &                \\ 
T02-030      &\object{SN 2002gl}& Ia & Si~II      & 0.510   & \object{SN 1989B}  -5 days &                \\ 
T02-047$^1$  &-     & -          & -    & 0.489   &                & Supernova-like light curve.  \\ 
\hline
SuF02-002    & \object{SN 2002kq} & -   & -         & 0.823   &                & Supernova-like light curve.               \\ 
SuF02-005    &-       & -         & -   & 0.863   &                & Small percentage increase.  \\ 
SuF02-007    &-       & -        & -    & 1.16?   & \object{SN 1981B} Max.     & No host.$^3$ Broad spectral features and \\
             &        &          &    &         &                   & supernova-like light curve.\\ 
SuF02-012    & \object{SN 2002lc} & -  & -          & 1.3?   & \object{SN 1999aa} -3 days & Broad spectral features. Supernova-like light curve.\\ 
SuF02-017    & \object{SN 2002kn} & Ia$^*$ & SF  & 1.03    & \object{SN 1999bm} +3 days & Faint host. Supernova-like light curve. \\ 
SuF02-025    & \object{SN 2002km} & Ia & Si~II    & 0.606   & \object{SN 1990N}  -7 days &                \\ 
SuF02-026    &-       & -            & -  & -     &                & Two unidentified lines.  Supernova-like light curve. \\ 
SuF02-028    & \object{SN 2002kz} & -   & -         & 0.347   &                & Host dominated.  Supernova-like light curve.               \\ 
SuF02-051    &-       & -            & -    & -   &                & Featureless. No detectable host$^3$ and\\
             &        &              &      &    &                   & supernova-like light curve.  \\ 
SuF02-060    & \object{SN 2002kr} & Ia$^*$ & SF  & 1.063   & \object{SN 1981B} Max.     & Host dominated. Small percentage increase.\\
             &           &              &     &     &                   & Supernova-like light curve. \\ 
SuF02-065    & \object{SN 2002ks} & Ia &  Si~II    & 1.181   & \object{SN 1981B} Max.     &                \\ 
SuF02-081    &-       & -       & -     & 1.478   &                & Narrow light curve     \\ 
SuF02-083    & \object{SN 2002kx}  & -     & -            & 1.272   &                & Small percentage increase. Supernova-like \\
             &        &              &     &    &                   & light curve.\\ 
 
\hline
\end{tabular}
\\$^1$ T02-047 was observed several months after maximum light\\
$^2$ These candidates were discovered at the CFHT. The remainder of the candidates with the
prefix ``S01'' were discovered at CTIO.\\
$^3$ This refers to the reference images. On deeper images, a host might become visible.\\
\end{table*}
				   

\begin{figure}
\centering

\resizebox{\hsize}{!}{\includegraphics{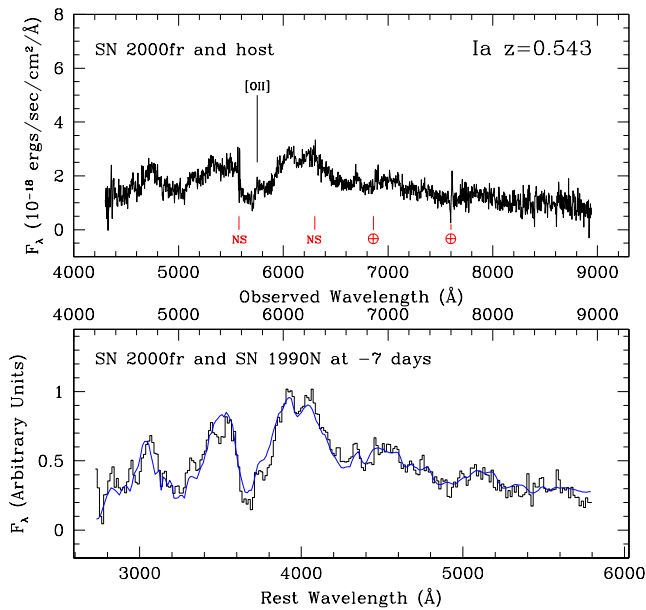}}

\caption{A spectrum of \object{SN 2000fr}, a SN~Ia at $z=0.543$ with
unambiguous detection of Si~II at 4000~\AA. In the upper spectrum, the
unbinned spectrum of the candidate is plotted in the observer's frame
and is uncorrected for host galaxy light. Night sky subtraction
residuals are marked with the letters ``NS'' and telluric absorption
features are marked with the symbol $\oplus$. In the lower spectrum,
contamination from the host is removed and the spectrum is
rescaled and rebinned by 20~\AA. This spectrum is plotted in black and it
is plotted in both the rest frame (lower axis) and the observer's
frame (upper axis). For comparison, the best fitting nearby supernova
is plotted in blue. \object{SN 2000fr} was first identified as a SN~Ia
in a very early spectrum that was taken with LRIS on the KeckII
telescope on 2001 May 4th and was subsequently observed again with
FORS1. The FORS1 spectrum has one of the highest S/N ratios of all
securely identified supernovae in this paper. \object{SN 2000fr} was
followed in the J-band with ISAAC \cite{Nobili04} and
in the R- and I-bands with HST and ground-based telescopes (Knop et
al. \cite{Knop03}). The J-band observations, which corresponds to the
rest-frame I-band, show a clear second maximum about 30 days after the
first maximum. A spectrum of the host galaxy (not shown here)
shows emission in [OII] and [OIII] as well as Balmer absorption
lines.}
\label{fig:Spec_C00-008}
\end{figure}               

\section{Discussion}

In terms of classifying candidates from the spectra alone, there is a
clear correlation between the number of candidates that are classified
as SNe~Ia and the redshift at which SNe~Ia were targeted. In searches
1, 2, 3 and 5, (See Table. \ref{tab:campaigns}), where SNe~Ia at
$z\sim0.5$ were targeted for VLT spectroscopic follow-up, 13 out of 16
candidates (excluding T02-049) are classified as SNe~Ia.  In search 8,
where SNe~Ia with $z > 1$ were targeted for VLT spectroscopic
follow-up, 4 out of 13 candidates are classified as SNe~Ia.

There are multiple reasons for the large difference. The aim of
search 8 was to find several $z > 1$ SN Ia and the strategy of the
spectroscopic follow-up was tuned to make the best use of the time
that was available. In general, each candidate was first observed for
one hour. Candidates that were found to have $z < 1$ were no longer
observed. This included SuF02-002, SuF02-005, \object{SN 2002km} and
SuF02-028. In one case (\object{SN 2002km}) a secure classification
could be made. In the other three cases, a supernova might have been
identified if we had chosen to integrate longer. Alternatively, if the
candidate showed evidence for broad features or if the redshift from
host galaxy lines (in particular [OII]) placed the host at $z >1$,
the candidates were re-observed during later nights. As the amount of
allocated time was limited, not all promising candidates could be
followed. These factors led to a lower overall yield at $z < 1$, but they
also enabled us to confirm several $z>1$ SNe~Ia and to obtain their
redshifts.

Nevertheless, the spectroscopic confirmation of $z>1$ SNe Ia is
challenging.  At $z \sim 1$, SNe Ia are about 1.5 magnitudes fainter
than at $z \sim 0.5$.  Additionally, the spectral features that one
uses for classification shift further and further into the red where
sky subtraction can be difficult because of variable night-sky
emission and detector fringing. This can be partially compensated by
integrating longer and using instruments and telescopes that are
efficient in the 600 to 1000 nm spectral region. Although the spectra
of SNe~Ia show significant features short-ward of the broad CaII
feature at 3900~\AA\ that could be used to aid the classification, the
lack of good quality UV spectra for nearby supernovae of all types
 means that these features cannot be used without using
features that are further into the red.

For $z>1$, which have peak magnitudes near $\rm I\sim25$, an additional
source of ambiguity appears. Given the typical signal-to-noise that
one can achieve with state-of-the-art instrumentation, one can
sometimes match the spectra equally well with  SNe~Ia at two different
redshifts. Fortunately, host galaxy lines, either [OII] or H and K or
sometimes all three, can be used to measure  a precise redshift in
most cases. In this paper, SNe 2002fe, 2002gi, 2002kn, 2002kr and
2002ks fall into this category. However, in other cases, such as SuF02-007
and \object{SN 2002lc}, there are no clear galaxy lines, even 
though the spectra of these candidates show broad features.

In terms of classifying candidates from the spectra alone, there is
also a correspondence between the selection criteria that are used to
select candidates and the percentage of candidates that are spectrally
identified as SNe~Ia.  In the rolling search with the CFHT, none of
the 5 candidates could be spectrally confirmed as a SN~Ia. For
comparison, in the Spring 2002 search with CTIO, all four candidates
(excluding T02-049) were confirmed as SNe~Ia.  Although the
numbers are small, they are significant. The search area and candidate
selection criteria of the rolling search were such that the search was
also sensitive to relatively fainter supernovae (Type II or
SN~1991bg-like supernovae) on relatively brighter hosts. The spectra
confirm this as many of the candidates from the rolling search are
dominated by the light of the host galaxy.

\begin{figure}
\centering
\includegraphics[width=8cm]{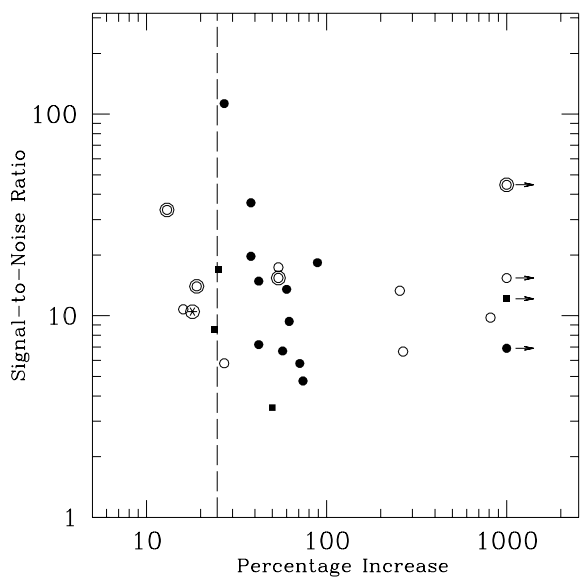}
\caption{The percentage increase versus the  signal-to-noise ratio
for candidates that were brighter than ${\rm I}=24.7$ at the time of
discovery. Candidates that were classified as Ia (Ia$^*$) are plotted
with solid circles (squares), C02-031 - a possible SN~II - is plotted
as a star and unclassified candidates are plotted with open circles.
Candidates from the CFHT 2002 search are highlighted with large
circles.  Candidates with arrows have percentage increases that are
greater than 1000\%, which means that the host was considerably
fainter than the candidate and perhaps undetected.  The dashed line
marks the region where the percentage increase is 25\% or
less. Candidates in this region are difficult to classify spectrally.}
\label{fig:increase}
\end{figure}

 The contribution from the host galaxy can be approximately
quantified with the percentage increase in the flux of the candidate
between the search and reference images. A small increase usually
means a significant amount of host contamination. A very large or
formally infinite increase usually means little or no host
contamination. The flux is measured over a fixed aperture whose
diameter depends on the seeing. The signal-to-noise ratio of the
detection over the same fixed aperture provides a measure of the
significance of the detection. A low signal-to-noise ratio usually
means that the candidate is faint, and this could mean that the
candidate, if it is a SN~Ia, is either very distant or has been
caught very early. In Fig.  \ref{fig:increase} the signal-to-noise
ratio of the detection is plotted against the percentage increase for
candidates that were brighter  than ${\rm I}=24.7$ at the time of
discovery.\footnote{For the surveys done in R-band, we use ${\rm
R}=24.7$} Candidates from the CFHT 2002 search are highlighted with
large circles.  The figure shows that classification from spectroscopy
is generally not successful if the percentage increase is below $\sim
25\%$.  The boundary of this region is marked with a dashed line in
Fig.  \ref{fig:increase}. Candidates in which the percentage increase
is less than 25\% are thus noted in Table \ref{tab:results}.

Not one of the candidates has broad emission lines that would
indicate that it is an AGN. This demonstrates that our selection
strategy, which selects against candidates having small intensity
variations that are also precisely centered on the host galaxy, is
quite effective in rejecting AGNs.

In this paper, we have strictly used only the spectra for
classification purposes and all the classifications listed in Table
\ref{tab:results} are based on the spectra alone.  However, in
searches like the CFHT Spring 2002 search and the Subaru Fall 2002
search, where spectroscopic confirmation is difficult because the
candidates are near relatively bright hosts or because the candidates
are relatively faint, additional information such as the light curve
or the colour of the candidate can become part of the criteria used
for classification. The strategy of these two searches was such that
most of the candidates were also monitored during the subsequent weeks
and months. Of the 18 candidates that were observed in these two
surveys, 5 were spectrally classified as either Ia, Ia$^*$ or II?. Of
the remaining 13 candidates,  10 were followed with sufficient
coverage (more than 4 light curve points) and 9 have supernova-like
light curves. This includes \object{SN 2002fr}, \object{SN 2002fm},
\object{SN 2002kq}, SuF02-007, \object{SN 2002lr},  SuF02-026,
\object{SN 2002kz}, SuF02-051 and \object{SN 2002kx}. These cases are
noted in Table \ref{tab:results} and in the comments on individual
candidates.

\section{Summary}

We have presented VLT FORS1 and FORS2 spectra of 39 candidate
high-redshift supernovae that were discovered as part of a program to
discover SNe~Ia over a wide range of
redshifts. By comparing these spectra with the spectra of nearby
SNe~Ia, 20 candidates have been identified as SNe~Ia with redshifts
ranging from $z=0.212$ to $z=1.181$.

Of the remaining 19 candidates that cannot be spectrally identified as
SNe~Ia, one candidate might be a Type II supernova at $z=0.541$ and 11
candidates exhibit broad supernova-like spectral features and/or have
supernova-like light curves. Of the final 7 candidates that cannot be
confirmed as supernova, (either from the light curves or the spectra),
5 possess a galaxy component, from which redshifts ranging from
$z=0.347$ to $z=1.478$ have been been measured, and 2 show featureless
blue continua.

\begin{acknowledgements}

This work would not have been possible without the dedicated efforts
of the daytime and nighttime support staff at the Cerro Paranal
Observatory. We thank ECT$^*$ (European Centre for Theoretical Studies
in Nuclear Physics and Related Areas) for the support they provided
during the preparation of this paper. The Subaru searches were
supported in part with a scientific research grant (15204012) from the
Ministry of Education, Science, Culture, and Sports of Japan, and in
part by the Japanese Society for the Promotion of Science (a Bilateral
Research Program between Japan and USA). The CFHT is operated by the
National Research Council of Canada, the Centre National de la 
Recherche Scientifique of France and the University of Hawaii. The
authors would like to thank the CFHT queue team for the efficient
operation of the CFHT12k camera.  This work was supported in part by
the Director, Office of Science, Office of High Energy and Nuclear
Physics, of the U.S. Department of Energy under Contract
No. DE-AC03-76SF00098.  Support for this work was provided by NASA
through grants HST-GO-08346.01-A , HST-GO-08585.14-A ,
HST-GO-09075.01-A , from the Space Telescope Science Institute, which
is operated by the Association of Universities for Research in
Astronomy, Inc., under NASA contract NAS 5-26555.

\end{acknowledgements}


\appendix

\section{Finding charts, spectra and notes on individual candidates}

This section contains finding charts and spectra of all candidates
except \object{SN 2000fr}, which are shown in
Figs. \ref{fig:FC_C00-008} and \ref{fig:Spec_C00-008}.  The candidates
are labelled with either their IAU names or their internal SCP names if
no IAU name was assigned. 

In the finding charts, North is up and East is to the left. The
candidate is marked with a cross and bright pivot stars are marked
with either a box or a hexagon. Fainter pivot stars are circled and
labelled alphabetically. The pivot star that was used during the
acquisition is recorded in Table \ref{tab:VLTsummary1}. The finding
charts were created from the images that were taken during the
reference and search runs. Regions that appear blank are regions that
are outside the field of view.

In general, the spectrum of the candidate is plotted twice. In the
upper spectrum, the unbinned spectrum of the candidate is plotted in
the observer's frame and is uncorrected for host galaxy light. Night
sky subtraction residuals are marked with the letters ``NS'' and
telluric absorption features are marked with the symbol $\oplus$. In
the lower spectrum, the spectrum is rescaled, contamination from the
host (if any) is removed, an extinction correction is applied and the
spectrum is re-binned, typically by 20~\AA. This spectrum is plotted in
black and it is plotted in both the rest frame (lower axis) and the
observer's frame (upper axis). For comparison, the best fitting nearby
supernova is plotted in blue. The extinction correction can correct
for extinction either in the host or the comparison spectrum. If the
candidate could not be classified, only the upper spectrum is plotted.

\begin{figure}
\centering
\resizebox{\hsize}{!}{\includegraphics{1504a01F.ps2}}
\resizebox{\hsize}{!}{\includegraphics{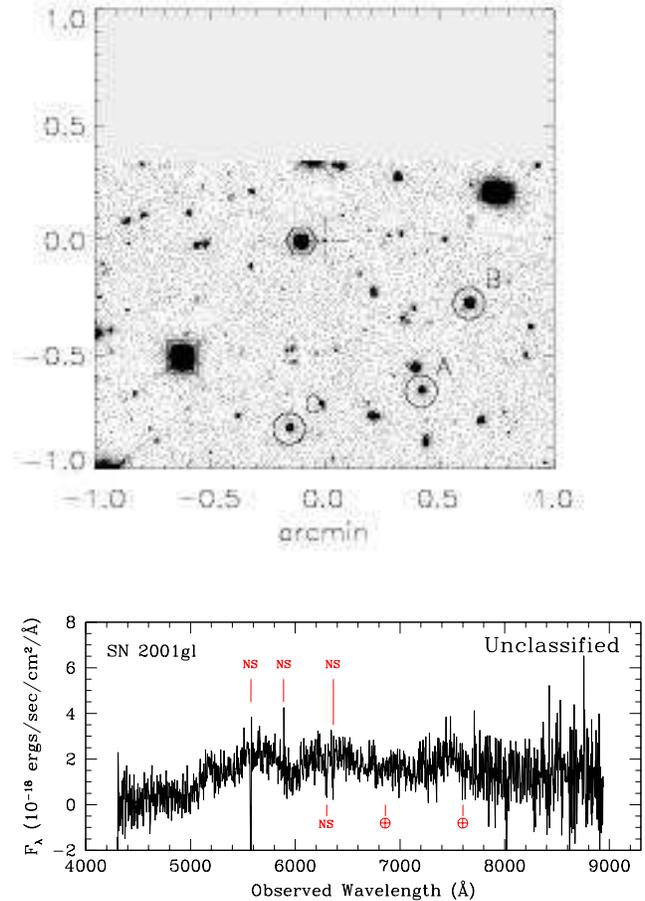}}
\vspace{-4cm}
\caption{Above, a finding chart centered on \object{SN 2001gl}
(S01-004), an unclassified candidate at an unknown redshift, and
below, the spectrum. This unusual candidate has very broad spectral features;
however, it was not possible to match this candidate with any of the
supernovae in our nearby catalog. No host was detectable in the
reference image, and the search images, which were taken 16 and 20 days
after the reference image, indicate that the candidate was real and
stationary, implying that it was not a solar system body. The spectrum
was taken 21 days after the reference images.}
\label{fig:FC_S01-004}
\end{figure}               

\begin{figure}
\centering
\resizebox{\hsize}{!}{\includegraphics{1504a02F.ps2}}
\resizebox{\hsize}{!}{\includegraphics{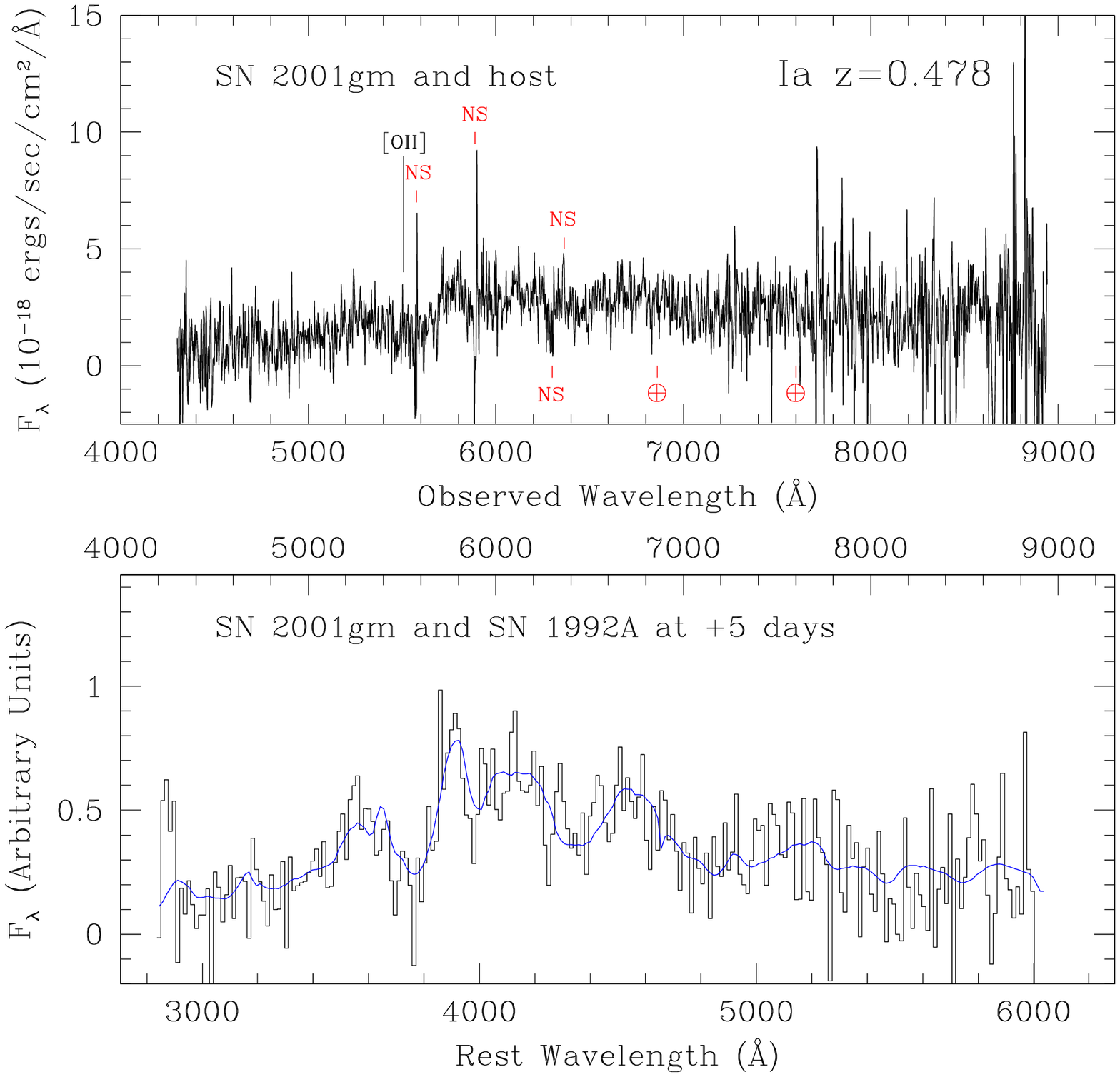}}
\caption{Above, a finding chart centered on \object{SN 2001gm}
(S01-005), a SN~Ia at $z=0.478$ and below, the spectrum. Although a
bright night sky line contaminates the 4000~\AA\ region, the Si~II
feature at 4000~\AA\ is clearly detected.  A separate spectrum of the
host (not shown here) shows weak [OII] emission.}
\label{fig:FC_S01-005}
\end{figure}               

\begin{figure}
\centering
\resizebox{\hsize}{!}{\includegraphics{1504a03F.ps2}}
\resizebox{\hsize}{!}{\includegraphics{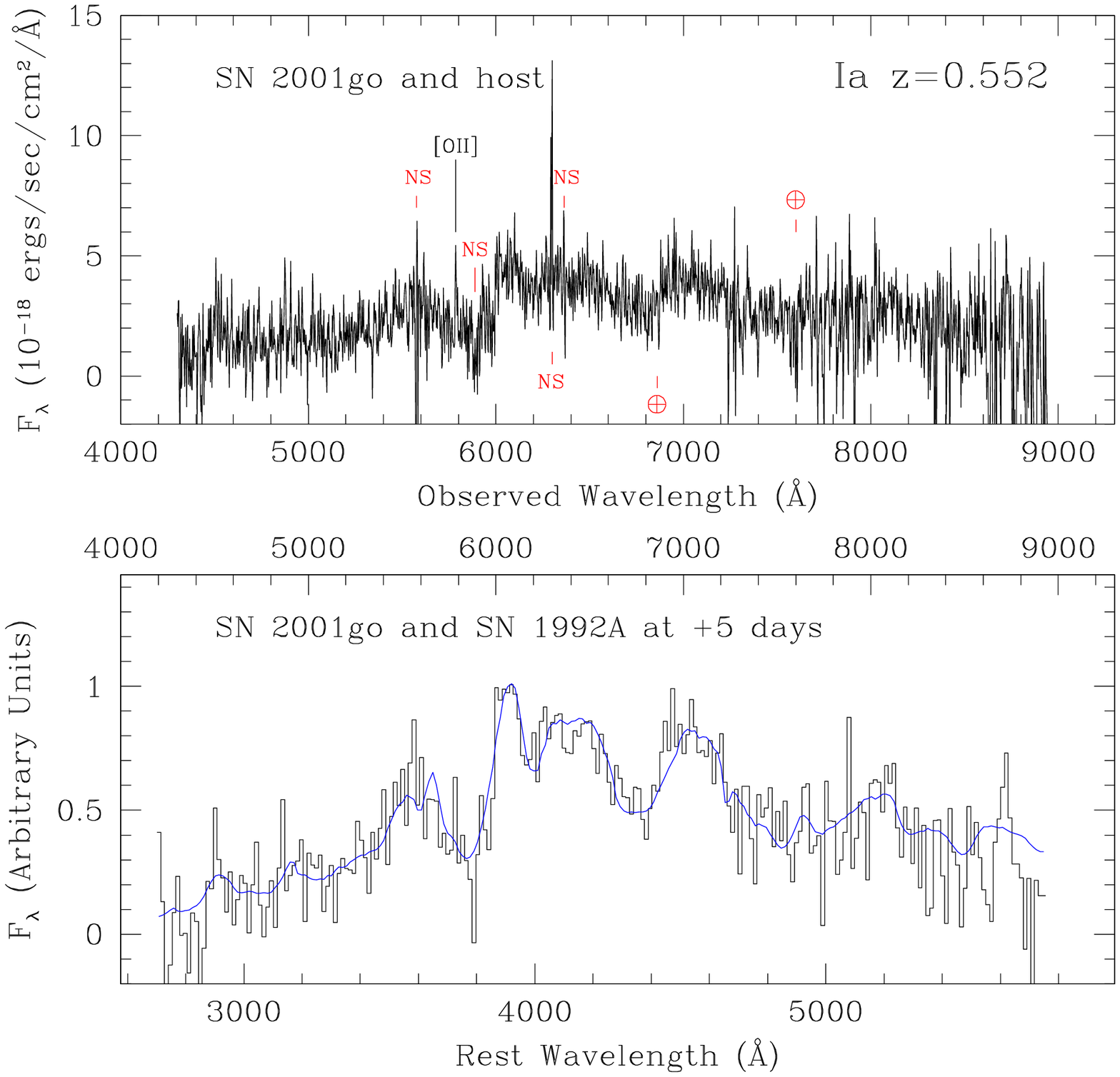}}
\caption{Above, a finding chart centered on \object{SN 2001go}
(S01-007), a SN~Ia at $z=0.552$, and below, the spectrum. This
candidate was observed at three epochs. The initial confirmation
spectrum (shown here) was taken on 2001 May 21. The Si~II feature at
4000~\AA\ feature can be clearly seen. Additional deeper spectra (not
shown here) were taken 6 and 37 observer-frame days later (Garavini et
al. in preparation).}
\label{fig:FC_S01-007}
\end{figure}               

\begin{figure}
\centering
\resizebox{\hsize}{!}{\includegraphics{1504a04F.ps2}}
\resizebox{\hsize}{!}{\includegraphics{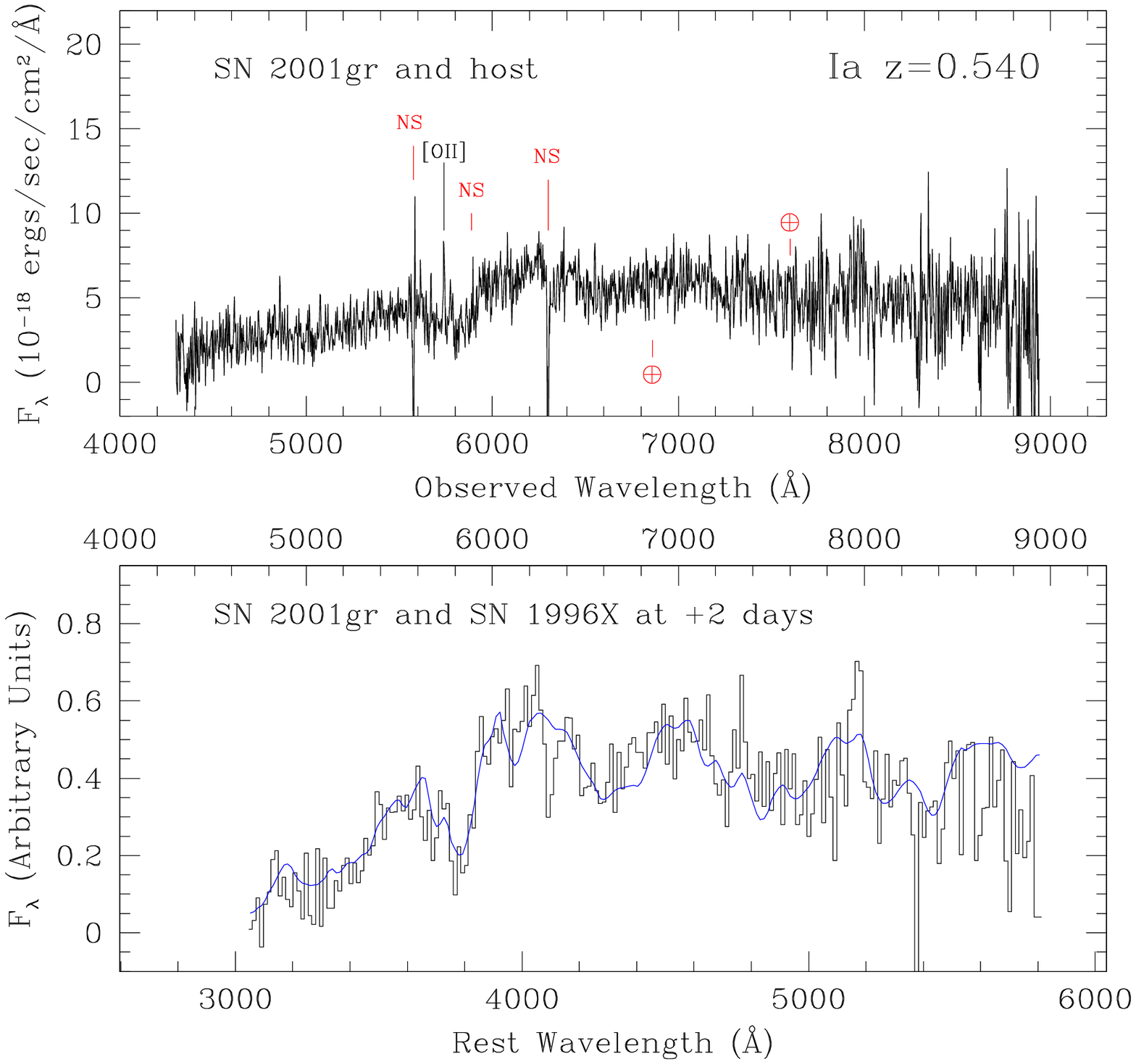}}
\caption{Above, a finding chart centered on \object{SN 2001gr}
(S01-017), a SN~Ia at $z=0.540$, and below, the spectrum. Although
Si~II feature at 4000~\AA\ is not clearly detected in this candidate, the data 
are significantly better fit with SN~Ia spectra than with the spectra of
other types.}
\label{fig:FC_S01-017}
\end{figure}

\begin{figure}
\centering
\resizebox{\hsize}{!}{\includegraphics{1504a05F.ps2}}
\resizebox{\hsize}{!}{\includegraphics{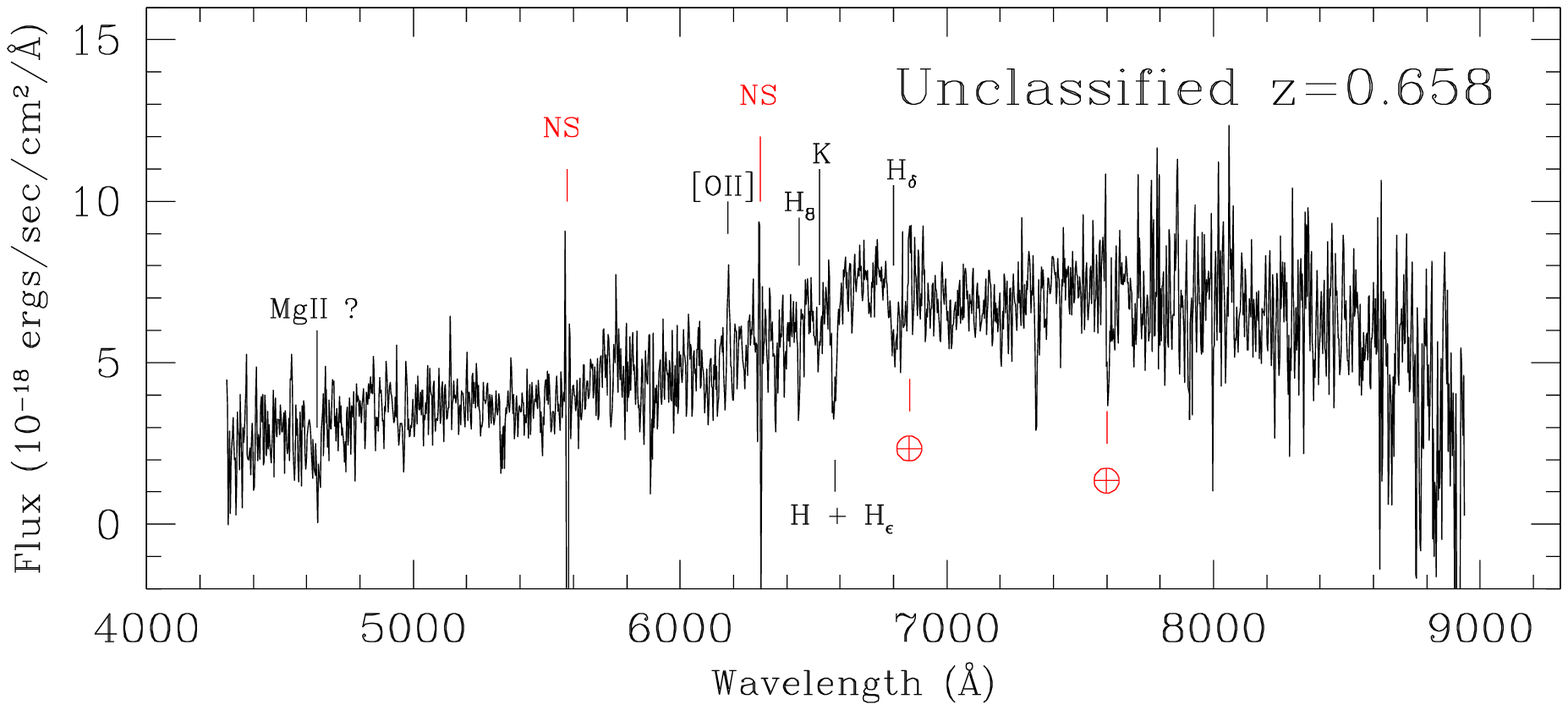}}
\vspace{-4cm}
\caption{Above, a finding chart centered on \object{SN 2001gs}
(S01-028), an unclassified candidate at $z=0.658$, and below, the
spectrum. This is a faint candidate on a bright host that was observed
during a period of relatively poor seeing. The percentage increase in
the flux was only 27\%, so most of the light in the spectrum is from
the host, which has several Balmer absorption lines.}
\label{fig:FC_S01-028}
\end{figure}               

\clearpage

\begin{figure}
\centering
\resizebox{\hsize}{!}{\includegraphics{1504a06F.ps2}}
\resizebox{\hsize}{!}{\includegraphics{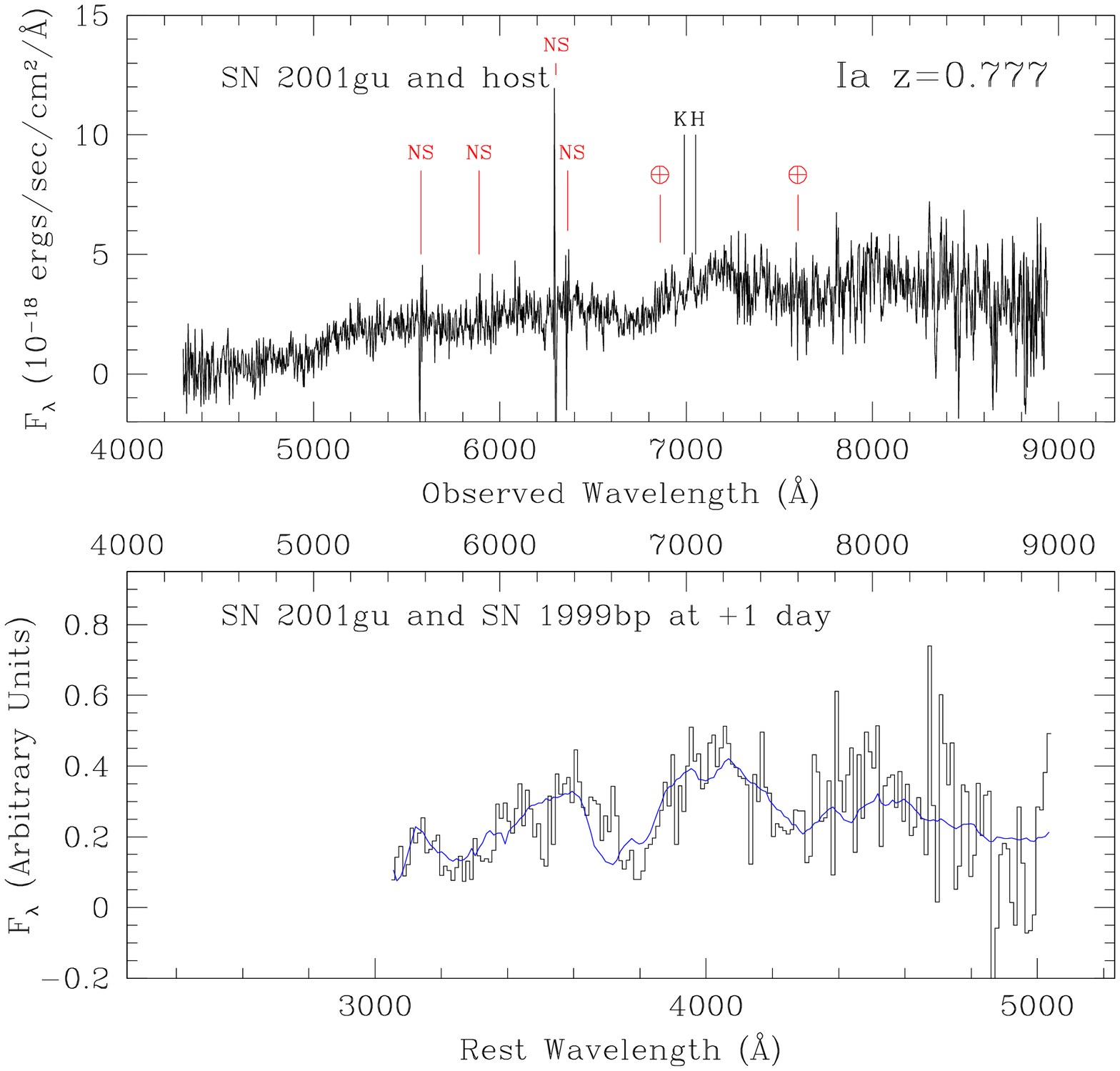}}
\caption{Above, a finding chart centered on \object{SN 2001gu}
(S01-031), a SN~Ia at $z=0.777$, and below, the spectrum. The host
shows CaII H and K absorption lines and no detectable [OII] emission
which suggests an early-type host. Since the Si~II feature at
4000~\AA\ is weak and contaminated by the H and K lines of the host,
the classification is based on the fit. The redshift of the fit was
constrained to that of the host. The wavelength coverage of the best
matching nearby SN~Ia, SN 1999bp, is restricted to rest frame
wavelengths that are greater than 3000~\AA, so the comparison is
limited to these wavelengths.}
\label{fig:FC_S01-031}
\end{figure}               

\begin{figure}
\centering
\resizebox{\hsize}{!}{\includegraphics{1504a07F.ps2}}
\resizebox{\hsize}{!}{\includegraphics{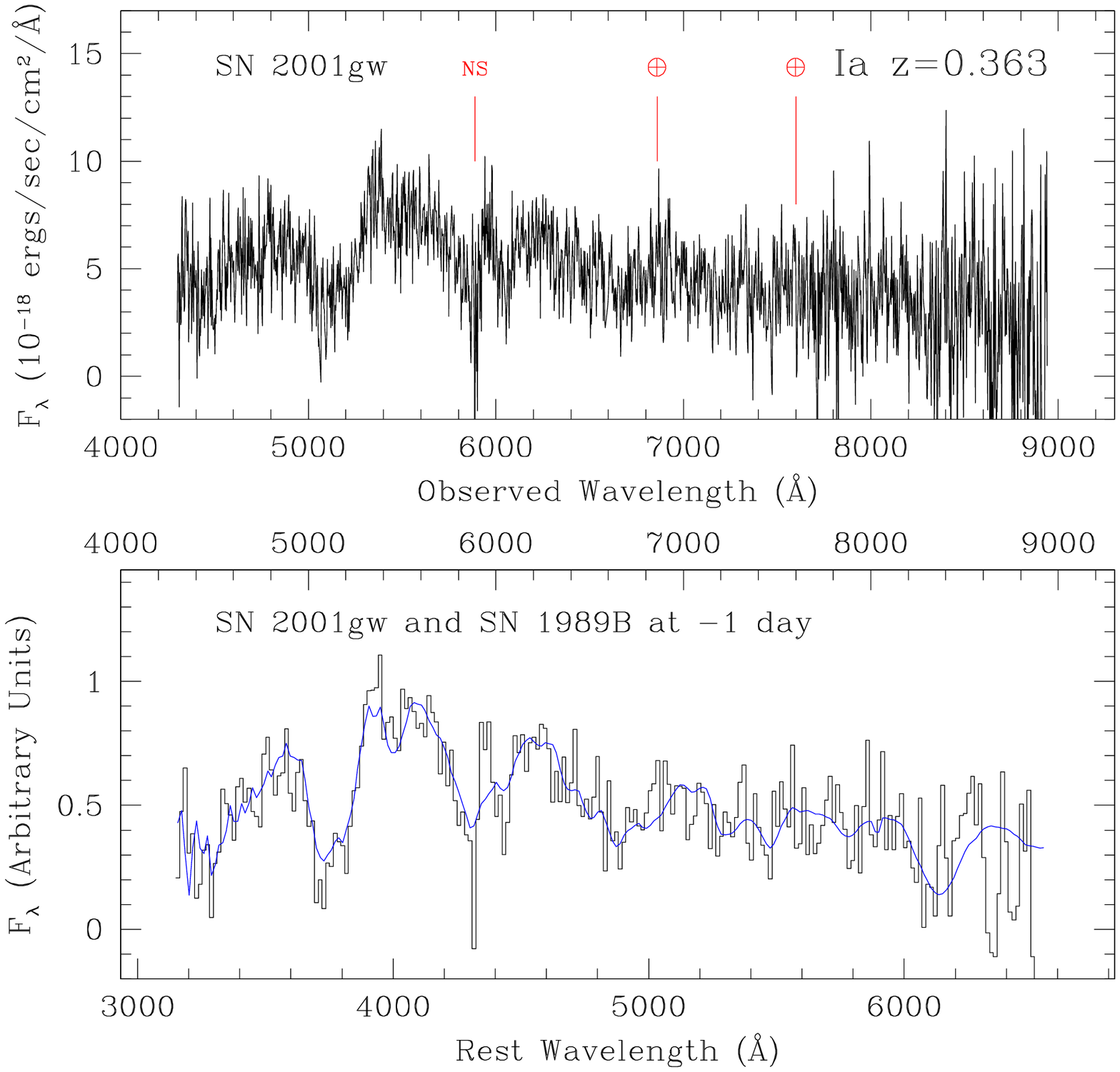}}
\caption{Above, a finding chart centered on \object{SN 2001gw}
(S01-033), a SN~Ia at $z=0.363$, and below, the spectrum. In addition
to the Si~II feature at 4000~\AA, the Si~II at 6150~\AA\ is also visible. The 
redshift is derived from an [OII] emission line in the spectrum of
the host galaxy (not shown here).}
\label{fig:FC_S01-033}
\end{figure}

\begin{figure}
\centering
\resizebox{\hsize}{!}{\includegraphics{1504a08F.ps2}}
\resizebox{\hsize}{!}{\includegraphics{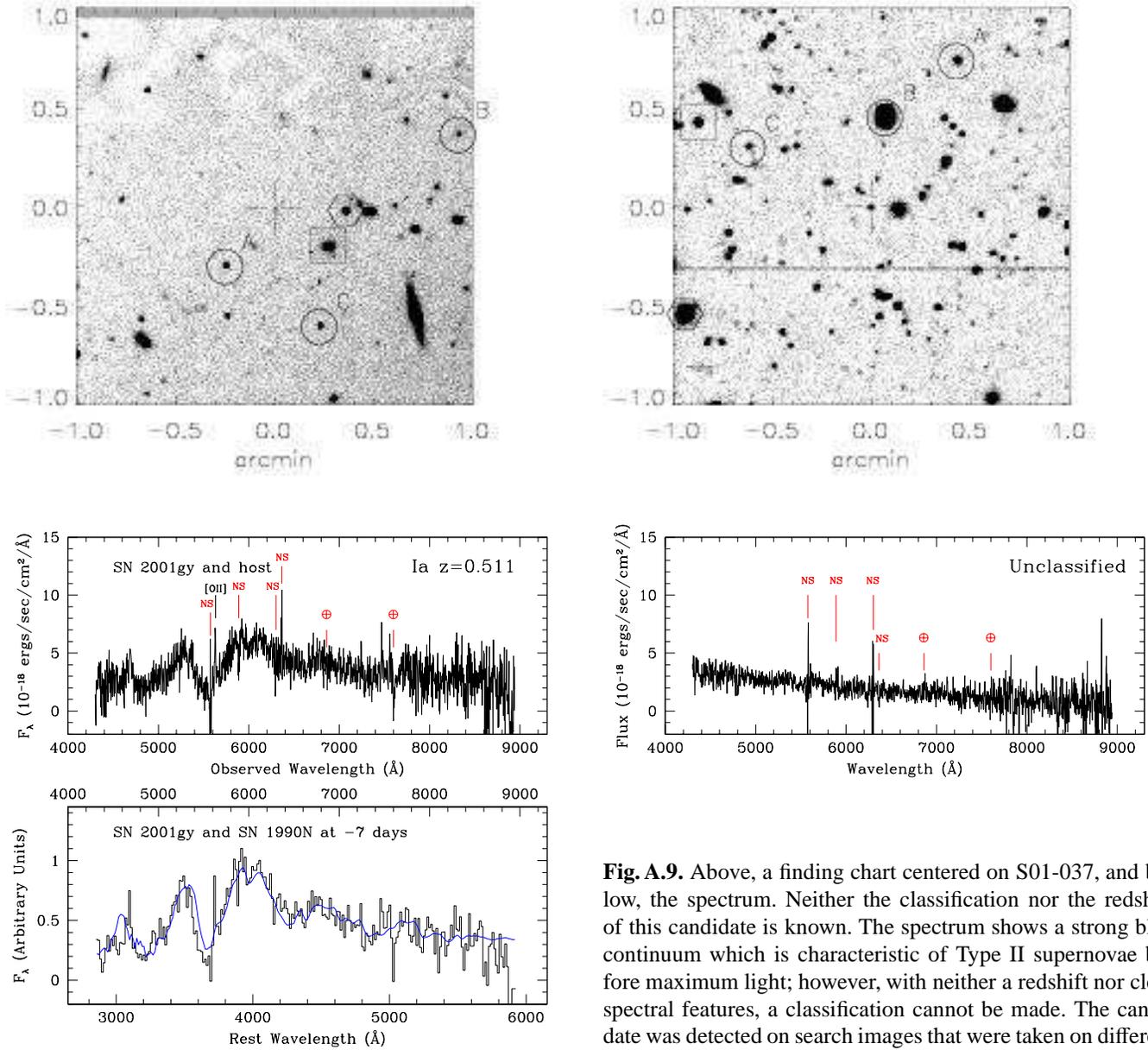}}
\caption{Above, a finding chart centered on \object{SN 2001gy}
(S01-036), a SN~Ia at $z=0.511$, and below, the spectrum. The Si~II
feature at 4000~\AA\ is clearly detected.}
\label{fig:FC_S01-036}
\end{figure}

\begin{figure}
\centering
\resizebox{\hsize}{!}{\includegraphics{1504a09F.ps2}}
\resizebox{\hsize}{!}{\includegraphics{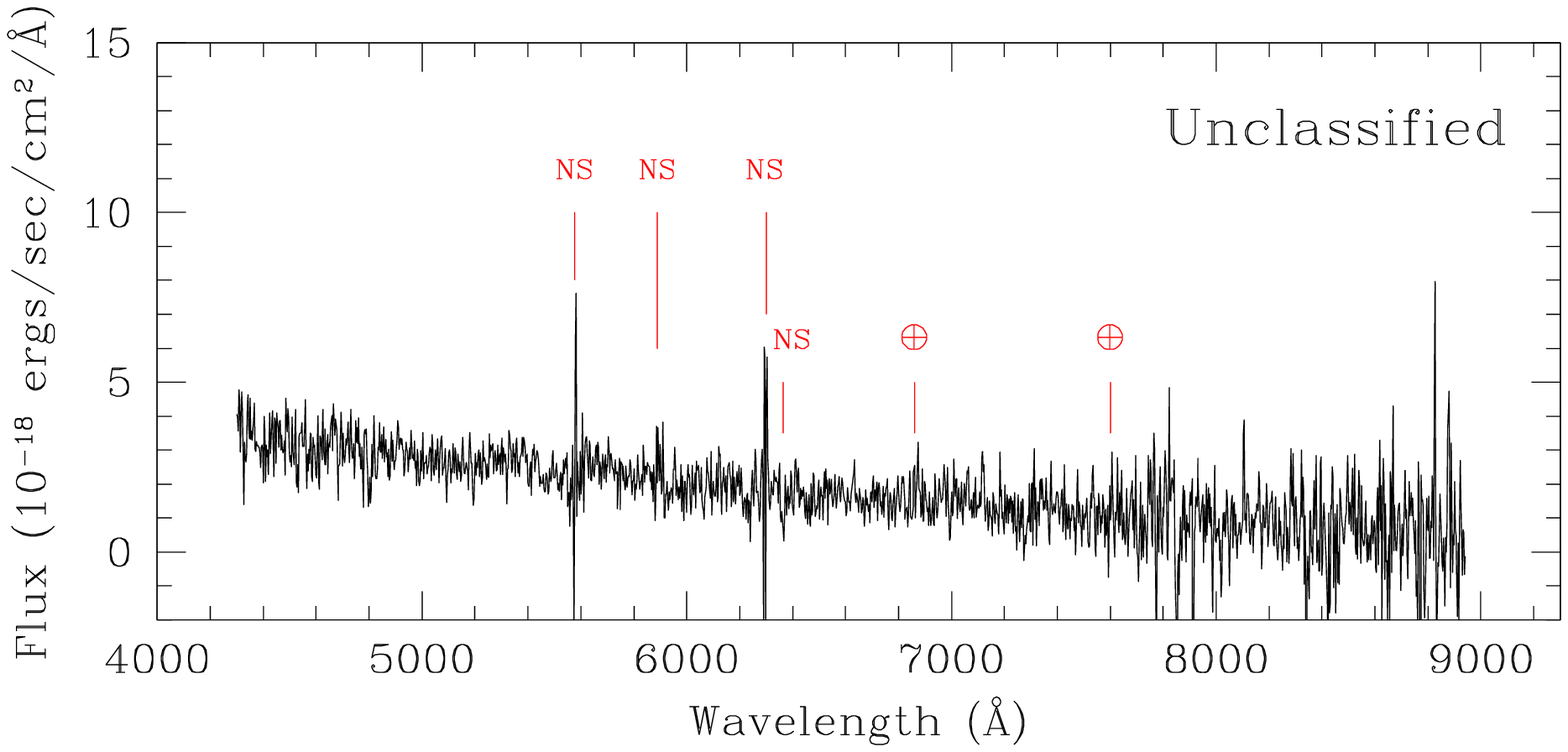}}
\vspace{-4cm}
\caption{Above, a finding chart centered on S01-037, and below, the
spectrum. Neither the classification nor the redshift of this
candidate is known. The spectrum shows a strong blue continuum which
is characteristic of Type II supernovae before maximum light; however,
with neither a redshift nor clear spectral features, a classification
cannot be made. The candidate was detected on search images that were
taken on different dates and is stationary, so it is not an asteroid
nor an artifact.}
\label{fig:FC_S01-037}
\end{figure}               

\clearpage

\begin{figure}
\centering
\resizebox{\hsize}{!}{\includegraphics{1504a10F.ps2}}
\resizebox{\hsize}{!}{\includegraphics{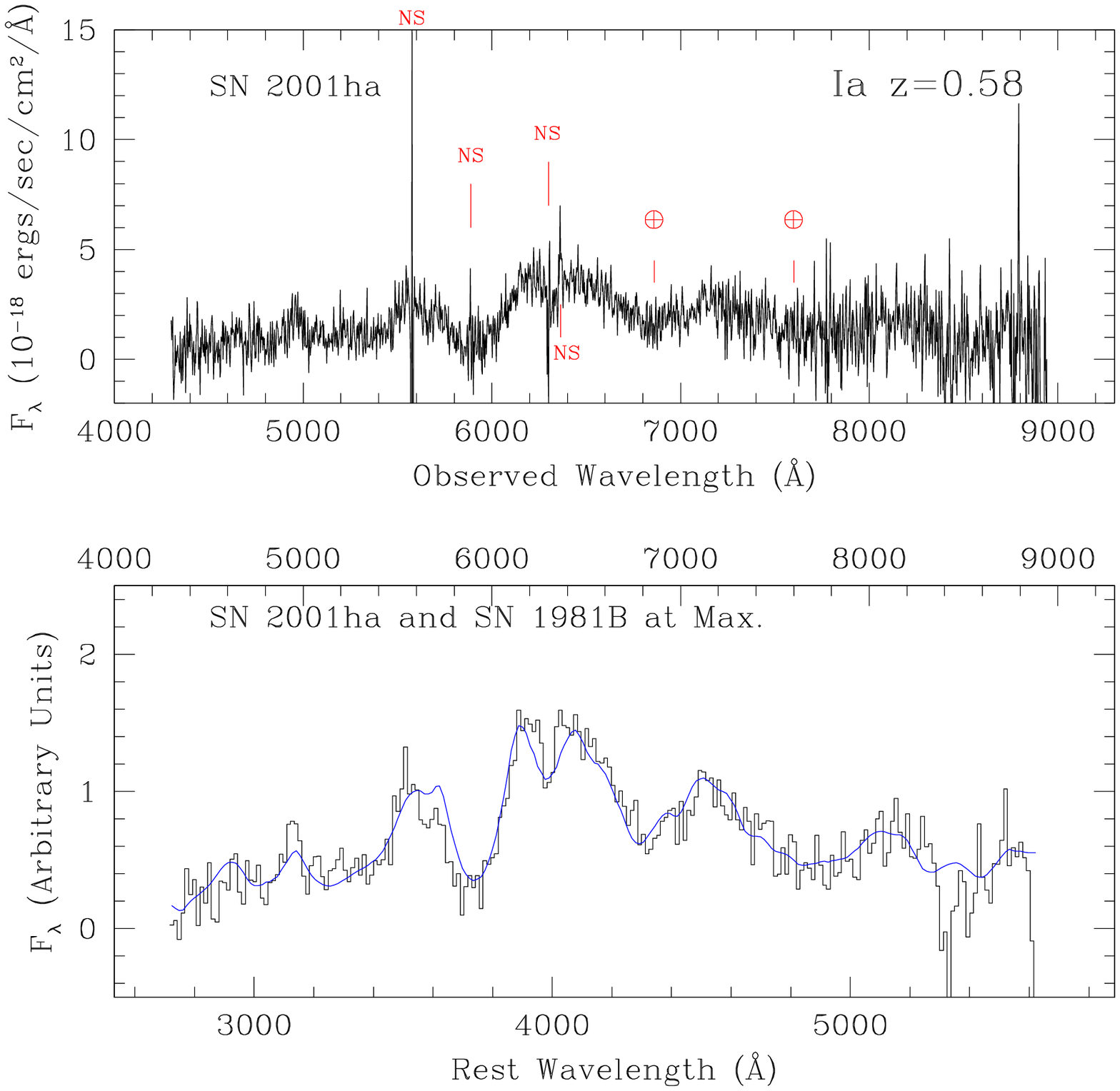}}
\caption{Above, a finding chart centered on \object{SN 2001ha}
(S01-054), a SN~Ia at $z=0.58$, and below, the spectrum. There are no
spectral features from the host and a host is not visible in the
reference image, so the redshift is determined from the fit. The Si~II
feature at 4000~\AA\ is clearly detected. }
\label{fig:FC_S01-054}
\end{figure}

\begin{figure}
\centering
\resizebox{\hsize}{!}{\includegraphics{1504a11F.ps2}}
\resizebox{\hsize}{!}{\includegraphics{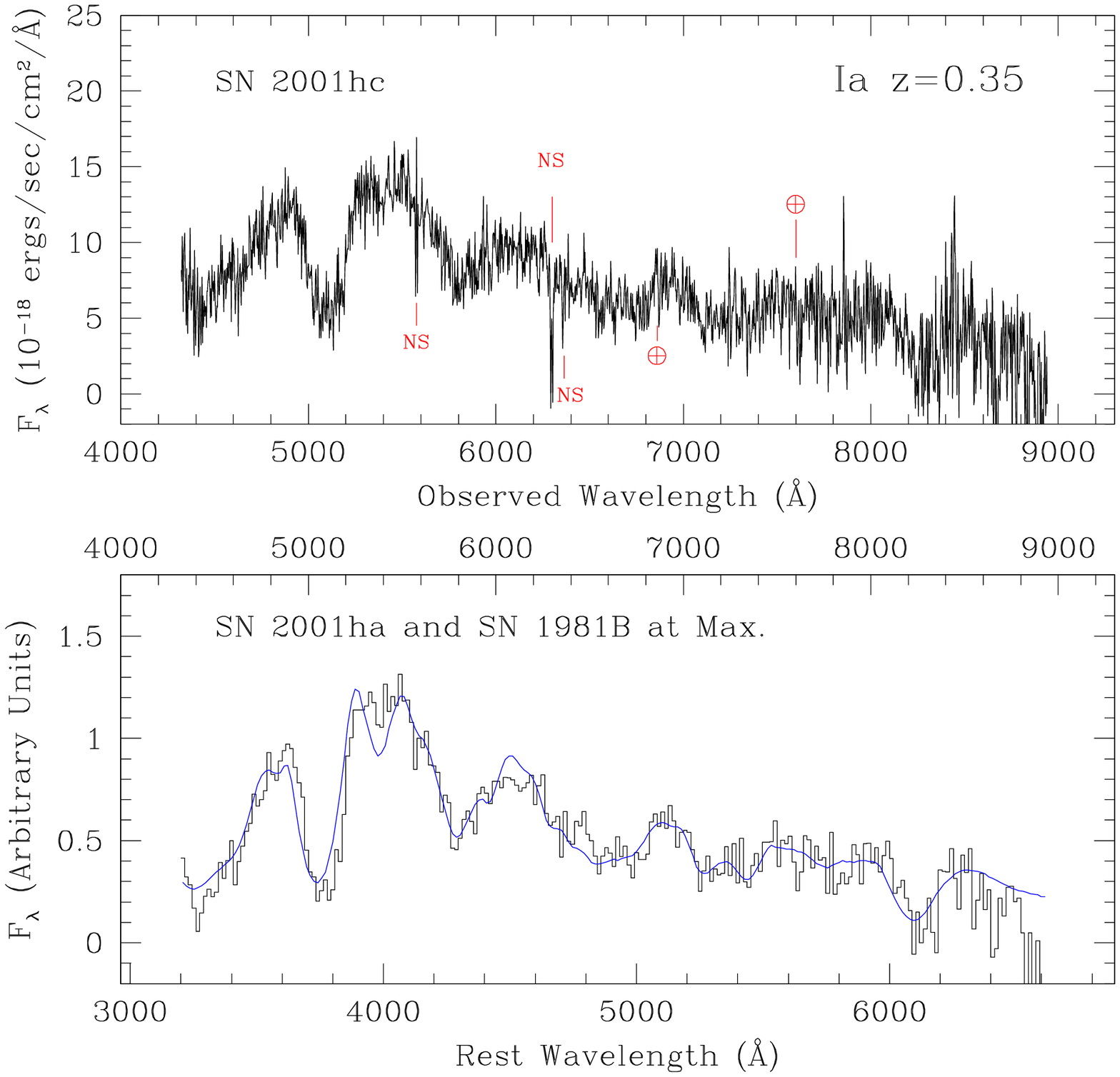}}
\caption{Above, a finding chart centered on \object{SN 2001hc}
(S01-065), a SN~Ia at $z=0.35$, and below, the spectrum. This relatively
nearby candidate has Si~II at 6150~\AA, S~II at 5400~\AA, and Si~II at
4000~\AA. There are no spectral features from the host, so the redshift
is determined from the fit. In the reference image, a very faint host
is visible.}
\label{fig:FC_S01-065}
\end{figure}              

\begin{figure}
\centering
\resizebox{\hsize}{!}{\includegraphics{1504a12F.ps2}}
\resizebox{\hsize}{!}{\includegraphics{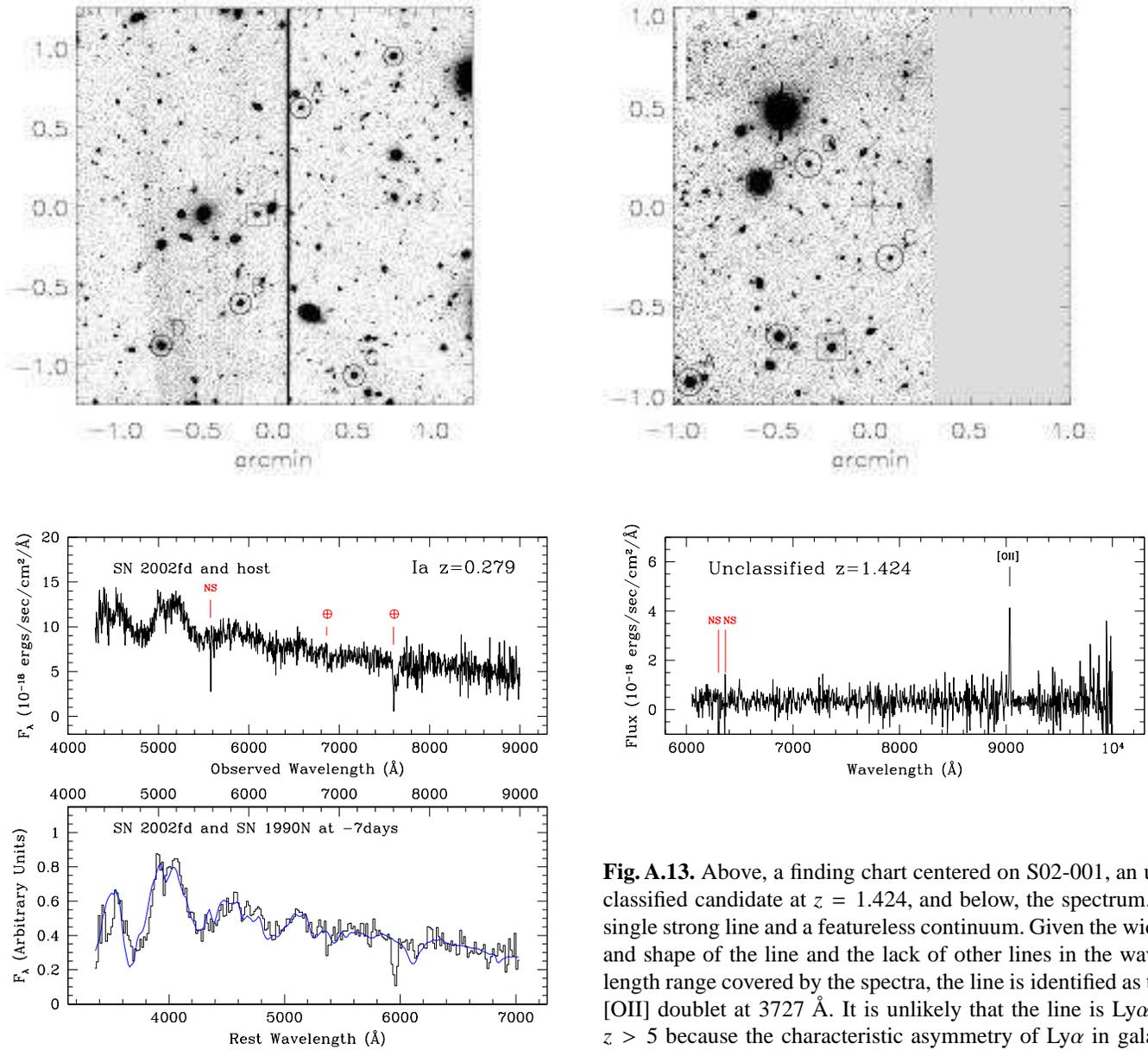}}
\caption{Above, a finding chart centered on \object{SN 2002fd}
(S02-000), a SN~Ia at $z=0.279$, and below, the spectrum. The
Si~II feature at 4000~\AA\ is clearly detected, but the Si~II feature
at 6150~\AA\ is relatively weak.}
\label{fig:FC_S02-000}
\end{figure}

\begin{figure}
\centering
\resizebox{\hsize}{!}{\includegraphics{1504a13F.ps2}}
\resizebox{\hsize}{!}{\includegraphics{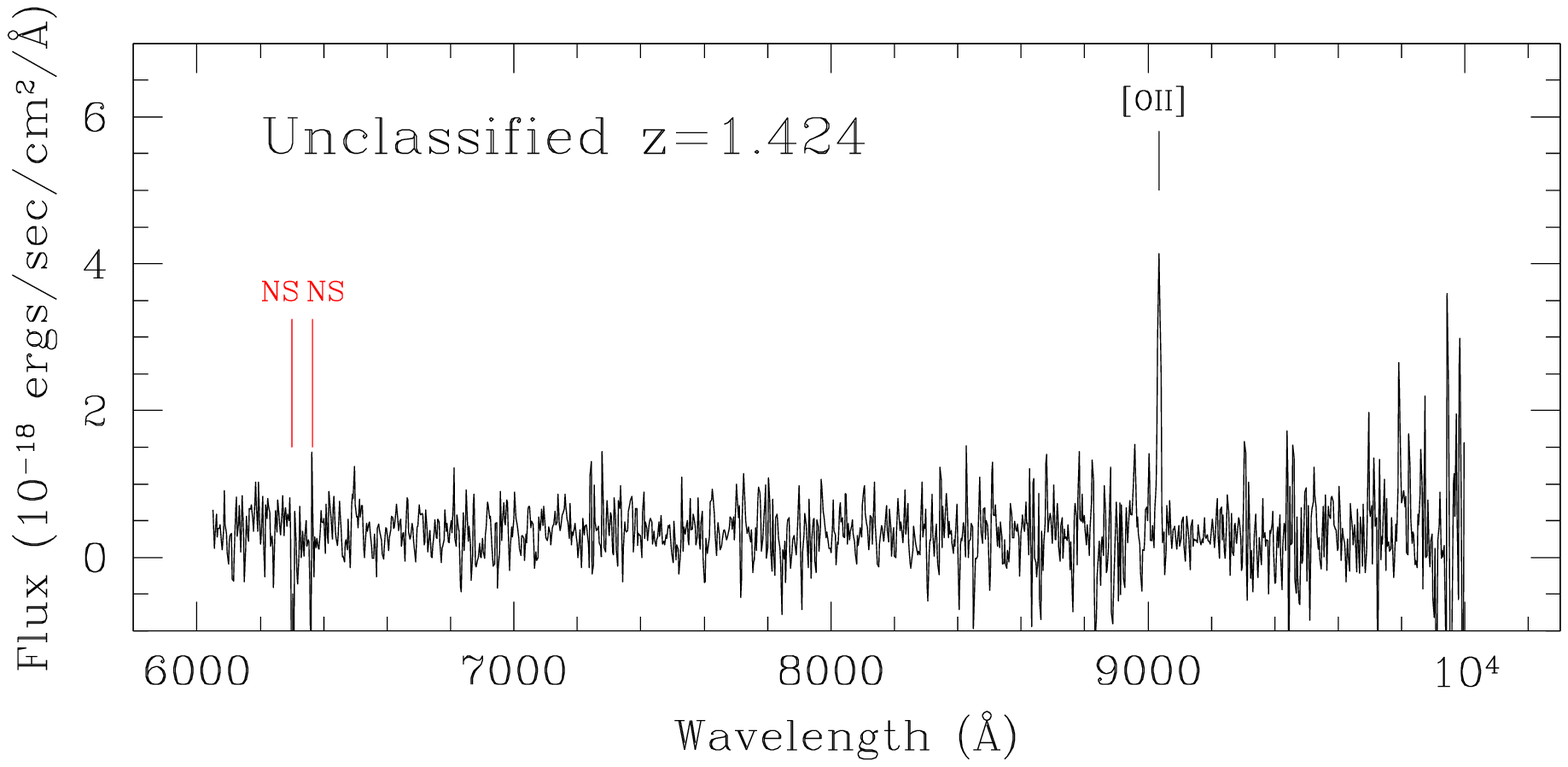}}
\vspace{-4cm}
\caption{Above, a finding chart centered on S02-001, an unclassified
candidate at $z=1.424$, and below, the spectrum. A single strong line
and a featureless continuum. Given the width and shape of the line and
the lack of other lines in the wavelength range covered by the
spectra, the line is identified as the [OII] doublet at 3727~\AA. It
is unlikely that the line is Ly$\alpha$ at $z>5$ because the
characteristic asymmetry of Ly$\alpha$ in galaxies at $z>5$ (Stern et
al. \cite{Stern00}) and the jump in the continuum across the line are
not evident in these spectra. Nor is the line likely to be ${\rm
H}\alpha$, as neither ${\rm H}\beta$ nor [OIII] are detected. The
equivalent width of the line is greater than 50~\AA\, so the host galaxy
would be classified as an emission line galaxy (ELG) if the line were
${\rm H}\alpha$ (Kniazev et al. \cite{Kniazev04}). In ELGs, ${\rm
H}\beta$ is typically three times weaker than ${\rm H}\beta$ and the
strengths of [OIII] and ${\rm H}\alpha$ are roughly equivalent. Given
the strength of the detected line in these spectra, both [OIII] and ${\rm
H}\beta$ should have been detected if the line was ${\rm H}\alpha$.}
\label{fig:FC_S02-001}
\end{figure}               

\clearpage

\begin{figure}
\centering
\resizebox{\hsize}{!}{\includegraphics{1504a14F.ps2}}
\resizebox{\hsize}{!}{\includegraphics{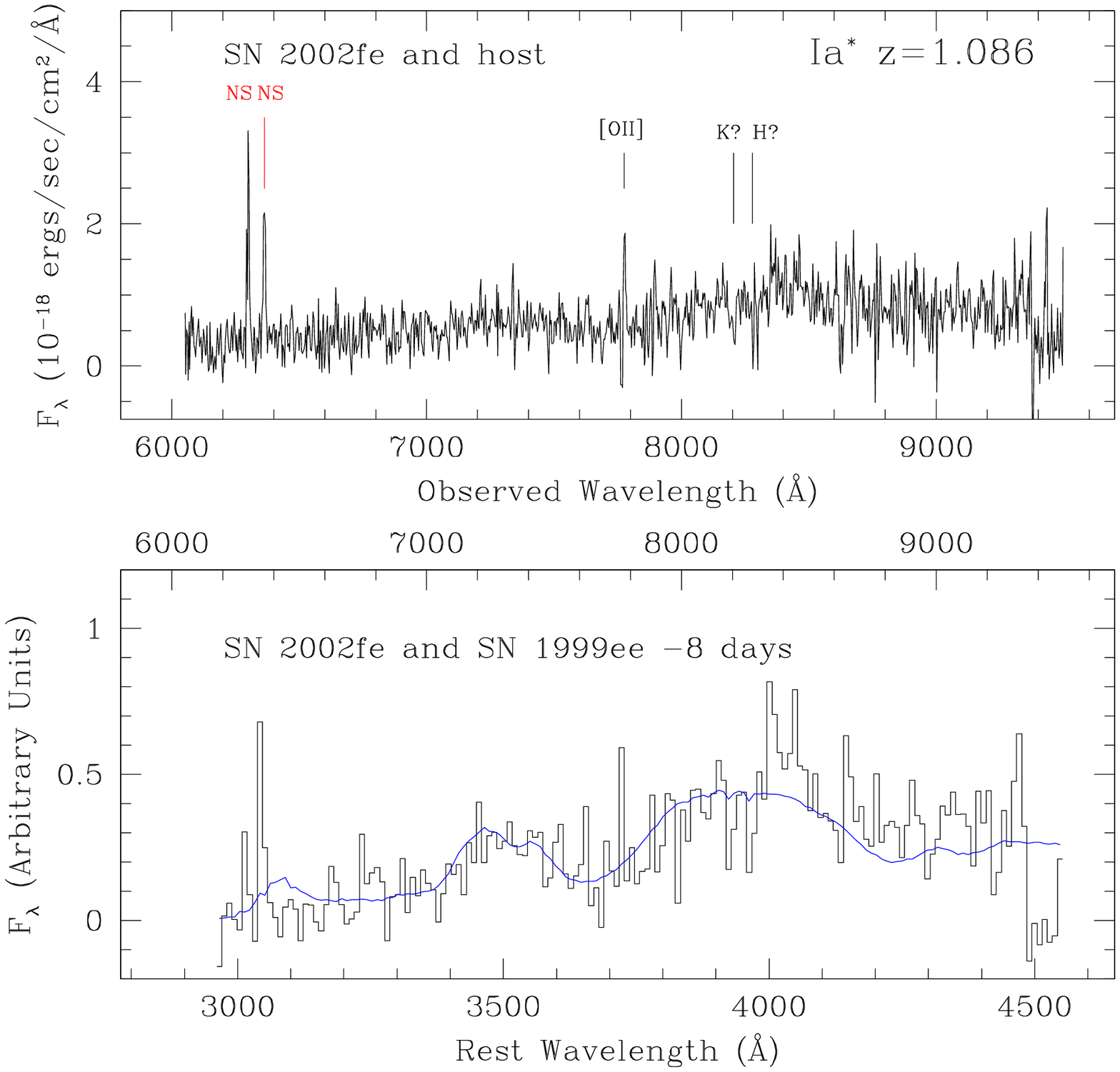}}
\caption{Above, a finding chart centered on \object{SN 2002fe}
(S02-002), a SN~Ia$^*$ at $z=1.086$, and below, the spectrum. The
profile of the line that we have identified as [OII] is affected by a
nearby bright night sky line; however, the line is clearly detected in
the 2-dimensional spectrum. This line, together with the probable
detection of the H and K Ca~II lines in the host, enables us to measure
a secure redshift.  The signal-to-noise ratio of the spectrum is
relatively low and the Si~II feature at 4000~\AA\ is not detected, so
the the classification is qualified with an asterisk. In some nearby SNe~Ia 
that are observed one to two weeks before maximum light, the Si~II feature
is absent. The best fit nearby SN~Ia, \object{SN 1999ee}, shows no
Si~II at 4000~\AA.}
\label{fig:FC_S02-002}
\end{figure}

\begin{figure}
\centering
\resizebox{\hsize}{!}{\includegraphics{1504a15F.ps2}}
\resizebox{\hsize}{!}{\includegraphics{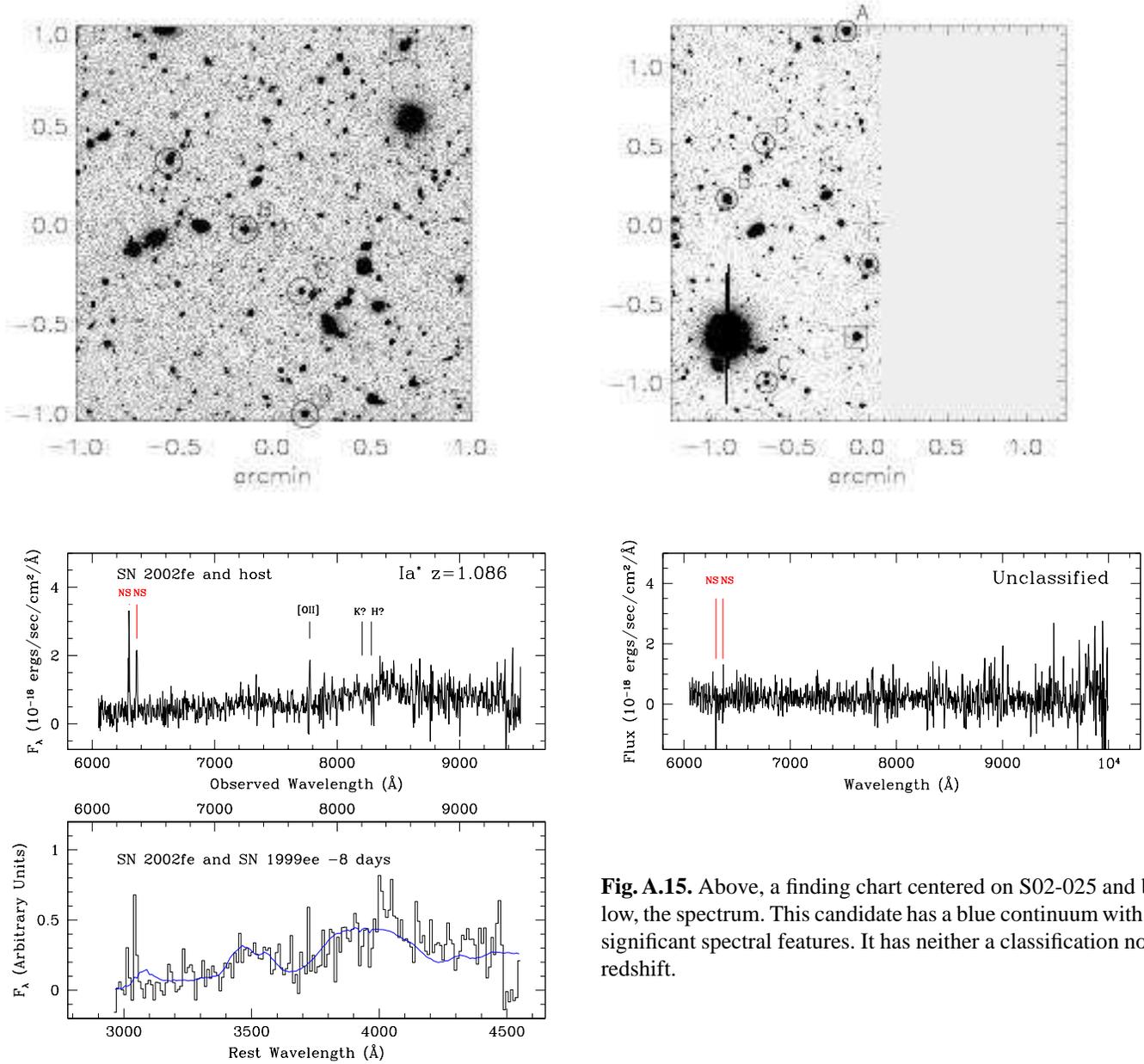}}
\vspace{-4cm}
\caption{Above, a finding chart centered on S02-025 and below, the
spectrum. This candidate has a blue continuum with no significant
spectral features. It has neither a classification nor a redshift.}
\label{fig:FC_S02-025}
\end{figure}              

\clearpage

\begin{figure}
\centering
\resizebox{\hsize}{!}{\includegraphics{1504a16F.ps2}}
\resizebox{\hsize}{!}{\includegraphics{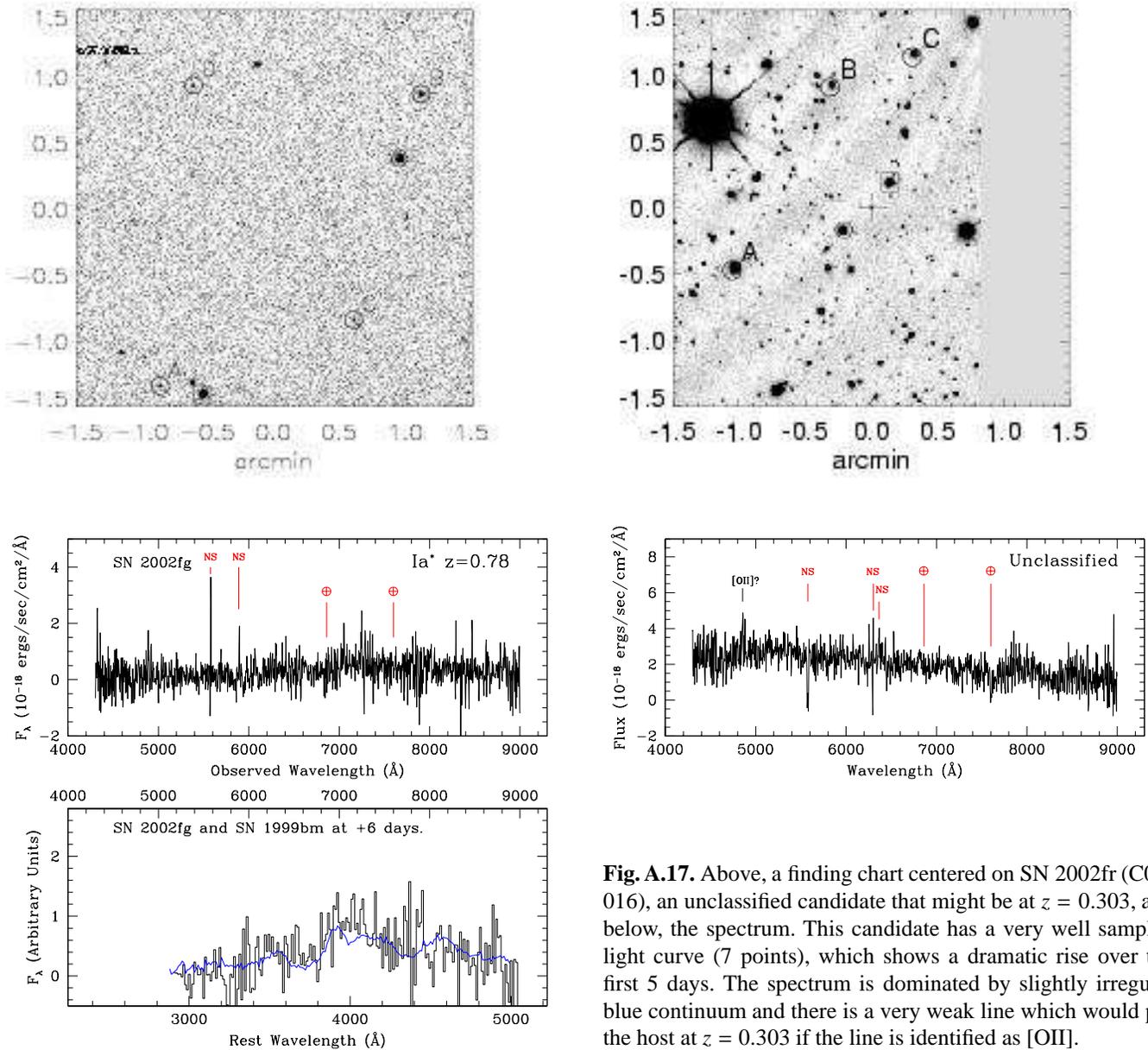}}
\caption{Above, a finding chart centered on \object{SN 2002fg}
(S02-075), a SN~Ia$^*$ at $z=0.78$, and below, the spectrum. Since the
Si~II feature at 4000~\AA\ is not clearly detected in this candidate
and since there are no spectral features from the host, the
classification and the redshift are derived from the fit. The
candidate was observed several weeks after it was discovered, so it is
likely that the spectrum was taken past maximum light. The best
matching nearby SN~Ia is \object{SN 1999bm} at 6 days past maximum
light. The signal-to-noise ratio is also relatively low and the SiII
feature at 4000~\AA\ is not detected, so the classification is
qualified with an asterisk.}
\label{fig:FC_S02-075}
\end{figure}               

\begin{figure}
\centering
\resizebox{\hsize}{!}{\includegraphics{1504a17F.ps2}}
\resizebox{\hsize}{!}{\includegraphics{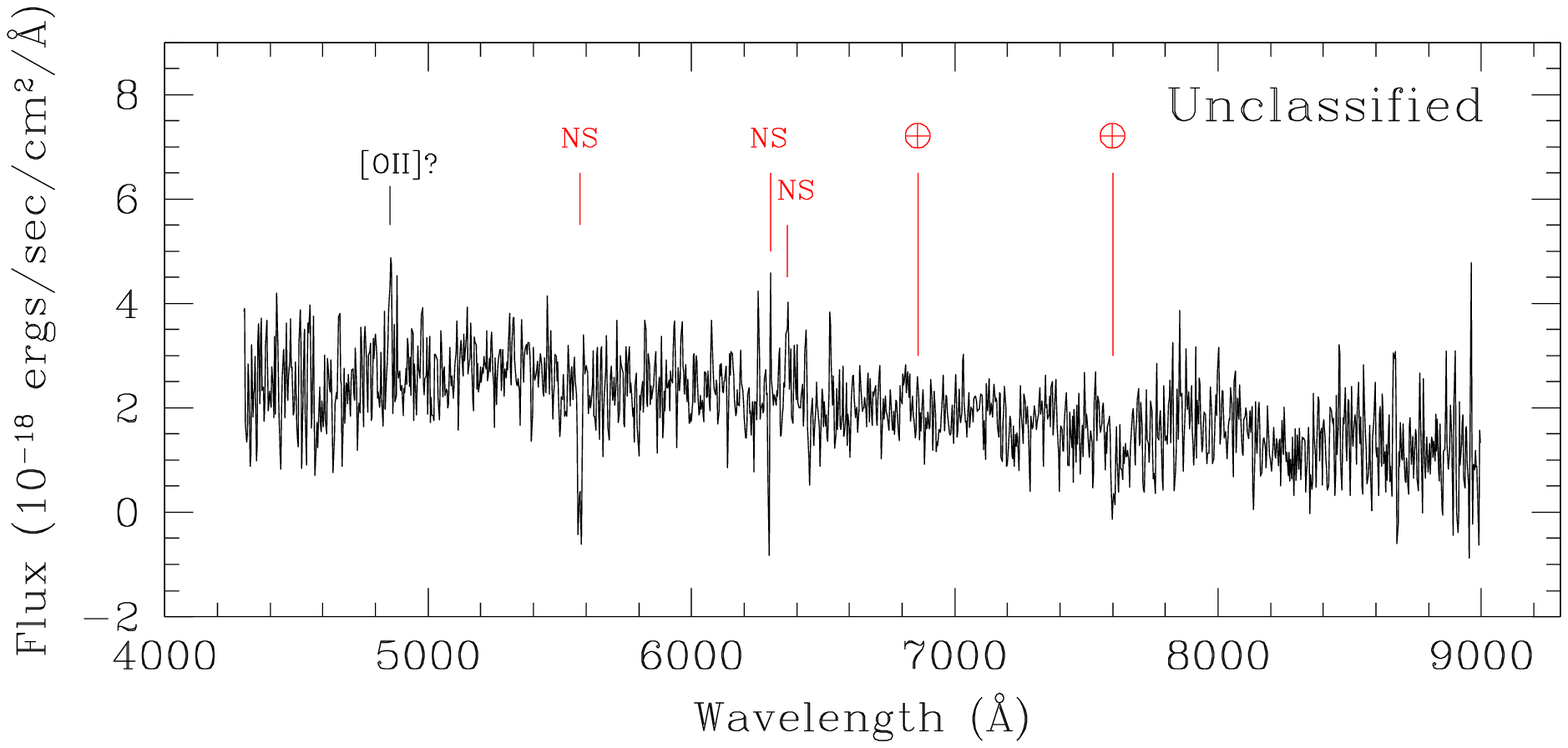}}
\vspace{-4cm}
\caption{Above, a finding chart centered on \object{SN 2002fr}
(C02-016), an unclassified candidate that might be at $z=0.303$, and
below, the spectrum. This candidate has a very well sampled light
curve (7 points), which shows a dramatic rise over the first 5
days. The spectrum is dominated by slightly irregular blue continuum
and there is a very weak line which would put the host at $z=0.303$ if
the line is identified as [OII].}
\label{fig:FC_C02-016}
\end{figure}               

\clearpage

\begin{figure}
\centering
\resizebox{\hsize}{!}{\includegraphics{1504a18F.ps2}}
\resizebox{\hsize}{!}{\includegraphics{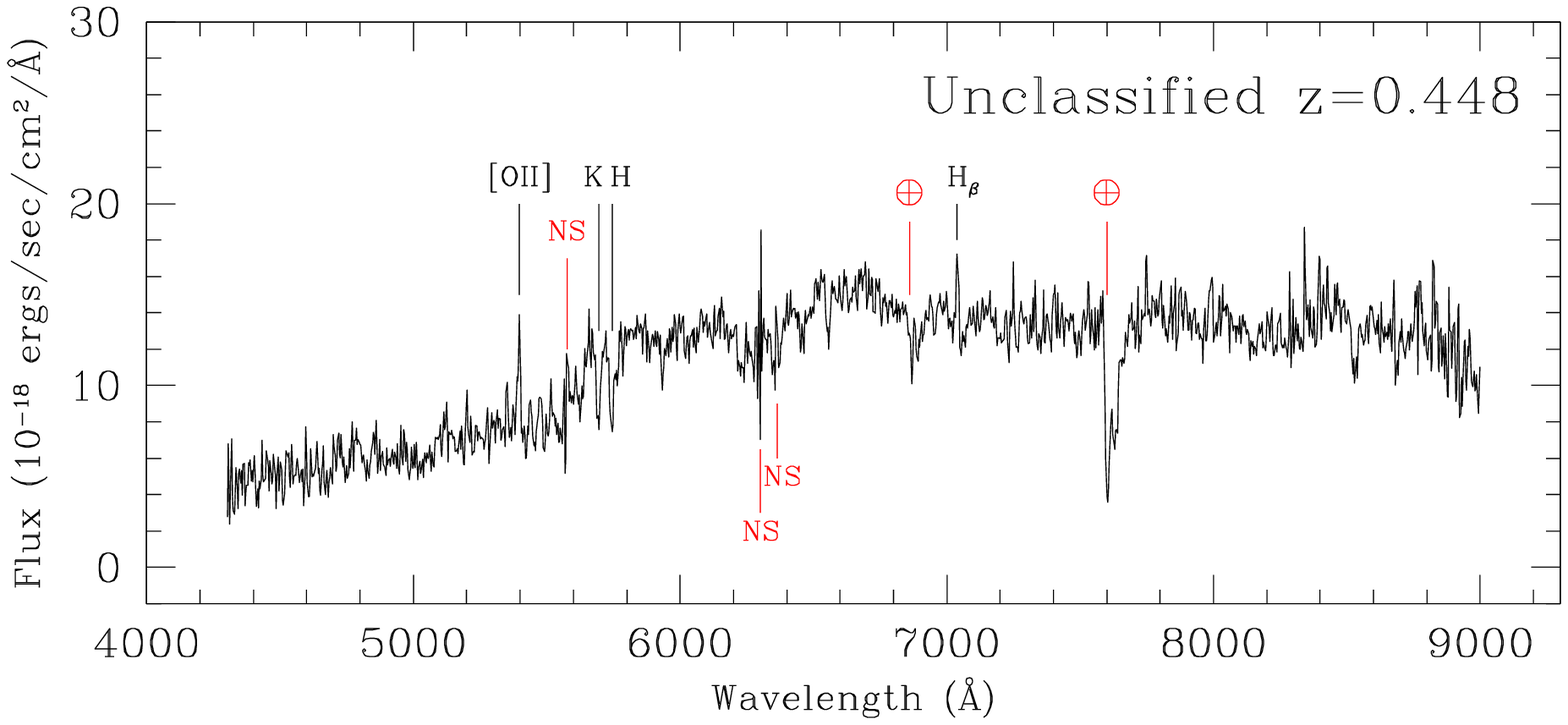}}
\vspace{-4cm}
\caption{Above, a finding chart centered on \object{SN 2002fm}
(C02-028), an unclassified candidate at $z=0.448$, and below, the
spectrum. The percentage increase in this candidate was very small,
only 13\%, and the spectrum is dominated by the light from the host
galaxy. However, there is excess flux at 6600~\AA\ and 5600~\AA\ that
might be from a supernova. Unfortunately, an acceptable fit with a
nearby SN~Ia was not possible. In such cases, the fit depends critically on
how well the galaxy template matches the spectrum of the host
galaxy. Relatively small errors can leave significant residuals which
can make the matching difficult. The most secure way of fitting this
candidate will be take a spectrum of the host after the supernova has
faded. The candidate is offset from the center of the host and the
light curve is well sampled with four points before maximum light and
four points after maximum light.}
\label{fig:FC_C02-028}
\end{figure}               

\begin{figure}
\centering
\resizebox{\hsize}{!}{\includegraphics{1504a19F.ps2}}
\resizebox{\hsize}{!}{\includegraphics{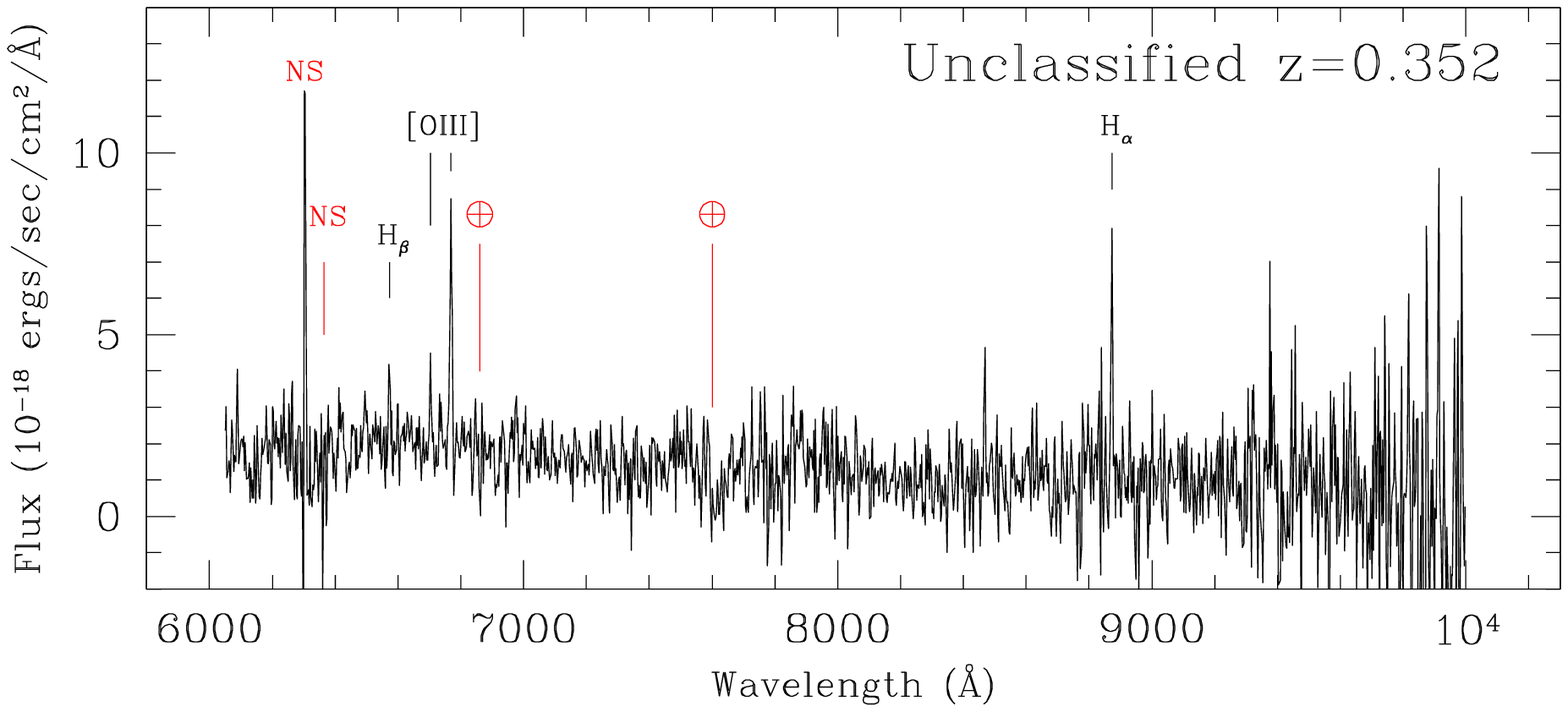}}
\vspace{-4cm}
\caption{Above, a finding chart centered on \object{SN 2002fp}
(C02-030), an unclassified candidate at $z=0.352$, and below, the
spectrum. The host galaxy has emission lines in [OIII] and ${\rm
H}_\beta$. The continuum is blue and, at this signal-to-noise ratio,
featureless. This candidate might be a SN~II, since the pre-maximum
spectra of SNe~II are generally featureless and blue; however, without
clear features in the continuum, we cannot assign a classification.}
\label{fig:FC_C02-030}
\end{figure}               

\clearpage

\begin{figure}
\centering
\resizebox{\hsize}{!}{\includegraphics{1504a20F.ps2}}
\resizebox{\hsize}{!}{\includegraphics{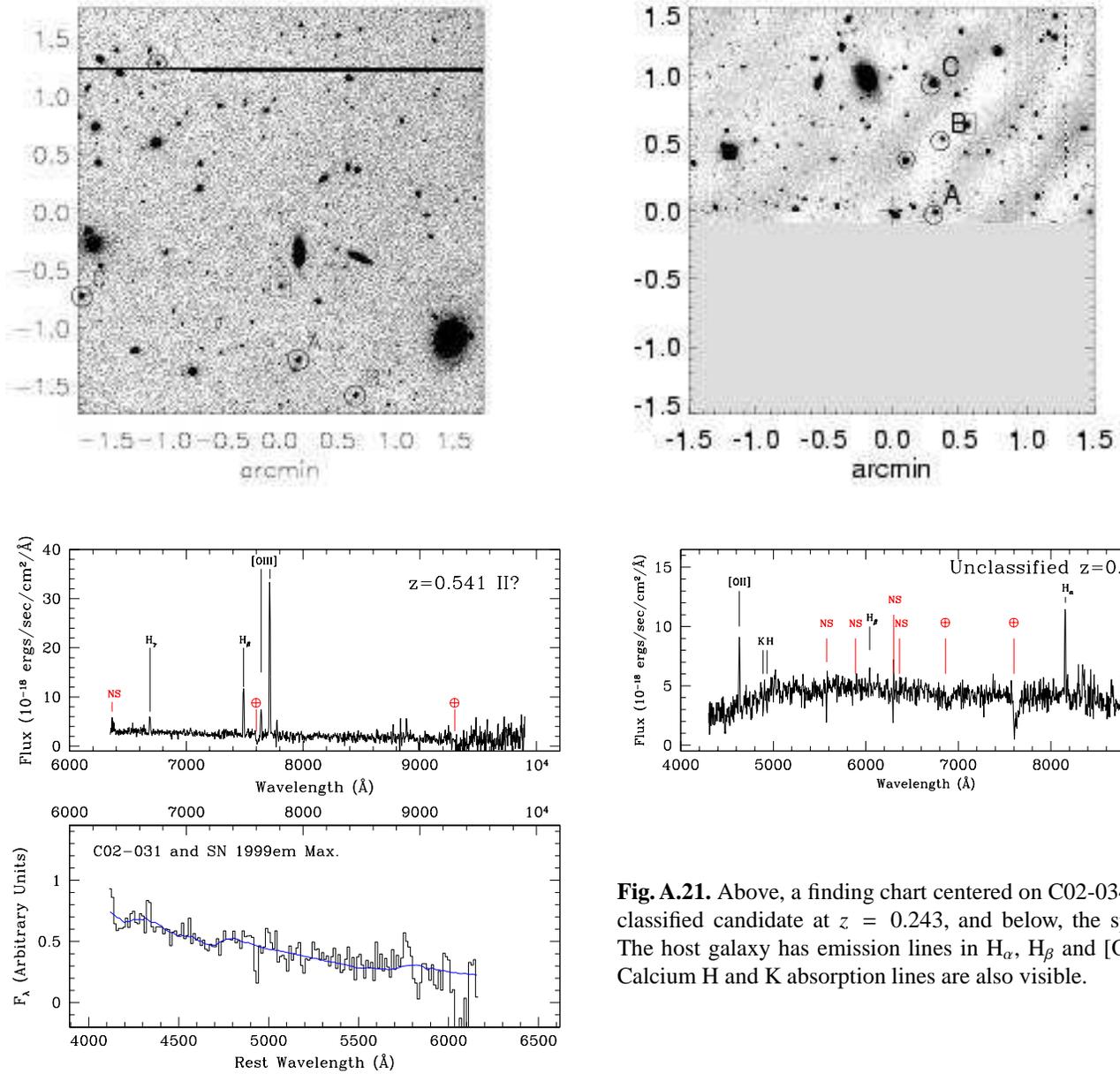}}
\caption{Above, a finding chart centered on C02-031, a possible Type
II supernova at $z=0.541$ and below, the spectrum. The host galaxy has
emission lines in [OIII], ${\rm H}_\beta$ and ${\rm H}_\gamma$.  The
tentative classification is based on the blue continuum and a weak
H-beta line with a P-Cygni profile. }
\label{fig:FC_C02-031}
\end{figure}               

\begin{figure}
\centering
\resizebox{\hsize}{!}{\includegraphics{1504a21F.ps2}}
\resizebox{\hsize}{!}{\includegraphics{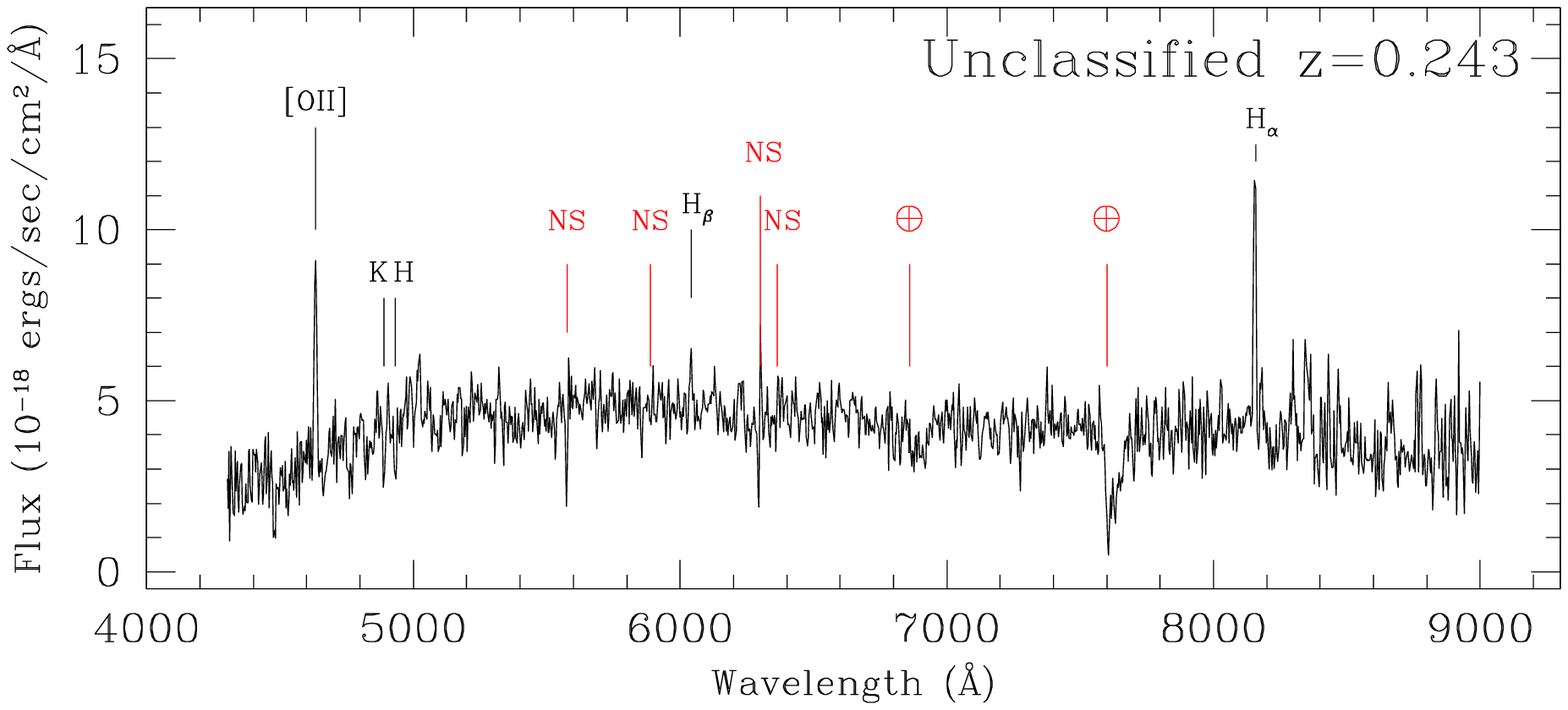}}
\vspace{-4cm}
\caption{Above, a finding chart centered on C02-034, an unclassified
candidate at $z=0.243$, and below, the spectrum. The host galaxy has
emission lines in ${\rm H}_\alpha$, ${\rm H}_\beta$ and [OII]. The
Calcium H and K absorption lines are also visible.}
\label{fig:FC_C02-034}
\end{figure}               

\clearpage

\begin{figure}
\centering
\resizebox{\hsize}{!}{\includegraphics{1504a22F.ps2}}
\resizebox{\hsize}{!}{\includegraphics{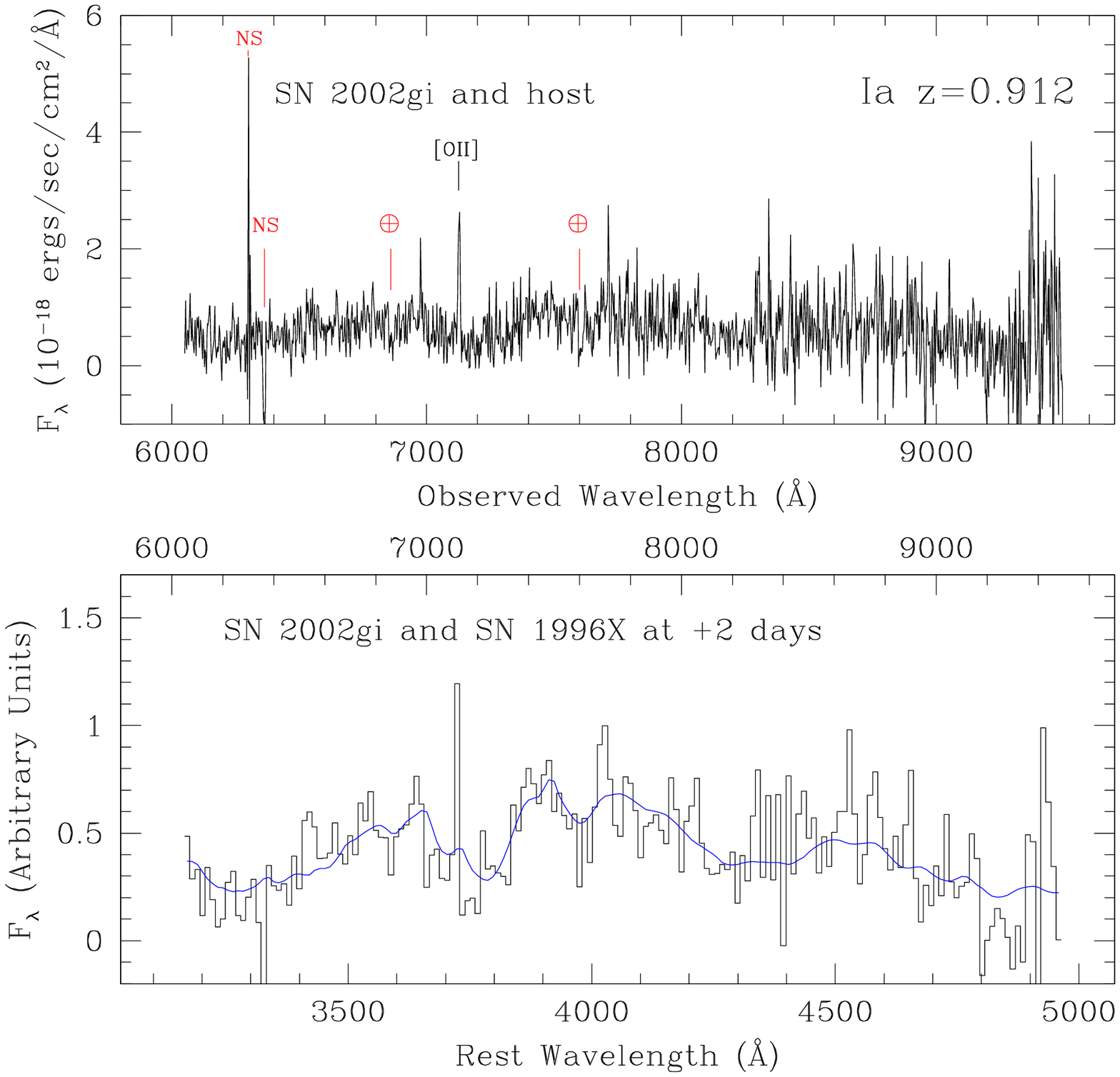}}
\caption{Above, a finding chart centered on \object{SN 2002gi}
(T02-015), a SN~Ia at $z=0.912$, and below, the spectrum. The Si~II
4000~\AA\ feature is clearly detected in this high redshift
candidate. This SN~Ia has the highest redshift of all securely
classified SNe~Ia that were observed with FORS1.}
\label{fig:FC_T02-015}
\end{figure}               

\begin{figure}
\centering 
\resizebox{\hsize}{!}{\includegraphics{1504a23F.ps2}}
\resizebox{\hsize}{!}{\includegraphics{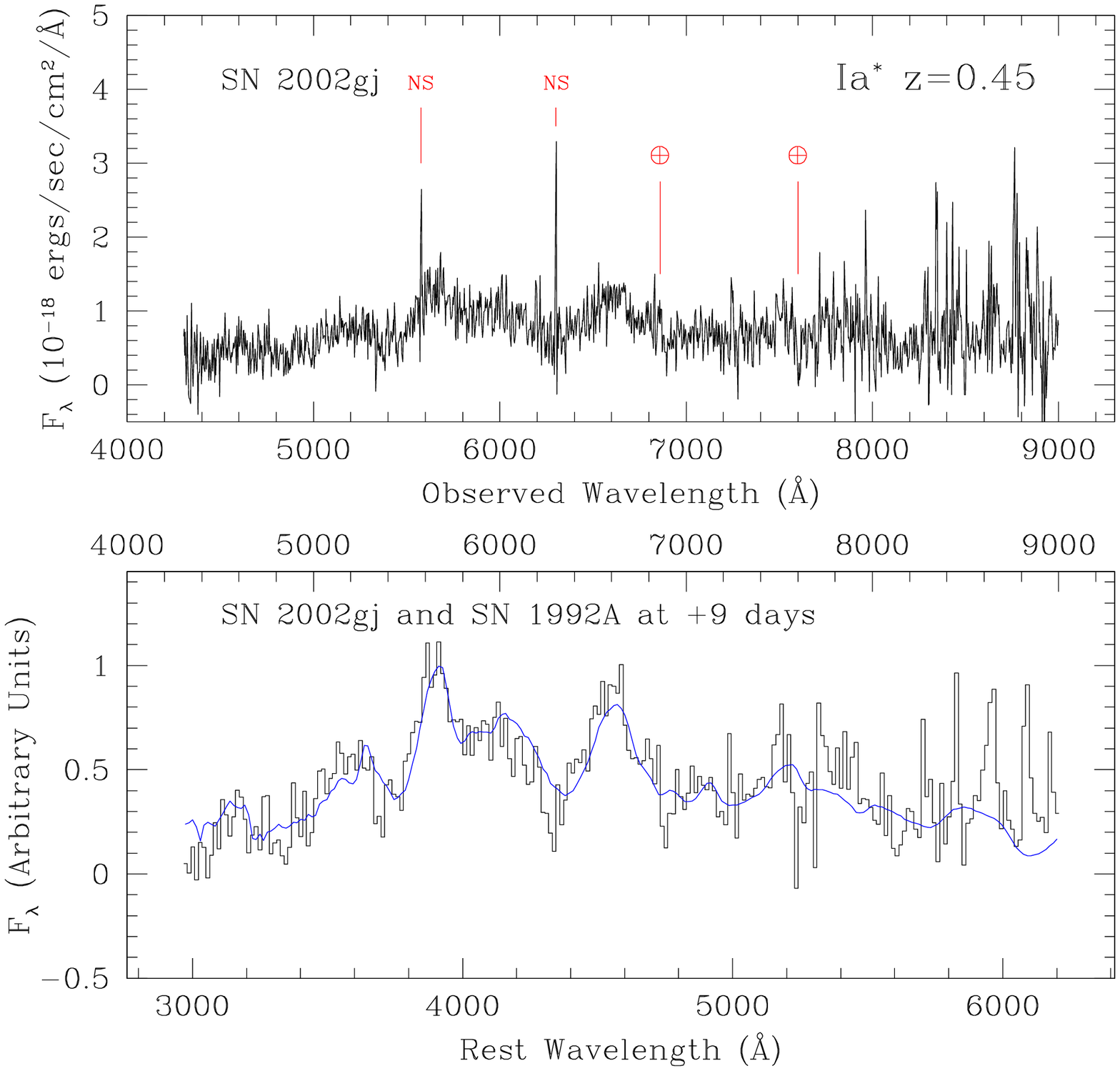}}
\caption{Above, a finding chart centered on \object{SN 2002gj}
(T02-028), a SN~Ia$^*$ at $z=0.45$, and below, the spectrum. From the
spectrum alone, this candidate can be matched with either a SN~Ia at
10 days after maximum light or with a SN~Ic near maximum light, so
the classification is qualified with an asterisk. The time of maximum
that is derived from the light curve shows that the spectrum was taken
about 10 rest frame days after maximum light, so the the candidate is
very likely to be a SN~Ia.}
\label{fig:FC_T02-028}
\end{figure}               

\clearpage

\begin{figure}
\centering
\resizebox{\hsize}{!}{\includegraphics{1504a24F.ps2}}
\resizebox{\hsize}{!}{\includegraphics{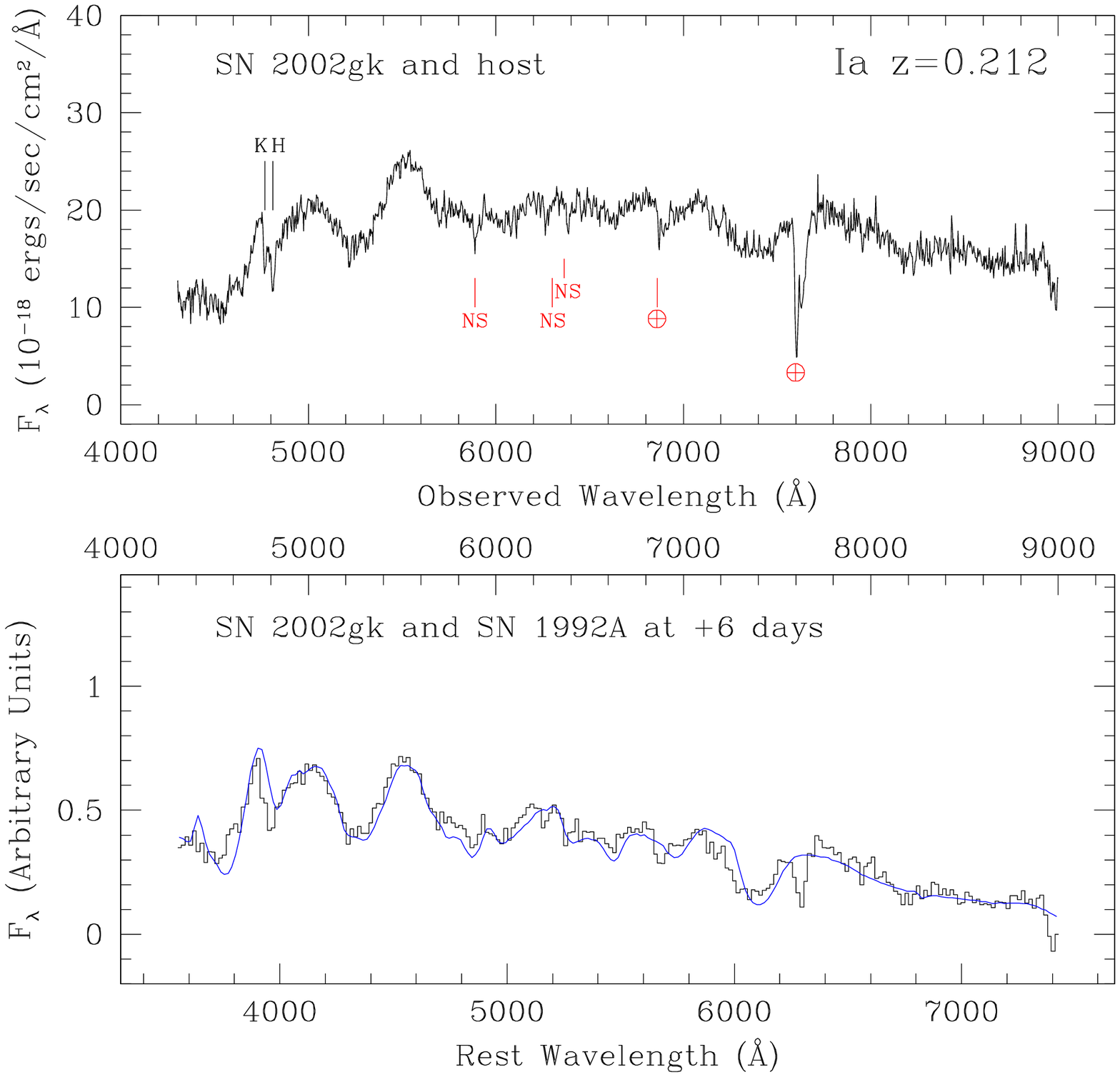}}
\caption{Above, a finding chart centered on \object{SN 2002gk}
(T02-029), a SN~Ia at $z=0.212$, and below, the spectrum. This SN~Ia
has the lowest redshift and the spectrum has the highest
signal-to-noise ratio of all candidates. Si~II at 4000~\AA\ and 6150
~\AA\ and S~II at 5400~\AA\ are all clearly detected.}
\label{fig:FC_T02-029}
\end{figure}               

\begin{figure}
\centering
\resizebox{\hsize}{!}{\includegraphics{1504a25F.ps2}}
\resizebox{\hsize}{!}{\includegraphics{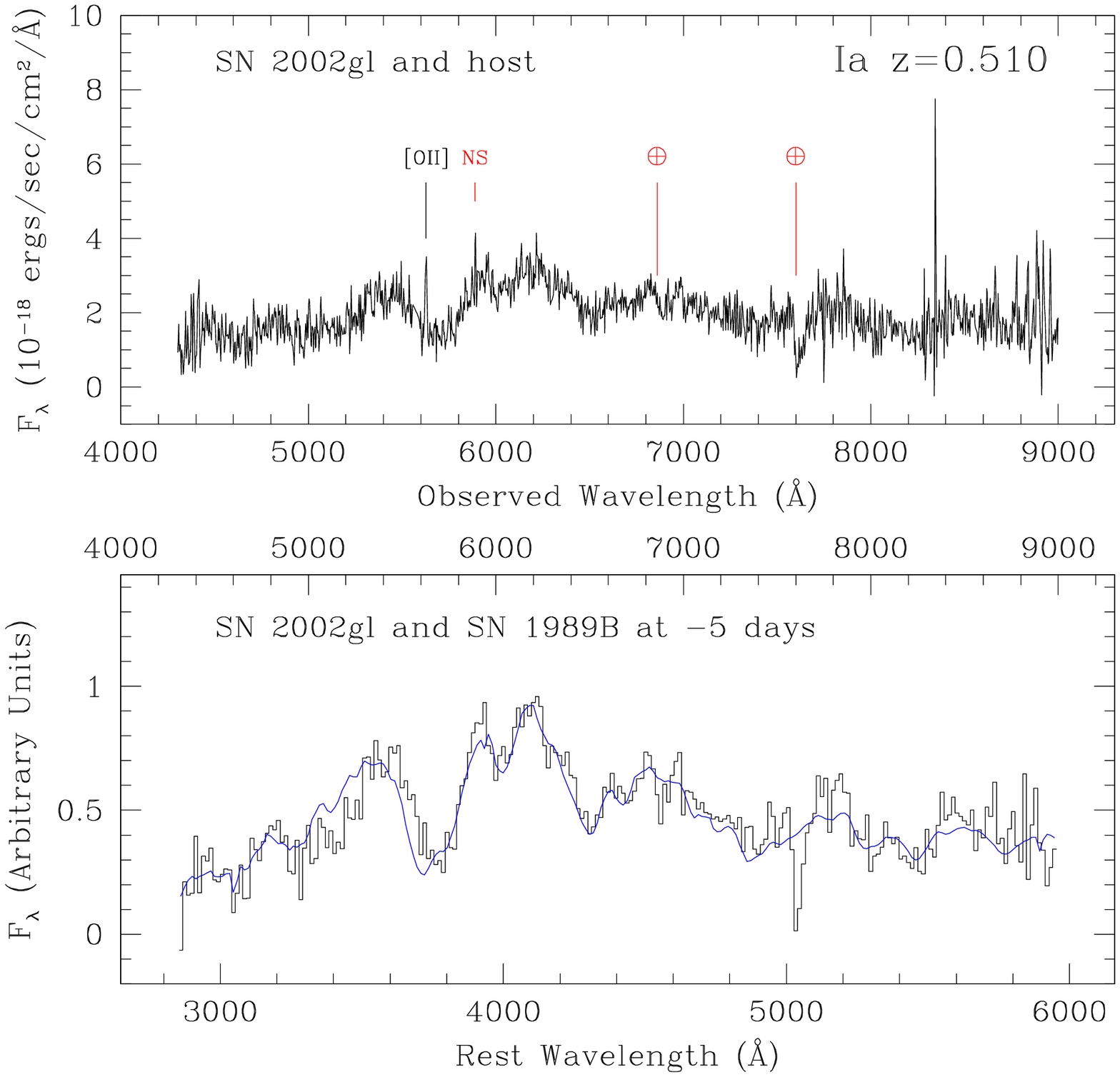}}
\caption{Above, a finding chart centered on \object{SN 2002gl}
(T02-030), a SN~Ia at $z=0.510$, and below, the spectrum. Si~II at
4000~\AA\ and S~II at 5400~\AA\ are clearly detected in this candidate.}
\label{fig:FC_T02-030}
\end{figure}               

\clearpage

\begin{figure}
\centering
\resizebox{\hsize}{!}{\includegraphics{1504a26F.ps2}}
\resizebox{\hsize}{!}{\includegraphics{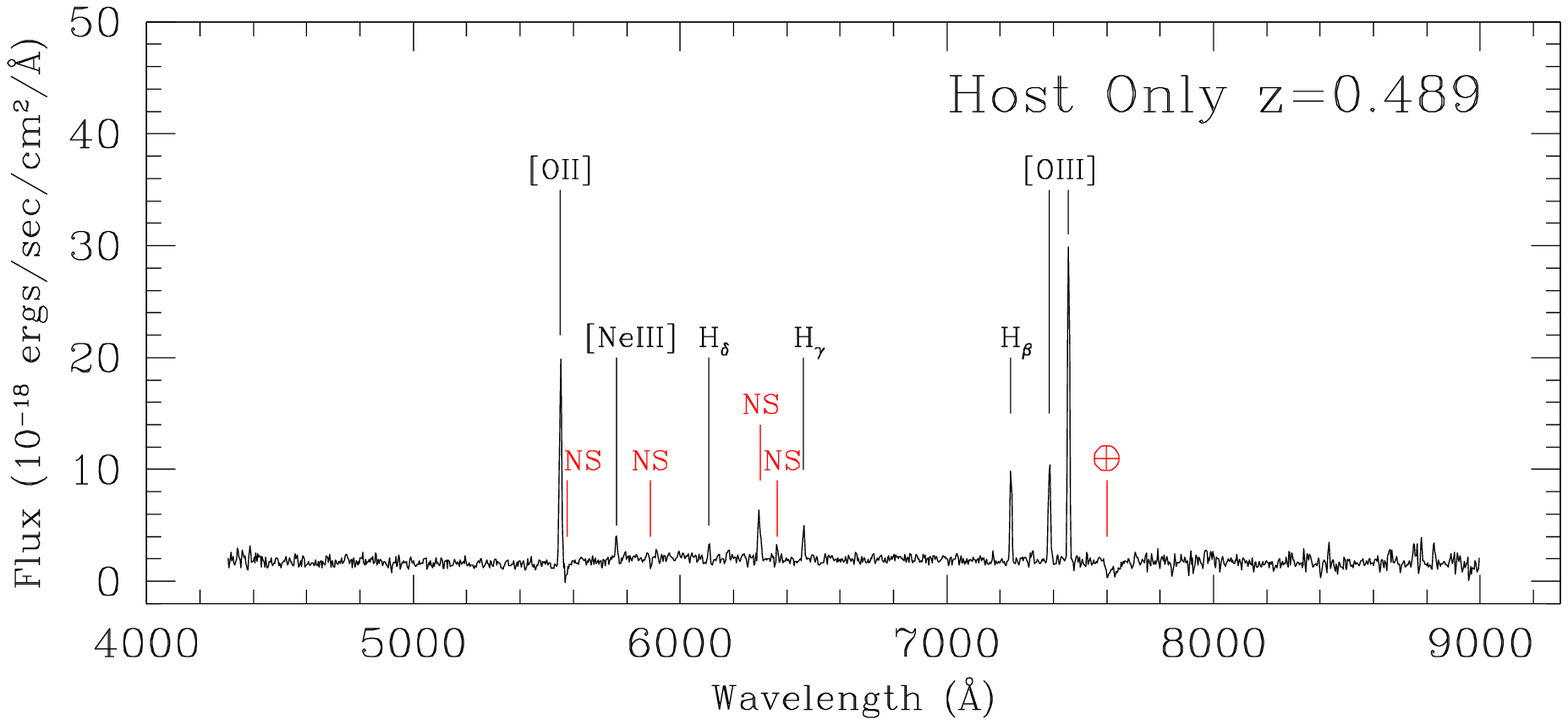}}
\vspace{-4cm}
\caption{Above, a finding chart centered on T02-047, a probable
supernova at $z=0.489$, and below, the spectrum. The spectrum of the
host was taken a couple of months after the candidate had faded and is
rich in emission lines. Although a spectrum of the candidate was not
obtained, the well-sampled light curve indicates that it is probably a
supernova.}
\label{fig:FC_T02-047}
\end{figure}               

\begin{figure}
\centering
\resizebox{\hsize}{!}{\includegraphics{1504a27F.ps2}}
\resizebox{\hsize}{!}{\includegraphics{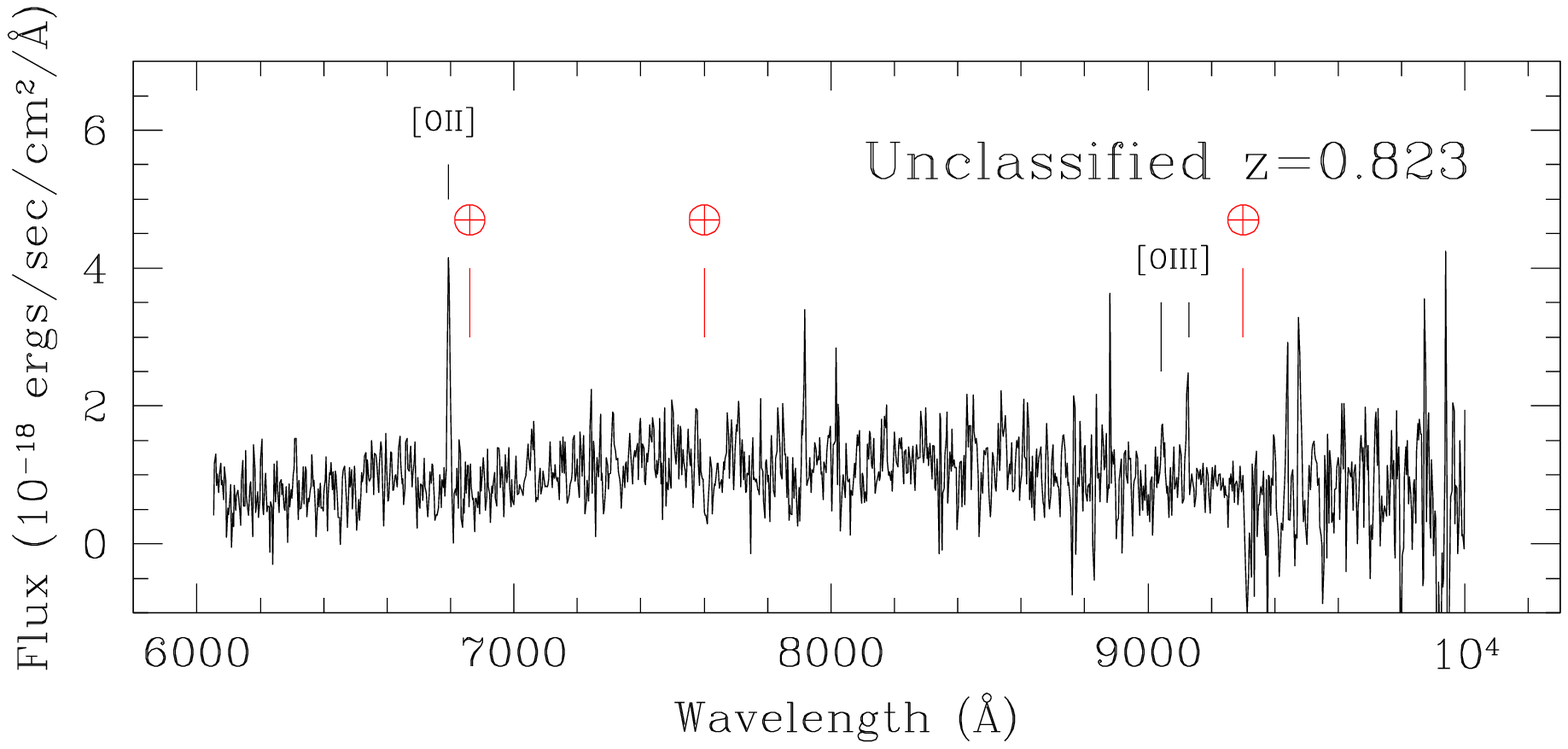}}
\vspace{-4cm}
\caption{Above, a finding chart centered on \object{SN 2002kq}
(SuF02-002), an unclassified candidate at $z=0.823$, and below, the
spectrum. This candidate was photometrically monitored and it has a
supernova-like light curve.}
\label{fig:FC_SuF02-002}
\end{figure}               

\clearpage

\begin{figure}
\centering
\resizebox{\hsize}{!}{\includegraphics{1504a28F.ps2}}
\resizebox{\hsize}{!}{\includegraphics{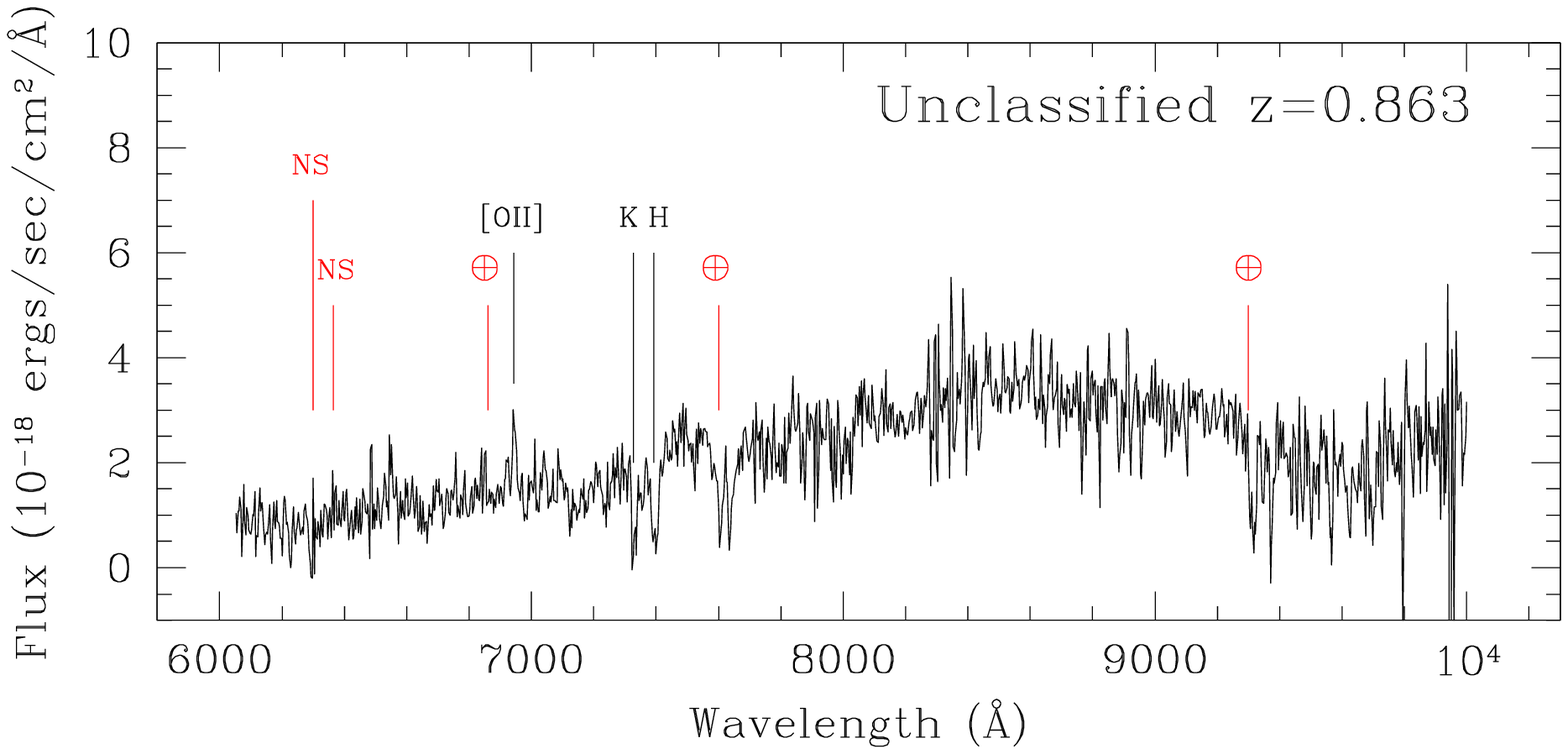}}
\vspace{-4cm}
\caption{Above, a finding chart centered on SuF02-005, an unclassified
candidate at $z=0.863$, and below, the spectrum. This candidate has an
extremely broad bump at 8500~\AA. Since we observed the pivot star
(object ``A'' in the finding chart) simultaneously with the candidate,
we can use the flux-calibrated spectrum of the pivot star to check the
calibration procedure. The flux-calibrated spectrum of star A does not
have the broad feature that can be seen in the candidate, so the broad
feature at 8500~\AA\ is real.}
\label{fig:FC_SuF02-005}
\end{figure}               

\begin{figure}
\centering
\resizebox{\hsize}{!}{\includegraphics{1504a29F.ps2}}
\resizebox{\hsize}{!}{\includegraphics{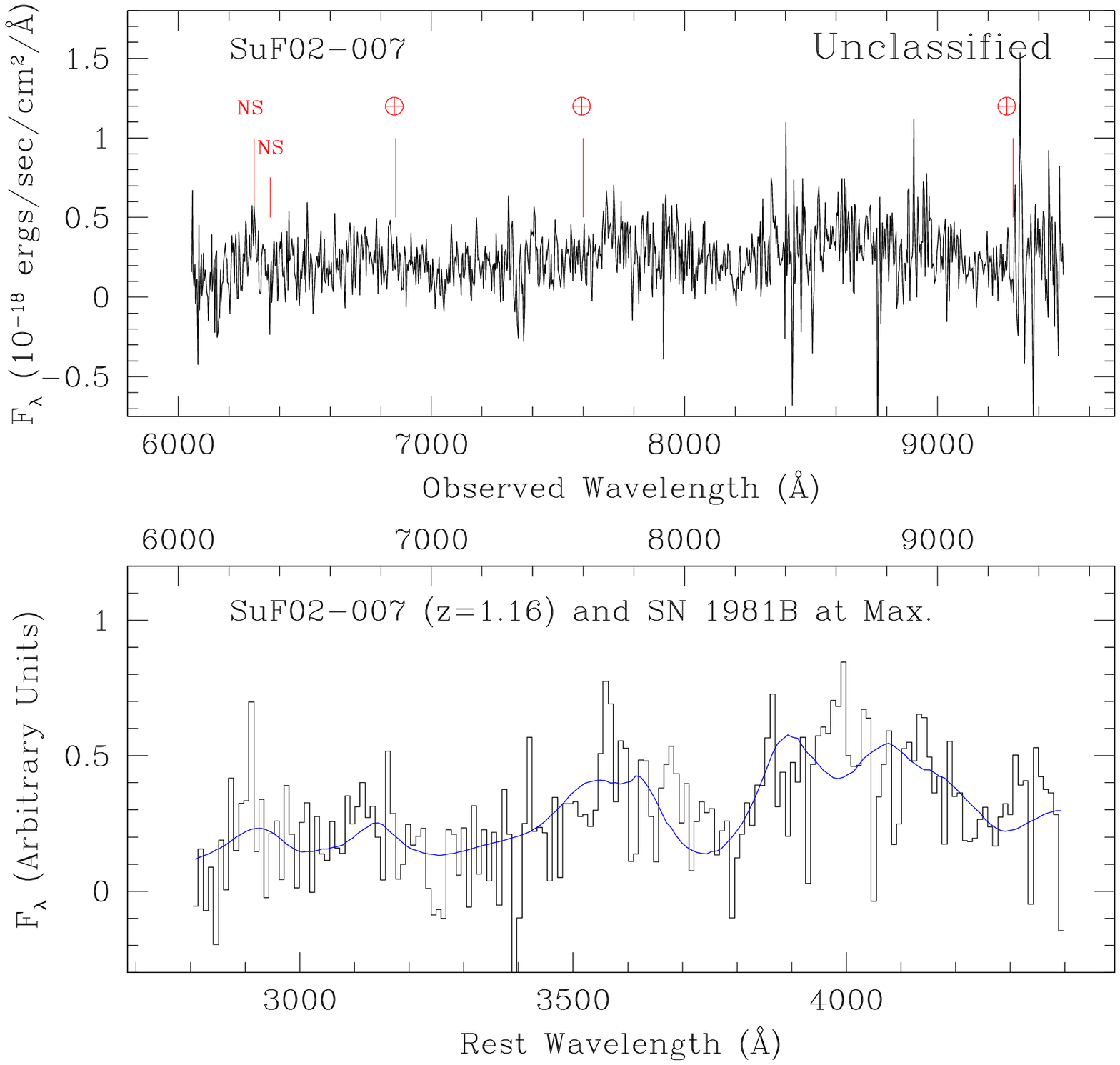}}
\caption{Above, a finding chart centered on SuF02-007 and below, the
spectrum. The binned  spectrum shows broad features that are
consistent with a SN~Ia at $z=1.16$; however, the signal-to-noise
ratio is too low for this candidate to be classified as a SN~Ia from
the  spectrum alone.  This candidate was photometrically
monitored and has a supernova-like light curve.}
\label{fig:FC_SuF02-007}
\end{figure}               

\clearpage

\begin{figure}
\centering
\resizebox{\hsize}{!}{\includegraphics{1504a30F.ps2}}
\resizebox{\hsize}{!}{\includegraphics{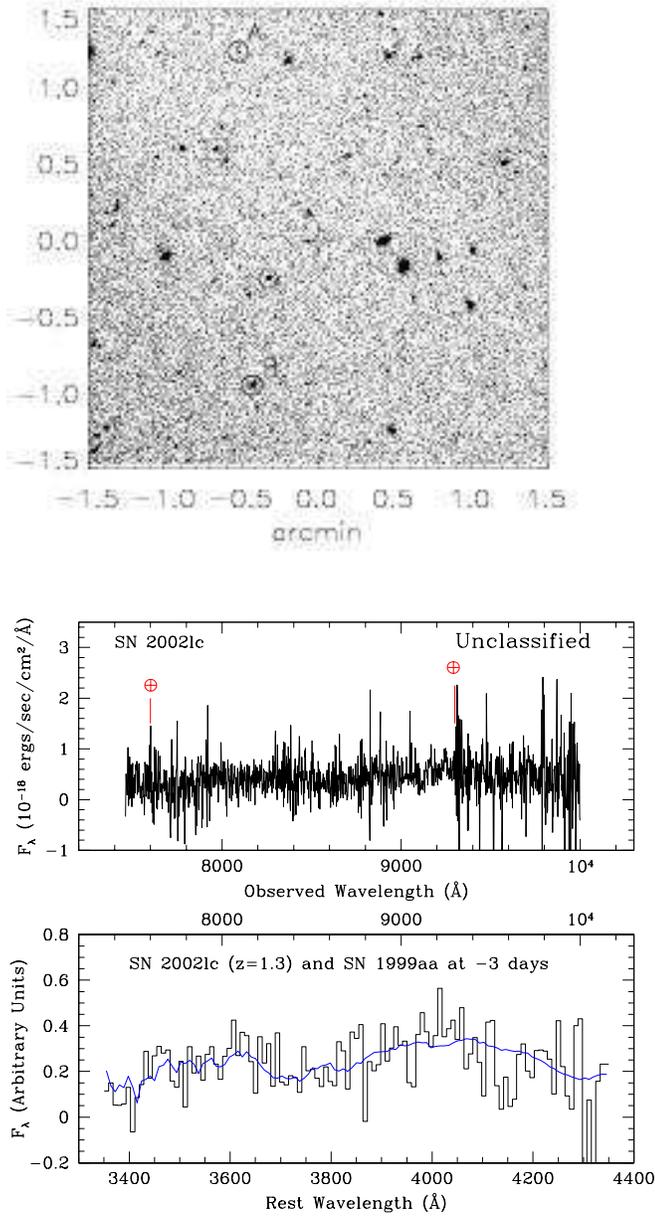}}
\caption{Above, a finding chart centered on \object{SN 2002lc}
(SuF02-012) and below, the spectrum. The binned  spectrum shows
broad SN~Ia like features which are consistent with a SN~Ia at
$z=1.3$, however the signal-to-noise ratio is too small for a spectral
classification. Hence, from the VLT spectrum alone it cannot be
classified. However, \object{SN 2002lc} was also observed with FOCAS
on Subaru, and the spectrum also shows similar broad features (Yasuda
et al. in preparation).  When added with the VLT data, a possible
match with a SN~Ia at $z=1.26$ emerges. Furthermore, a spectrum of
\object{SN 2002lc} was also taken with the ACS grism on HST. The
reduced ACS spectrum shows the same broad features as the ground-based
data. This candidate was photometrically monitored and has a
supernova-like light curve.}
\label{fig:FC_SuF02-012}
\end{figure}               

\begin{figure}
\centering 
\resizebox{\hsize}{!}{\includegraphics{1504a31F.ps2}}
\resizebox{\hsize}{!}{\includegraphics{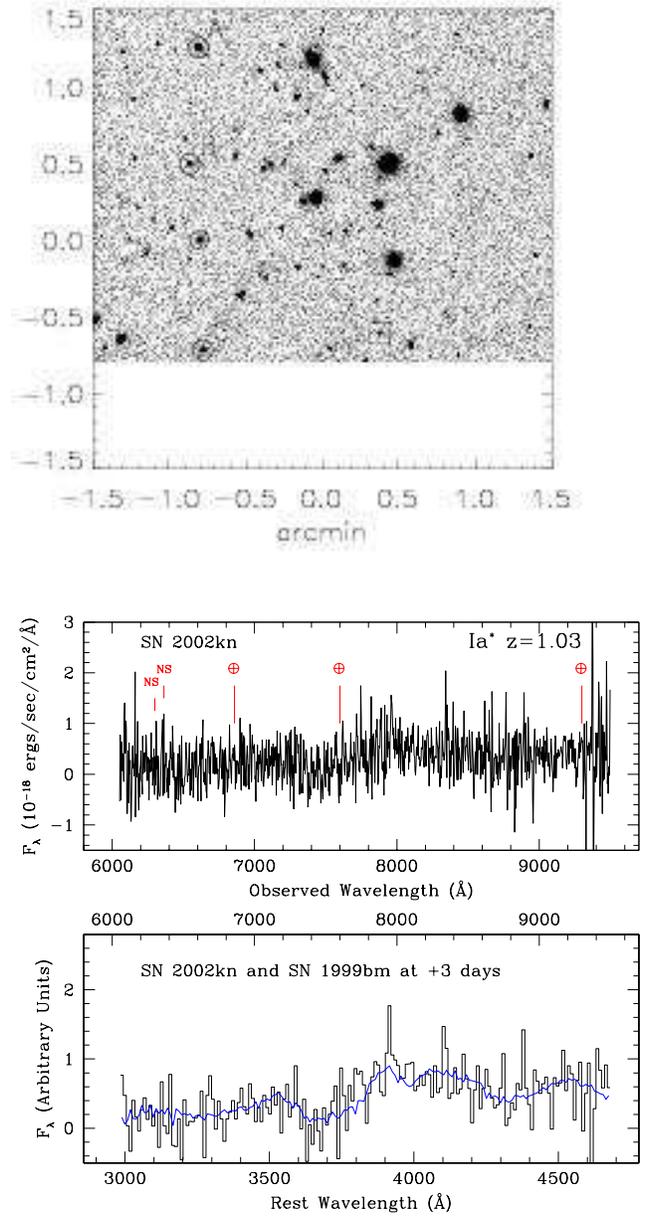}}
\caption{Above, a finding chart centered on \object{SN 2002kn}
(SuF02-017), a SN~Ia$^*$ at z=1.03, and below, the spectrum. The host
galaxy is approximately three magnitudes fainter than the candidate,
so the fraction of host galaxy light is set to zero in the fit. Since
the Si~II feature at 4000~\AA\ is not clearly detected in this
candidate and since there are no spectral features from the host, the
redshift and the classification are derived from the fit. The spectrum
can be fit equally well with a SN~Ic, so the classification is qualified
with an asterisk.}
\label{fig:FC_SuF02-017}
\end{figure}               

\clearpage

\begin{figure}
\centering
\resizebox{\hsize}{!}{\includegraphics{1504a32F.ps2}}
\resizebox{\hsize}{!}{\includegraphics{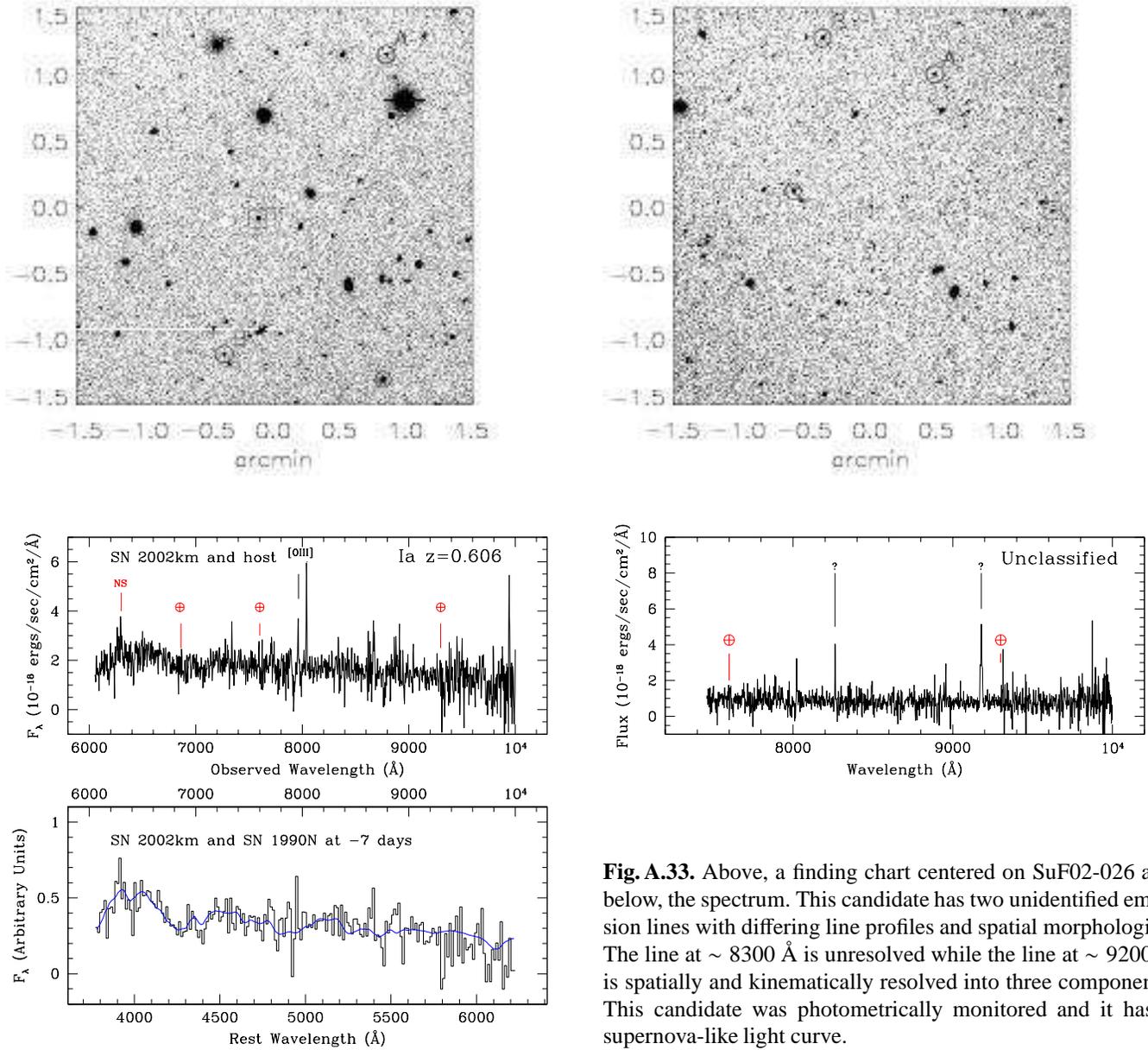}}
\caption{Above, a finding chart centered on \object{SN 2002km}
(SuF02-025), a SN~Ia at $z=0.606$, and below, the spectrum. The Si~II
line at 4000~\AA\ is clearly detected. There is a hint of the Si~II
line at 6150~\AA.}
\label{fig:FC_SuF02-025}
\end{figure}               

\begin{figure}
\centering
\resizebox{\hsize}{!}{\includegraphics{1504a33F.ps2}}
\resizebox{\hsize}{!}{\includegraphics{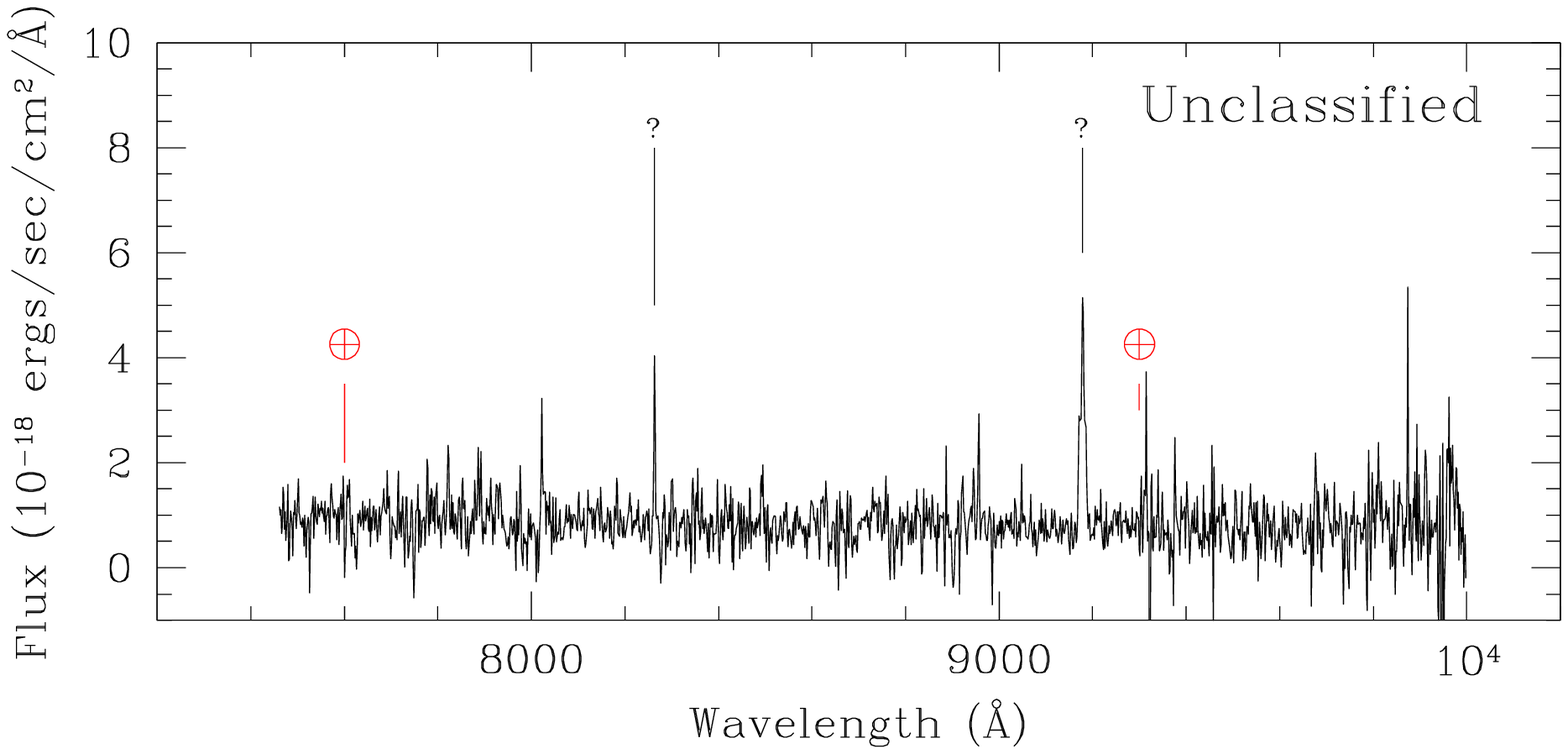}}
\vspace{-4cm}
\caption{Above, a finding chart centered on SuF02-026 and below, the
spectrum. This candidate has two unidentified emission lines with
differing line profiles and spatial morphologies. The line at $\sim
8300~\AA$ is unresolved while the line at $\sim 9200~\AA$ is spatially
and kinematically resolved into three components. This candidate was
photometrically monitored and it has a supernova-like light curve.}
\label{fig:FC_SuF02-026}
\end{figure}               

\clearpage

\begin{figure}
\centering
\resizebox{\hsize}{!}{\includegraphics{1504a34F.ps2}}
\resizebox{\hsize}{!}{\includegraphics{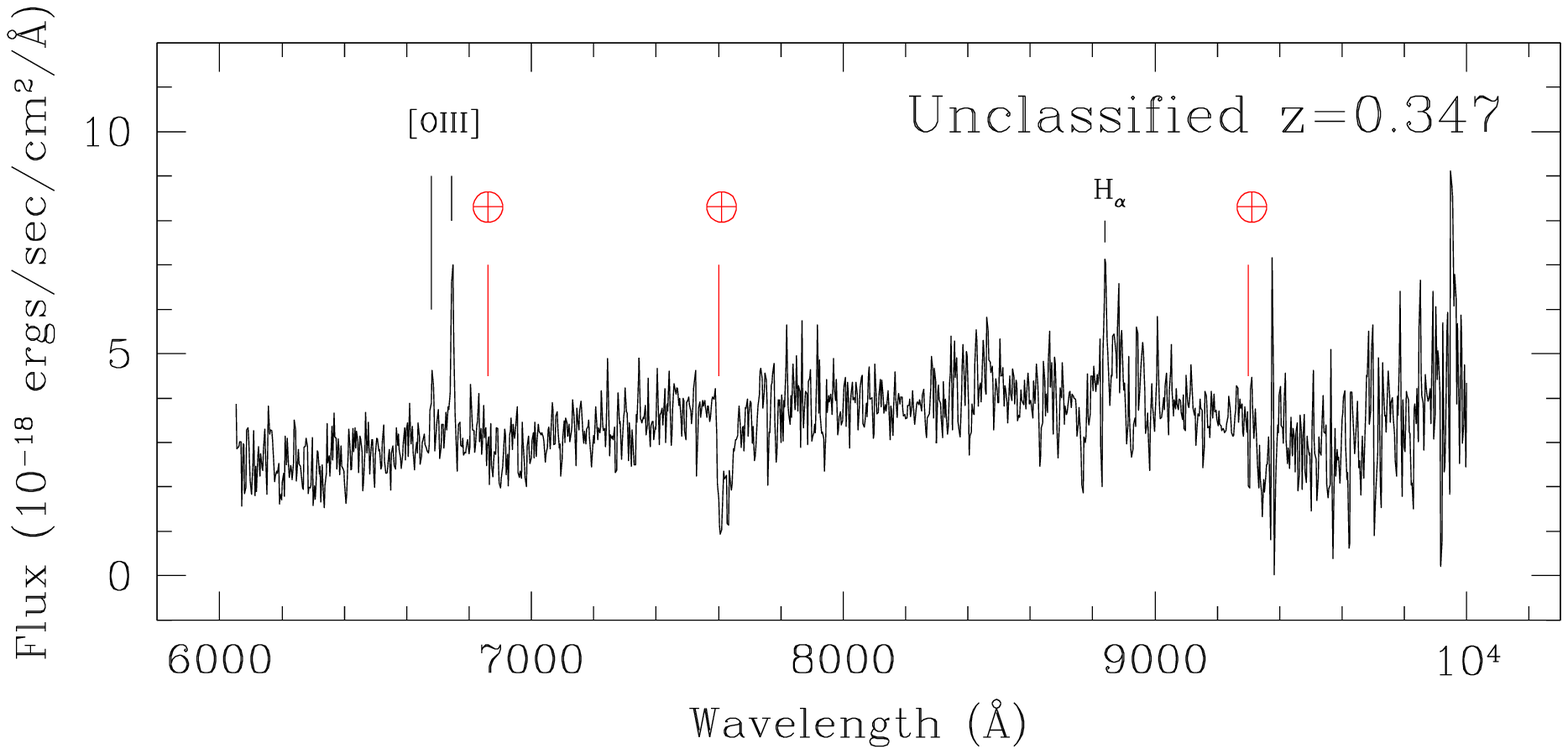}}
\vspace{-4cm}
\caption{Above, a finding chart centered on \object{SN 2002kn}
(SuF02-028), an unclassified candidate at $z=0.347$, and below, the
spectrum. This candidate is dominated by host galaxy light. It was
photometrically monitored and it has a supernova-like light curve.}
\label{fig:FC_SuF02-028}
\end{figure}               

\begin{figure}
\centering
\resizebox{\hsize}{!}{\includegraphics{1504a35F.ps2}}
\resizebox{\hsize}{!}{\includegraphics{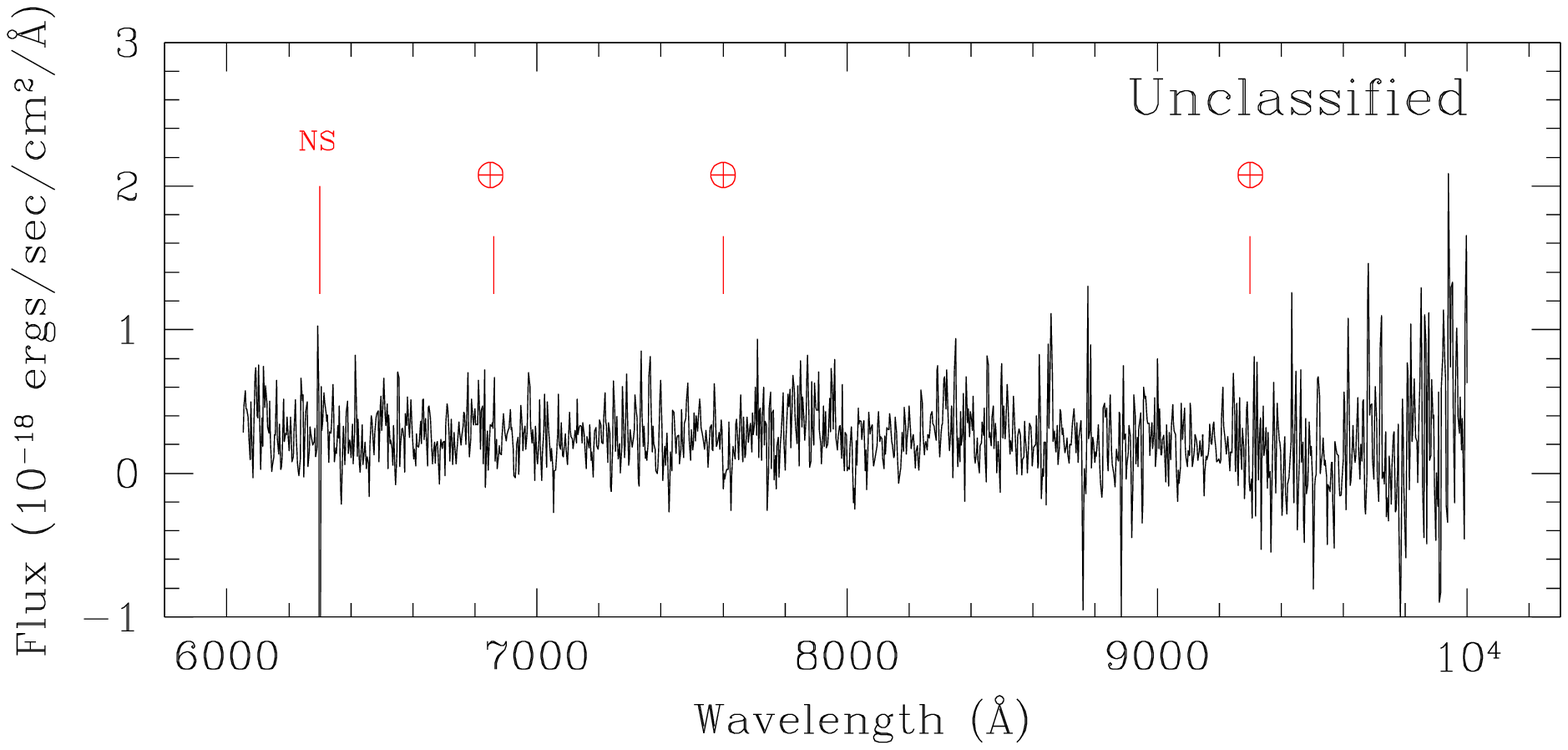}}
\vspace{-4cm}
\caption{Above, a finding chart centered on SuF02-051, an unclassified
candidate at an unknown redshift, and below, the spectrum. The
spectrum is a featureless, slightly blue continuum. This candidate was
photometrically monitored and it has a supernova-like light curve.}
\label{fig:FC_SuF02-051}
\end{figure}               

\clearpage

\begin{figure}
\centering
\resizebox{\hsize}{!}{\includegraphics{1504a36F.ps2}}
\resizebox{\hsize}{!}{\includegraphics{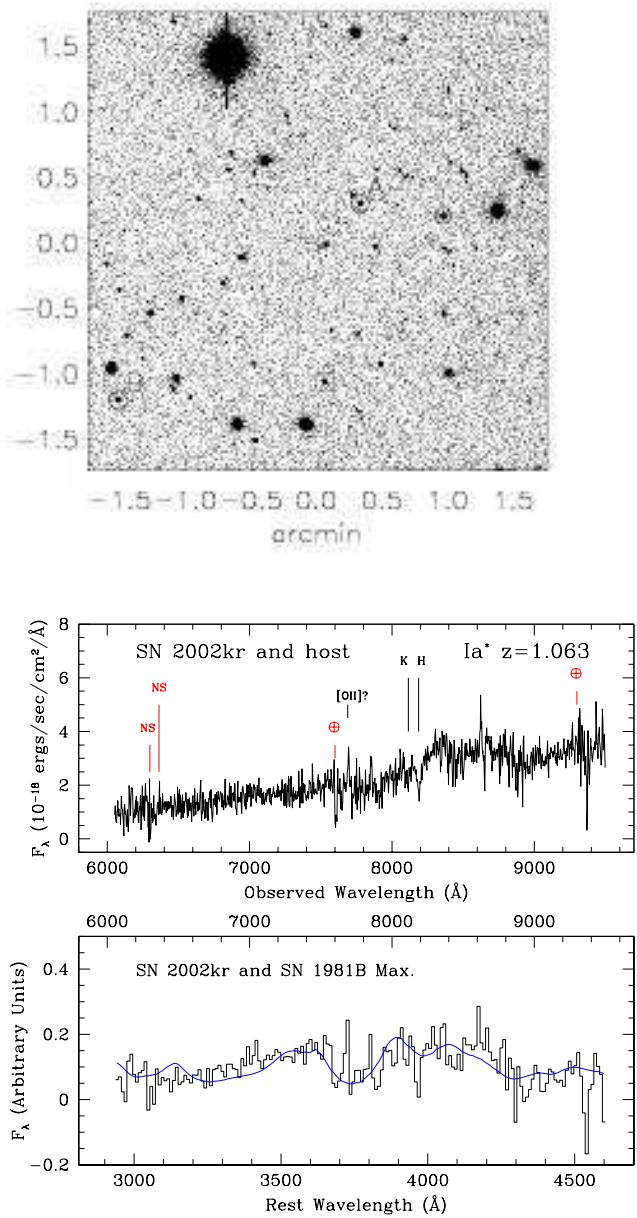}}
\caption{Above, a finding chart centered on \object{SN 2002kr}
(SuF02-060), a SN~Ia$^*$ at $z=1.063$, and below, the spectrum. The
redshift is determined from H and K Ca~II absorption lines in the
host. There is a hint of [OII] emission, but this is uncertain as the
[OII] line at this redshift lies very close to atmospheric A band. The
percentage increase was only 25\%, so the spectrum is dominated by the
host, which means that the host subtracted spectrum is sensitive to
the host spectrum used in the fit. Hence, the classification is
qualified with an asterisk. \object{SN 2002kr} was also observed with
the ACS grism on HST and the GMOS spectrograph on Gemini. Both the
Gemini and ACS show the same broad features as the VLT data.}
\label{fig:FC_SuF02-060}
\end{figure}               

\begin{figure}
\centering
\resizebox{\hsize}{!}{\includegraphics{1504a37F.ps2}}
\resizebox{\hsize}{!}{\includegraphics{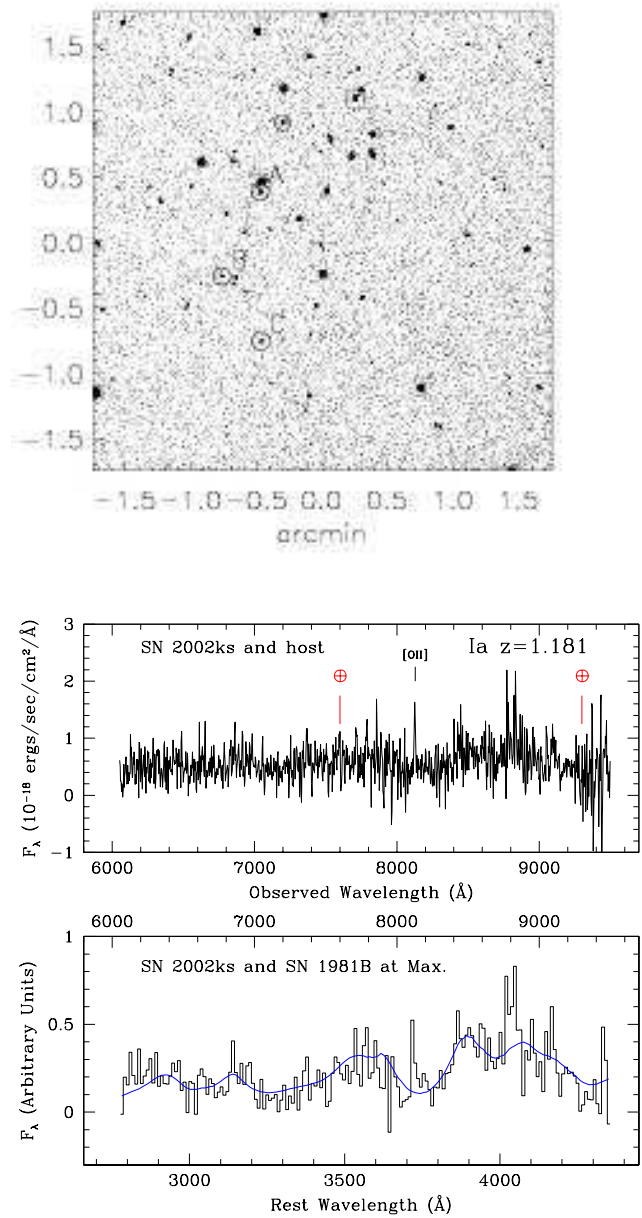}}
\caption{Above, a finding chart centered on \object{SN 2002ks}
(SuF02-065), a SN~Ia at $z=1.181$, and below, the spectrum. The Si~II
feature at 4000~\AA\  is clearly detected.  This SN~Ia has the highest
redshift of all securely classified SNe~Ia that were observed with the
ESO VLT.}
\label{fig:FC_SuF02-065}
\end{figure}               

\clearpage

\begin{figure}
\centering 
\resizebox{\hsize}{!}{\includegraphics{1504a38F.ps2}}
\resizebox{\hsize}{!}{\includegraphics{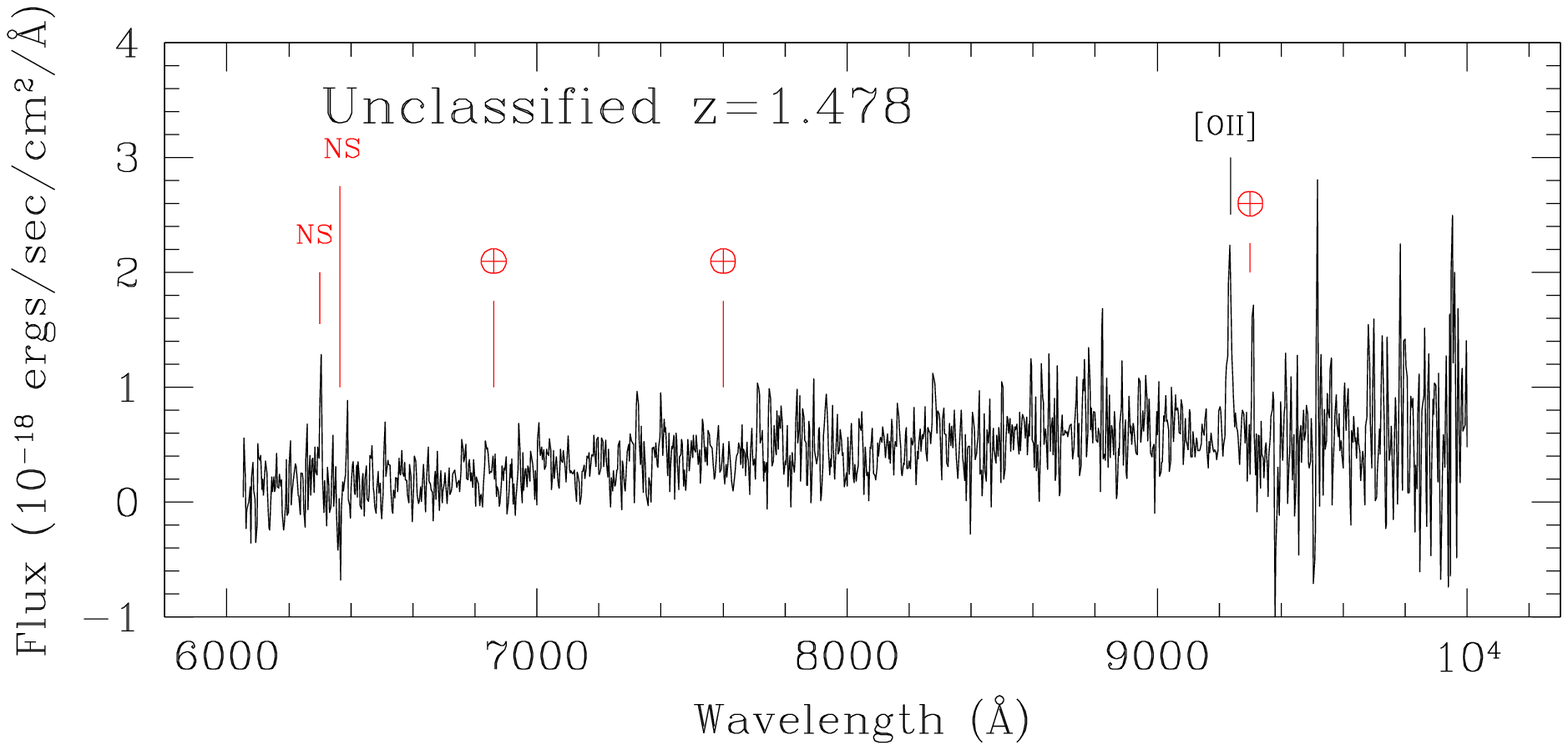}}
\vspace{-4cm}
\caption{Above, a finding chart centered on SuF02-081, an unclassified
candidate at $z=1.478$ and below, the spectrum. A single strong line
and a featureless red continuum. Like S02-001 and \object{SN 2003kx}
we identify this line as the [OII] doublet at 3727~\AA. This candidate
was photometrically monitored and it has a light curve that is too
narrow for it to be a SN~Ia.}
\label{fig:FC_SuF02-081}
\end{figure}               

\begin{figure}
\centering
\resizebox{\hsize}{!}{\includegraphics{1504a39F.ps2}}
\resizebox{\hsize}{!}{\includegraphics{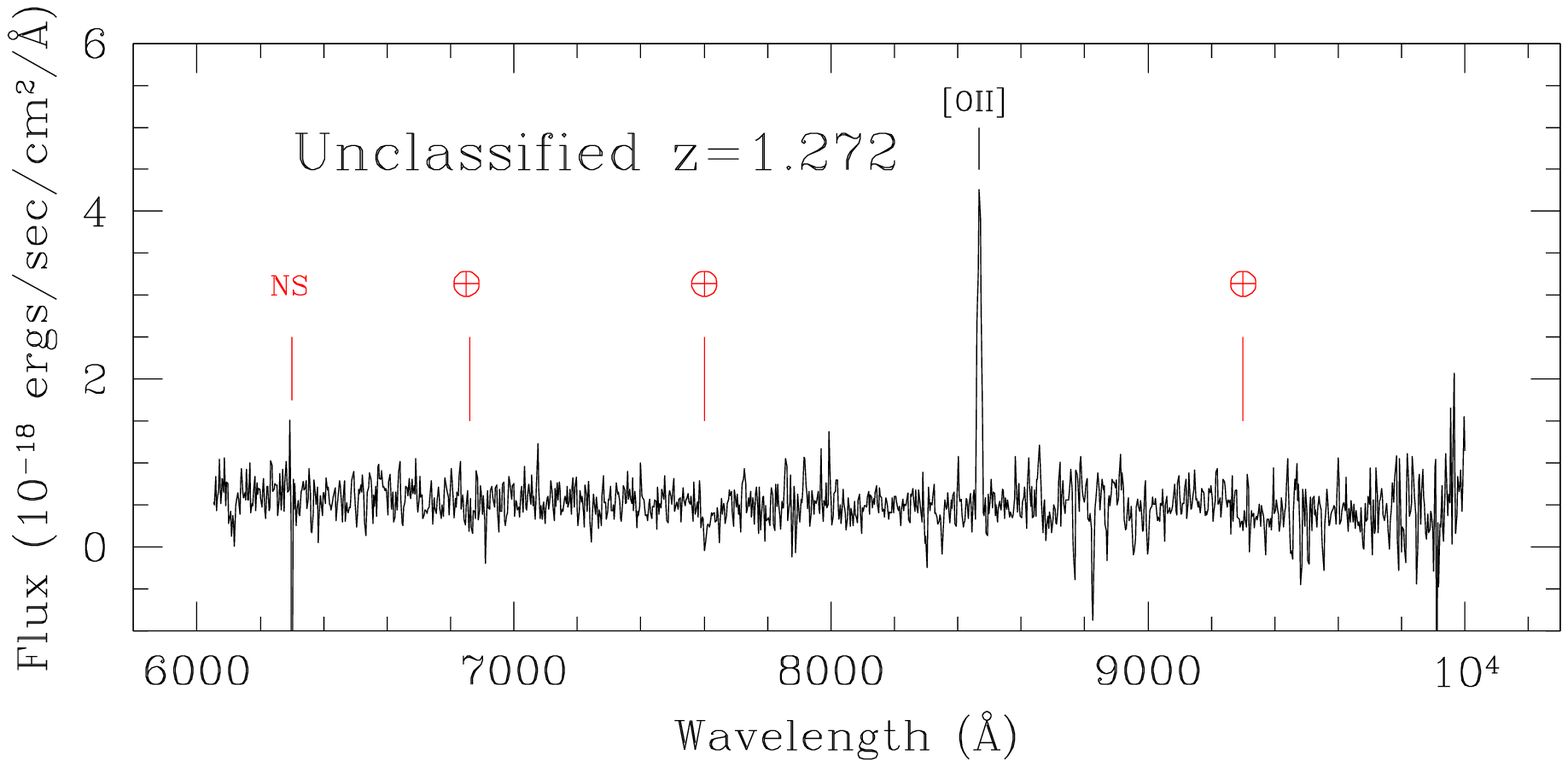}}
\vspace{-4cm}
\caption{Above, a finding chart centered on \object{SN 2003kx}
(SuF02-083), an unclassified candidate at $z=1.272$, and below, the
spectrum. A single strong line and a featureless continuum. Like
S02-001 and SuF02-083, we identify this line as the [OII] doublet at
3727~\AA.This candidate was photometrically monitored and it has a
supernova-like light curve.}
\label{fig:FC_SuF02-083}
\end{figure}               


\begin{thebibliography}{}

\bibitem[1998]{Aldering} Aldering, G. 1998, IAUC 7046

\bibitem[2002]{Allen02} Allen, A. W., Schmidt, R. W., \& Fabian, A. C. 2002, \mnras, 334, L11

\bibitem[1998]{Appenzeller98} Appenzeller, I., Fricke, K., F\"{u}rtig, W., et al. 1998, The Messenger, 94, 1

\bibitem[2004]{Barris04} Barris, B. J., Tonry, J. L., Blondin, S., .et al. 2004, \apj, 602, 571

\bibitem[2002]{Bennett03} Bennett, C. L., Halpen, M., Hinshaw, G., et al. 2003, \apjs, 148, 1

\bibitem[2001]{Borgani01} Borgani, S., Rosati, P., Tozzi, P., et al. 2001, \apj, 561, 13


\bibitem[1998]{Garnavich98} Garnavich, P. M., Kirshner, R. P., Challis, P., et al. 1998, \apjl, 493, L53

\bibitem[1995]{Goobar95} Goobar, A. \& Perlmutter, S. 1995, \apj, 450, 14

\bibitem[2003]{Hawkins02} Hawkins, E., Maddox, S., Cole, S., et al. 2003, \mnras, 346, 78


 \bibitem[2002]{Howell02} Howell, D. A. \& Wang, L.  2002, AAS, 201, 9103.

\bibitem[2001]{Jaffe01} Jaffe, A. H.. Ade, P., Balbi, A., et al. 2001, \prl, 86, 3475

 \bibitem[2004]{Kniazev04} Kniazev, A. Y., Pustilnik, S. A. Grebel, E., et al. 2004, \apjs, 153, 429

\bibitem[2003]{Kodaira03} Kodaira,K. Taniguchi,Y. Kashikawa,N. et al. 2003, PASJ, 55L, 17

\bibitem[2003]{Knop03} Knop, R. A., Aldering, G. A., Amanalluh, R., et al. 2003, \apj, 598, 102

\bibitem[1995]{Perlmutter95} Perlmutter, S., Pennypacker, C. R., Goldhaber, G., et al. 1995, \apjl, 440, L41

\bibitem[1997]{Perlmutter97} Perlmutter, S., Gabi, S., Goldhaber, G., et al. 1997, \apj, 483, 565

\bibitem[1998]{Perlmutter98} Perlmutter, S., Aldering, G., della Valle, M., et al. 1998, \nat, 391, 51

\bibitem[1999]{Perlmutter99} Perlmutter, S., Aldering, G., Goldhaber, G. et al. 1999, \apj, 517, 565

 \bibitem[2003]{Perlmutter03} Perlmutter, S. and Schmidt, B. 2003, in ``Supernovae \& Gamma Ray Bursts'', ed. Weller, K. (Springer)

\bibitem[1998]{Riess98} Riess, A. G., Filippenko, A. V., Challis, P. et al. 1998, \aj, 116, 1009

\bibitem[2004]{Riess04} Riess, A. G., Strolger, L.-G., Tonry, J. et al. 2004, \apj, 607, 665.

\bibitem[2004]{Nobili04} Nobili, S. 2004, Ph. D. thesis, Stockholm Univ.

\bibitem[1998]{Schmidt98} Schmidt, B. P., Suntzeff, N. B., Phillips, M. M., et al. 1998, \apj, 507, 46

\bibitem[2003]{Spergel03} Spergel, D. N., Verde, L., Peiris, H. V. et al. 2003, \apjs, 148, 175

\bibitem[2000]{Stern00} Stern, D. et al. 2000, \apj, 537, 73

\bibitem[2003]{Tonry03} Tonry, J. L., Schmidt, B. P., Barris, B., et al. 2003, \apj,  594, 1

\end{thebibliography}
\end{document}